\documentclass[aps, prl, twocolumn, notitlepage]{revtex4-2}
\usepackage{xspace}
\usepackage{graphicx}
\usepackage{dcolumn}
\usepackage[usenames]{color}
\usepackage{slashed}
\usepackage[utf8]{inputenc}
\usepackage{amsfonts}
\usepackage{amsthm}
\usepackage{amsmath}
\usepackage{amssymb}
\usepackage{mathrsfs}
\usepackage{hyperref}
\usepackage{booktabs}
\usepackage{diagbox}
\usepackage{latexsym}
\usepackage{color}
\usepackage{bm}

\usepackage{float}
\makeatletter
\let\newfloat\newfloat@ltx
\makeatother
\usepackage{algorithm}
\usepackage{algorithmicx}
\usepackage{algpseudocode}

\begin{document}

\title{Single-shot quantum measurements sketch quantum many-body states}
\author{Jia-Bao Wang}
\affiliation{International Center for Quantum Materials, School of Physics, Peking University, Beijing, 100871, China}
\author{Yi Zhang}
\email{frankzhangyi@gmail.com}
\affiliation{International Center for Quantum Materials, School of Physics, Peking University, Beijing, 100871, China}

\begin{abstract}
Quantum measurements are our eyes to the quantum many-body systems consisting of a multitude of microscopic degrees of freedom. However, the quantum uncertainty and the exponentially large Hilbert space pose natural barriers to simple interpretations of the quantum measurement outcomes. We propose a nonlinear ``measurement energy" based upon the measurement outcomes and a general approach akin to quantum machine learning to extract the most probable states (maximum likelihood estimates), naturally reconciling non-commuting observables and getting more out of the quantum measurements. Compatible with established quantum many-body ansatzes and efficient optimization, our strategy offers \emph{state-of-art} capacity with \emph{control} and \emph{full information}. We showcase the versatility and accuracy of our strategy on random long-range fermion and Kitaev quantum spin liquid models, where smoking-gun signatures were lacking.
\end{abstract}

\maketitle

\emph{Introduction} \textemdash Quantum many-body systems exhibit fascinating yet elusive quantum phenomena, such as quantum fluctuations, strong correlations \cite{Quintanilla2009}, quantum entanglements \cite{KitaevTee, WenTee, Chen2010prb, FrankTEE, FrankCSL, smat, Grover_2013, smat2}, quantum anomalies \cite{Haldane1988, LaughlinFQH, QSHE2005, TI2010, NIELSEN1983389, Yuan2020}, with no counterpart in the macroscopic world \cite{Keimer2017}. For example, nontrivial spin and electronic systems like quantum spin liquid (QSL) \cite{Kitaev20032, Kitaev2006, White2010, Banerjee2016}, superconductors \cite{Bednorz1986, FeSC1, FeSC2, Cao2018}, topological phases \cite{QHE1980, TsuiFQH, LaughlinFQH, Haldane1983, QHE1986, Wen1999gsd, Haldane2004, QSHE2005, Bernevig2006, 3DTI, ChenSPT}, form a modern-day scientific cornerstone. While scientists have made much progress and established physical pictures that are simple and beautiful, it is common that we scratch our heads over their complex behaviors when encountering the vast and intertwined microscopic or emergent degrees of freedom \cite{KivelsonIntertwine, Taillefer2019}.

Experiments on quantum many-body systems are our window to their microscopic worlds. However, analysis of quantum measurements is intrinsically difficult due to quantum fluctuations that whenever a general observable $\hat O= \sum_{\tau } a_{\tau } \hat P_{\tau }$ is measured, the outcome stochastically picks one eigenvalue $a_{\tau }$ with probability $\langle \hat P_{\tau }\rangle$, where $\hat{P}_{\tau }$ is the projection operators corresponding to the eigenvalue $a_\tau$ \cite{Sakurai}. Fortunately, if we measure the target state repeatedly, through either identical copies or relaxation, the resulting average converges to a non-stochastic and more physically interpretable expectation value $\langle \hat O\rangle$ \cite{Sakurai}. We may further facilitate the investigation with a phenomenological picture or microscopic model, whose predictions offer smoking-gun signatures that we can compare with the quantum measurements. However, by presuming a model or picture, we not only waste seemingly unrelated data but also risk biases consciously or unconsciously. In addition, exotic quantum matters such as QSLs lack definitive signatures, compelling scientists to resort to a negative-evidence stance \cite{White2010, Han2012} that may remain controversial and less controlled to a degree.

\begin{figure}
    \centering
    \includegraphics[width=0.4\textwidth]{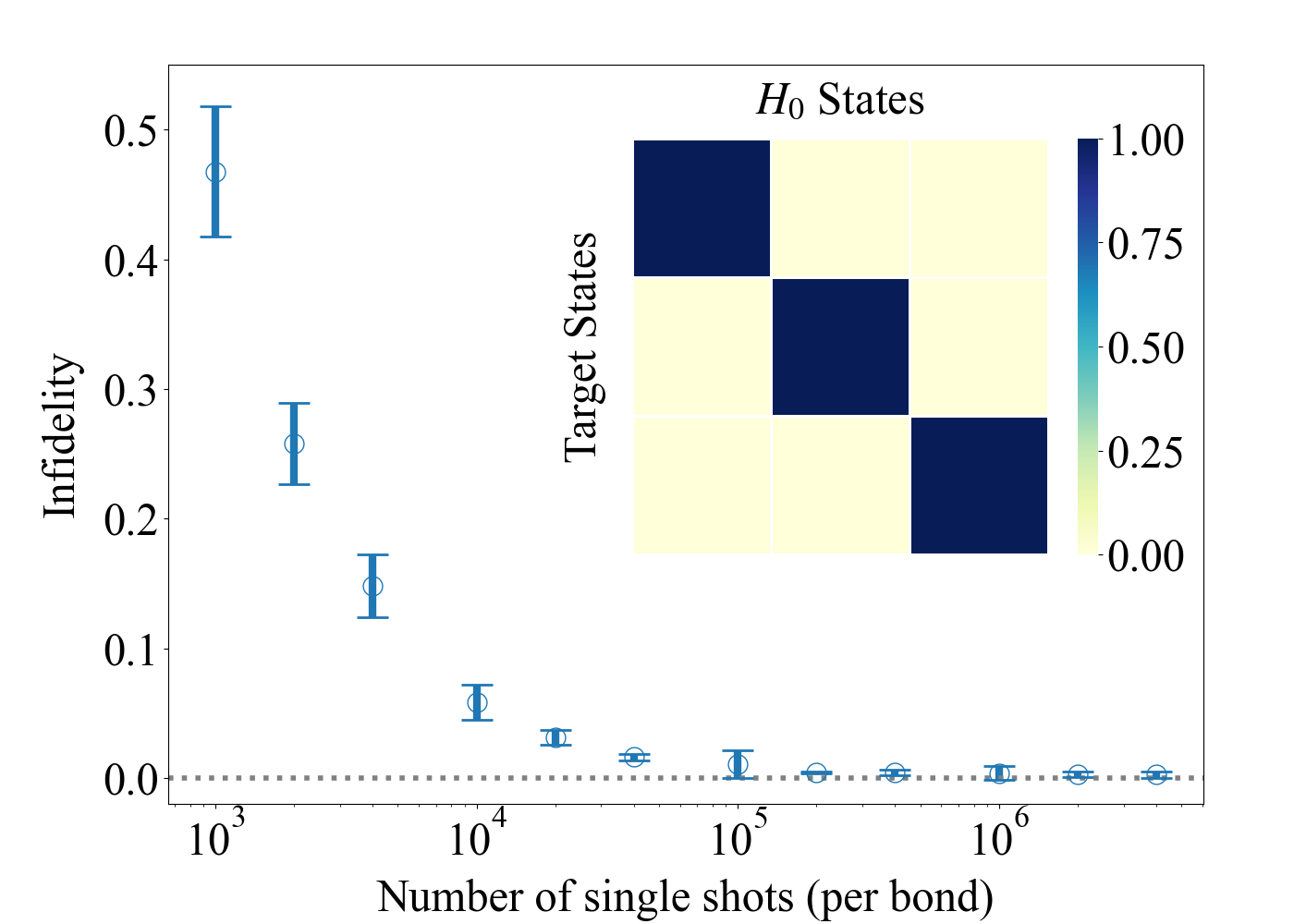}
    \caption{Single-shot quantum measurements $\sigma_i^{\hat n} \sigma_j^{\hat n}$ for random $\hat n$ directions yield a list of fluctuation-laden $\pm 1$. Based upon such single-shot outcomes from non-Abelian topologically ordered ground states with $K_{ij}=-1$, $J_{ij}\in [0,0.1]$, $\Gamma_{ij}\in[0,0.03]$ in the Kitaev model in Eq. \ref{eq:KitaevHam}, our strategy allows the MLE states to converge asymptotically well to the target states within 1500 iterations as the number of single shots increases. The error bars are based on ten different trials. The inset demonstrates the fidelity between the lowest three eigenstates of the iteration $\hat{H}_0$ at convergence and the topological degenerate states of the target system.}
    \label{fig:snap}
\end{figure}

In this letter, we discuss a general strategy to determine the most probable quantum many-body states given the quantum measurement data. We interpret the quantum measurements as nonlinear measurement energy and offer an iterative effective-Hamiltonian strategy to obtain the measurement outcomes' maximum likelihood estimate (MLE) states in the Hilbert space, which in turn, provide us with all information, including those unachievable directly, such as quantum entanglements \cite{KitaevTee, WenTee, Chen2010prb, FrankTEE, FrankCSL, smat, Grover_2013, smat2} and topological characters \cite{Haldane1988, LaughlinFQH, smat, Grover_2013, smat2}. In this way, we can utilize all measurement outcomes on a neutral and equal footing and remove the necessity of any presumed model or picture. We showcase the strategy's generality and effectiveness on random long-range fermion and Kitaev spin liquid models \cite{Kitaev2006}, which lack a smoking-gun signature for quantum measurements. Especially, our strategy can work wonders even for complex states such as the disordered Kitaev QSL even with only non-repeating single-shot quantum measurements, fully capturing its non-Abelian topological degeneracy (Fig. \ref{fig:snap}). Indeed, every single-shot quantum measurement matters, as its outcome carries information. On the other hand, quantum-state reconstruction through measurements, often named quantum state tomography, has been a long-standing topic in quantum physics \cite{QSTRMP2009}. The recent introduction of neural network quantum state tomography (NNQST) \cite{Torlai2018, Torlai2019} and shadow tomography \cite{Huang2020} have achieved practical efficiency over multiple qubits. In comparison, our strategy provides the full quantum states and even the topologically degenerate ground-state manifold, complementing shadow tomography, which estimates feasible physical quantities. Also, thanks to exceptional optimization efficiency \cite{Tianlun2022} and compatibility with various quantum many-body ansatzes, including the tensor network states and neural network states \cite{Carleo2016} (that NNQST based on), our strategy offers \emph{state-of-art} tomography capacity with \emph{control} and \emph{full information}. Further, unlike the previous tomography based on computational basis, our approach is more compatible with physical observables, applicable to a broader range of experiments.

\emph{The measurement energy} \textemdash Consider the a-priori probability distribution $p(\Phi)$ of all quantum states $|\Phi \rangle$ spanning the Hilbert space, if a single-shot measurement of observable $\hat O=\sum_{\tau } a_{\tau } \hat{P}_{\tau }$ yields an outcome, which is labeled as event $\gamma$. The posterior probability after this measurement(event) is:
\begin{equation}
p(\Phi|\gamma) = p(\gamma|\Phi)p(\Phi)/p(\gamma),
\end{equation}
where $p(\gamma|\Phi) = {\langle \Phi |\hat{P}_{\gamma} |\Phi \rangle}$ is the probability of $\gamma$ given the quantum state $|\Phi \rangle$, $\hat{P}_{\gamma}$ is the projection operator corresponding to event $\gamma$ and $p(\gamma)$ offers normalization \cite{huszar2012adaptive}.

As the measurements progress, we obtain a series of results $\mathcal{D} =\{\gamma_1,\gamma_2,... \}$ of single-shot measurements over observables $\{\hat O_1,\hat O_2,...\}$, and update the probability as:
\begin{equation}
p(\Phi|\mathcal{D}) \propto \prod_{\gamma \in \mathcal{D}} p(\gamma|\Phi)
=\prod _{\gamma \in \mathcal{D}} {\langle \Phi|\hat{P}_{\gamma}|\Phi \rangle}.
\end{equation}
We define the ``measurement energy" \footnote{Unlike the expectation value of a linear operator, the measurement energy $E(\Phi|\mathcal{D})$ is explicitly nonlinear due to the $\log$ function. Therefore, the probability distribution of a quantum state with measurement outcomes offers realizations of exotic nonlinear-operator Hamiltonian.}:

\begin{equation}
    E(\Phi|\mathcal{D}) = - \sum_{\gamma \in \mathcal{D}} \log{\langle \Phi| \hat{P}_{\gamma} |\Phi \rangle}, \label{eq:m_energy}
\end{equation}
so that $p(\Phi|\mathcal{D}) \propto \exp[-E(\Phi|\mathcal{D})]$ becomes analogous to a Boltzmann distribution with energy $E(\Phi|\mathcal{D})$ in unit of $k_B T$. The measurement energy also responds to the negative logarithm of the likelihood function in MLE studies \cite{hradil20043, ALTEPETER2005105, PhysRevA.55.R1561, PhysRevA.63.040303, PhysRevA.64.052312, PhysRevA.95.062336, PhysRevA.75.042108}. We will show a protocol to locate the MLE states with minimum $E(\Phi|\mathcal{D})$.

The statistical meaning of Eq. \ref{eq:m_energy} becomes clear in case of multiple measurements $N_{\hat{O}}$ on the same observable $\hat O$, yielding $N_{\tau}$ instances of $a_{\tau}$ outcomes. By binning them together, we re-express the measurement energy as:
\begin{equation}
E(\Phi|\mathcal{D}) = - \sum_{\hat{O}} \sum_{\tau }N_{\hat O}  f_{\tau } \log{\langle \Phi|\hat{P}_{\tau } |\Phi \rangle},
\label{eq:m_energy_bin}
\end{equation}
which describes the cross entropy between the expected probability given a quantum state $|\Phi \rangle$ and the measured frequency $f_{\tau } = N_{\tau } / N_{\hat O}$. Besides, the lower bound for measurement energy on given data is $\min E(\mathcal{D})=- \sum_{\hat{O}} \sum_{\tau } N_{\hat O} f_{\tau } \log f_{\tau }$, which makes $E(\Phi_0|\mathcal{D}) - \min E(\mathcal{D})$ a feasible indicator for satisfiability and convergence.

\emph{Measurement-energy minimums via iterative effective Hamiltonians} \textemdash For a generic nonlinear cost function $E = f(\langle \boldsymbol{\hat O} \rangle)$ defined for the expectation values $\langle \hat O_{\kappa}  \rangle = \langle \Phi|\hat O_{\kappa}|\Phi \rangle$, its functional derivative with respect to $|\Phi\rangle$ should vanish at its minimum:
\begin{equation}
\delta E = \sum_{\kappa} \left.\frac{\partial f(\langle \boldsymbol{\hat O} \rangle)}{\partial \langle {\hat O_{\kappa}} \rangle} \right|_{\langle \boldsymbol{\hat O} \rangle_{gs}} \cdot \delta \langle {\hat O_{\kappa}} \rangle_{gs} = 0, \label{eq:Ederivative}
\end{equation}
where $\langle {\hat O_{\kappa}} \rangle_{gs}$ are the expectation values at the minimum. We note that a Hamiltonian $\hat{H}_{eff}$ on the same Hilbert space:
\begin{equation}
    \hat{H}_{eff}=\sum_{\kappa} \alpha_{\kappa} \hat{O}_{\kappa},
    \label{eq:Heff}
\end{equation}
should possess a ground state $|\Phi'_{gs}\rangle$ that satisfies $\delta \langle \hat{H}_{eff}\rangle = \sum_{\kappa} \alpha_{\kappa} \cdot \delta \langle {\hat O_{\kappa}} \rangle_{gs} = 0$, which coincides with Eq. \ref{eq:Ederivative} if we set $|\Phi'_{gs}\rangle = |\Phi_{gs}\rangle$ and:
\begin{equation}
    \alpha_{\kappa} = \left.\frac{\partial f(\langle \boldsymbol{\hat O} \rangle)}{\partial \langle {\hat O_{\kappa}} \rangle} \right|_{\langle \boldsymbol{\hat O} \rangle_{gs}}.
    \label{eq:selfconsis}
\end{equation}
Eqs. \ref{eq:Heff} and \ref{eq:selfconsis} form a self-consistent equation for the minimum of measurement energy $E(\Phi|\mathcal{D})$.

\begin{figure}
    \centering
    \includegraphics[width=0.4\textwidth]{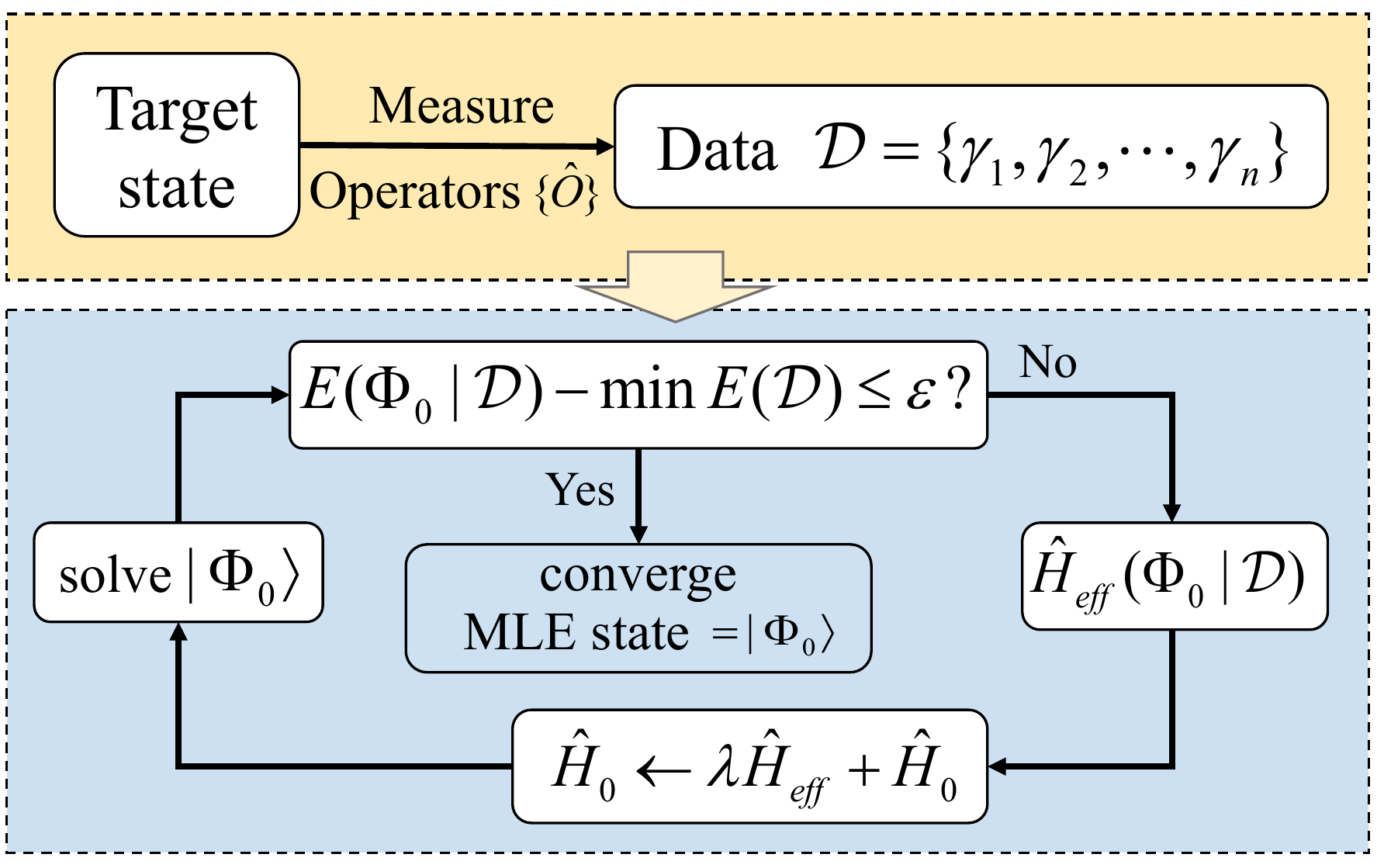} \caption{We outline our strategy for the MLE quantum state: given the quantum measurement results, we iteratively update $\hat{H}_0$ with $\hat H_{eff}$ and solve its ground state $|\Phi_0\rangle$, which converges to the MLE state $|\Phi_{gs}\rangle$. The measurement energy $E(\Phi_0|\mathcal{D})-\min E(\mathcal{D})$  serves as an indicator of convergence and also reveals whether additional measurements and/or observables are preferable.}
    \label{fig:algorithm}
    \end{figure}

Applying such protocol to the measurement energy in Eq. \ref{eq:m_energy_bin}, the effective Hamiltonian is:
\begin{equation}
    \hat{H}_{eff} = \sum_{\hat{O}}\sum_{\tau } N_{\hat{O}}\alpha_{\tau }  \hat{P}_{\tau }, \; \alpha_{\tau } = -\frac{f_{\tau }}{\langle \Phi_{gs}| \hat{P}_{\tau } | \Phi_{gs} \rangle} ,
    \label{eq:Heff_Menergy}
\end{equation}
$N_{\hat{O}}=1$ for single shots. Note that different observables may contribute to the same projection operator. Eq. \ref{eq:Heff_Menergy} is one of the main conclusions of this letter: given the quantum measurements, the self-consistent ground state $|\Phi_{gs}\rangle$ of Eq. \ref{eq:Heff_Menergy} is our MLE quantum state. 

However, as the iteration state approaches the target state, every $\langle \hat{P}_{\tau } \rangle \rightarrow f_{\tau }$, resulting a diminishing $\hat H_{eff}$ and unstable eigenstates. Inspired by supervised machine learning \cite{MLbook, Melko20161, qlt2016, mlstm2019}, we introduce an iteration Hamiltonian $\hat{H}_0$, which is initiated randomly and updated as $\hat H_0 \rightarrow \hat H_0 + \lambda \hat H_{eff}$, where $\lambda$ is the step size. The ground state $|\Phi_0\rangle$ of $\hat H_0$ moves closer and converges to the MLE state upon updates, while $\hat H_{eff}$'s noises average out over the iterations. We summarize the strategy in Fig. \ref{fig:algorithm}, and provide further details, rigorous proof, and generalizations to mixed states in Ref. \cite{SuppQMeasure, Tianlun2022}.

To see how $\hat H_{eff}$ performs as an optimizing gradient for $\hat H_0$, let's consider a toy model with a single qubit $|\Phi_t(\theta,\varphi)\rangle = \cos(\theta/2) |\hat z, +\rangle +\sin (\theta/2) e^{i\varphi} |\hat z, -\rangle$ as the target state. Among various measurements, let us focus on the $\hat S_z = \sigma^z/2$ measurements whose outcomes approach:
\begin{equation}
  \lim_{N_{\hat{O}} \to \infty}   N_{\pm}
  = N_{\hat{O}} \langle \Phi_t |
\hat{P}_{\pm}|\Phi_t \rangle
= N_{\hat{O}} \frac {(1 \pm \cos\theta)}{2},
\label{eq:Szterms}
\end{equation}
where $\hat{P}_{\pm}=(1 \pm \sigma^z)/2$ are the projection operators onto the $\sigma^z = \pm 1$ eigenspaces, respectively. Correspondingly, given an iteration state $|\Phi_0(\theta', \varphi')\rangle$, these $S_z$ measurements contributes to the next $\hat H_{eff}$ as follows:
\begin{equation}
    N_{\hat{O}} (\alpha_+ \hat P_+ + \alpha_- \hat P_- )= -N_{\hat{O}} \frac{\cos\theta -\cos\theta'}{1-\cos^2 \theta'} \sigma^z + const.
\label{eq:singleqbit}
\end{equation}
whose $\sigma^z$ coefficient is negative (positive) when $\theta' > \theta$ ($\theta' < \theta$), opting for a smaller (larger) $\theta'$ at the next iteration, and so on till convergence at $\theta$. As $\hat S_z$ measurements provide no information on $\varphi$, $\varphi'$ remains its initial value. Measurements of $\hat S_{\boldsymbol{n} \neq z}$ contribute additional terms to $\hat{H}_{eff}$ and a more comprehensive optimization of $\hat{H}_0$ and $|\Phi_0(\theta', \varphi')\rangle$.

Unlike previous tomography that faces costly direct parameterization of quantum states and challenging non-convex optimization, we encode $|\Phi_0\rangle$ intrinsically via $\hat H_0$, which holds several advantages: our strategy guarantees efficient descent and convergence \cite{Tianlun2022}, and also takes advantage of various established quantum many-body ansatzes, such as Lanczos, density-matrix renormalization group \cite{MPS1992, DMRG2005}, and quantum Monte Carlo methods \cite{QMCreview2001, Troyer2005}, neural network states \cite{Carleo2016}, or quantum simulators \cite{QAOA2014Farhi, QAOA2020PRX, QAOA2020PRApplied}. Essentially, the ansatz choice relies on a-priori knowledge, such as symmetries and localities, which allows us to conduct more relevant and efficient searches in Hilbert space sub-manifolds.

It is high time we discussed the choices of observables $\hat{O}$. If the a-priori knowledge about the target state is sufficient, we may choose the most physically relevant measurements, usually lower-order and/or local operators; otherwise, such observables still make a good starting point for tentative studies. In reality, we are often limited by experiments and data availability as well. Fortunately, our strategy can still locate the MLE state even under such circumstances and also tell whether the information is inadequate \footnote{A lack of observables may lead to misleading MLE states, which we can identify with signatures in measurement energy: it may constrain the search space leading to a sub-optimal MLE state as the measurement energy converges above its lower bound, or end up with different MLE states simultaneously consistent with the measurement outcomes.}, upon which one may decide to resort to additional operators or experiments. We illustrate such a procedure on Haar random quantum states without any a-priori knowledge in Ref. \cite{SuppQMeasure}.

\emph{Example: random long-range fermion model} \textemdash Let's consider the ground state of the following Hamiltonian:
\begin{equation}
    \hat{H}=-\sum_{ij} t_{ij} (c^\dagger_{i}c_j + c^\dagger_{j}c_i) - \sum_i \mu_i c^\dagger_i c_i,
\label{Eq: freefermion}
\end{equation}
where $1 \le i, j \le L$. We apply random $t_{ij} \in [0,1] $ between arbitrary sites and $\mu_i \in [-0.5, 0.5] $ to deny the system symmetries and locality. Still, our strategy can derive the target states, placed in a black box and tangible only via quantum measurements, even on relatively large systems. Two-point correlators are key to a fermion direct-product state, whose other properties are obtainable via Wick's theorem, thus we choose the observables $\hat{O}_i = c^\dagger_i c_i$ and $\hat{O}'_{ij} = (c^\dagger_i+c^\dagger_{j})(c_i+c_j)/2$, each with two eigenvalues \footnote{An observable with more eigenvalues acts as a double-edged sword: they may incur cost in post-processing $\hat{H}_{eff}$ due to more complex $\hat{P}_{\tau }$, but the distribution also offer more information than the average in a similar spirit to shot-noise studies \cite{Zhou2019, Sivre2019}. We can make an observable simpler to handle by binning together some outcomes and giving up some information, but not vice versa.}, making $\hat{P}_{\tau}$ and $\hat{H}_{eff}$ fermion-bilinear and the subsequent procedure straightforward.

\begin{figure}
    \centering
    \includegraphics[width=0.4\textwidth]{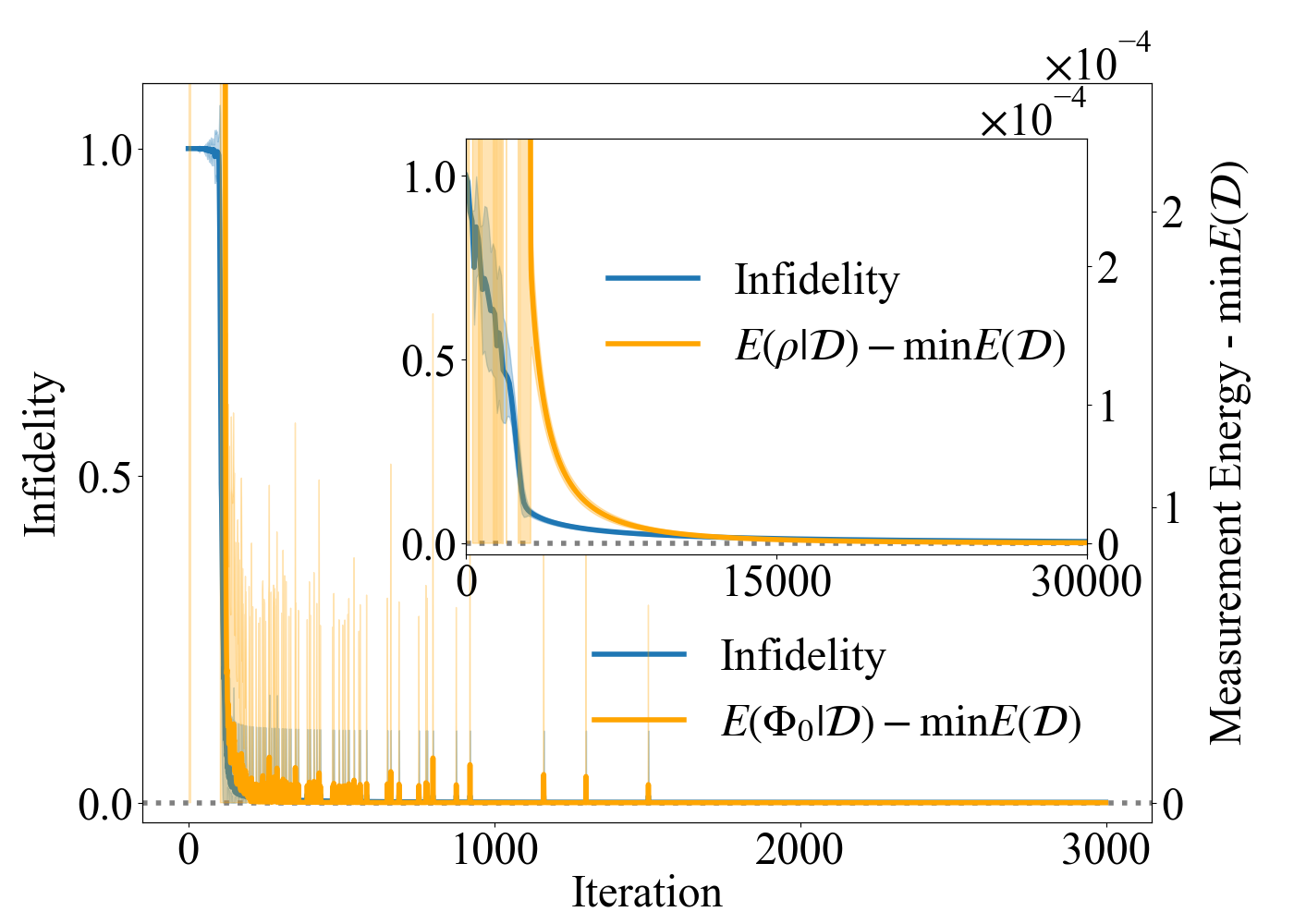}
    \caption{
    Following our strategy, the iteration state quickly converges to the target quantum state, and the average measurement energy converges to its lower bound. The infidelity for pure state and mixed state (shown in the inset) reaches $\sim 6 \times 10^{-4}$ and $\sim 5 \times 10^{-3}$, respectively. The shade is based upon 100 trials on different target states - the ground states and Gibbs states of random long-range fermion models with system size $L=100$.}
    \label{fig:freefermion}
\end{figure}

For simplicity, we measure each observable $\hat{O}=\sum_{\tau }\alpha_{\tau } \hat{P}_{\tau }$ on the target quantum state an equal number $N_{\hat{O}}\rightarrow\infty$ of times to suppress fluctuations. Putting these results on $L=100$ systems into the iterative process in Eq. \ref{eq:Heff_Menergy}, we obtain the results in Fig. \ref{fig:freefermion}. We observe a quick convergence of the iteration state $|\Phi_0\rangle$ towards the target state, its average measurement energy $E(|\Phi_0\rangle)$ towards lower bound \footnote{The spikes in the figure are mainly due to the inconsistent particle number the iteration state $|\Phi_0\rangle$ receives over the slight modifications. Better convergence largely suppresses such phenomena in later iterations.}. Our strategy also works for data laden with quantum fluctuations due to finite numbers of quantum measurements $N_{\hat O}$ and the numerical studies reveal that $N_{\hat O}$ necessary for a certain fidelity level scales polynomially to the system size \cite{SuppQMeasure}.

We also extend applications to mixed states: $\rho_0 \propto e^{-H_0}$ as the quantum state and $\mbox{tr}(\hat{\rho_0} \hat{P}_\tau)$ as the expectation value in Eq. \ref{eq:Heff_Menergy}. Based on $N_{\hat{O}}\rightarrow\infty$ measurements of $\hat{O}_i$ and $\hat{O}'_{ij} $ observables for target Gibbs states $\rho_{tar} = e^{-\beta \hat{H}}/ \mbox{tr}(e^{-\beta \hat{H}})$ on $\hat{H}$ in Eq. \ref{Eq: freefermion} with $L=100$ sites, we observe a quick and unambiguous convergence of the iteration $\rho_0$ towards their target (Fig. \ref{fig:freefermion} inset). Further details, examples, and proof for mixed states are in \cite{SuppQMeasure, Tianlun2022}.

\begin{figure}
    \centering
    \includegraphics[width=0.4\textwidth]{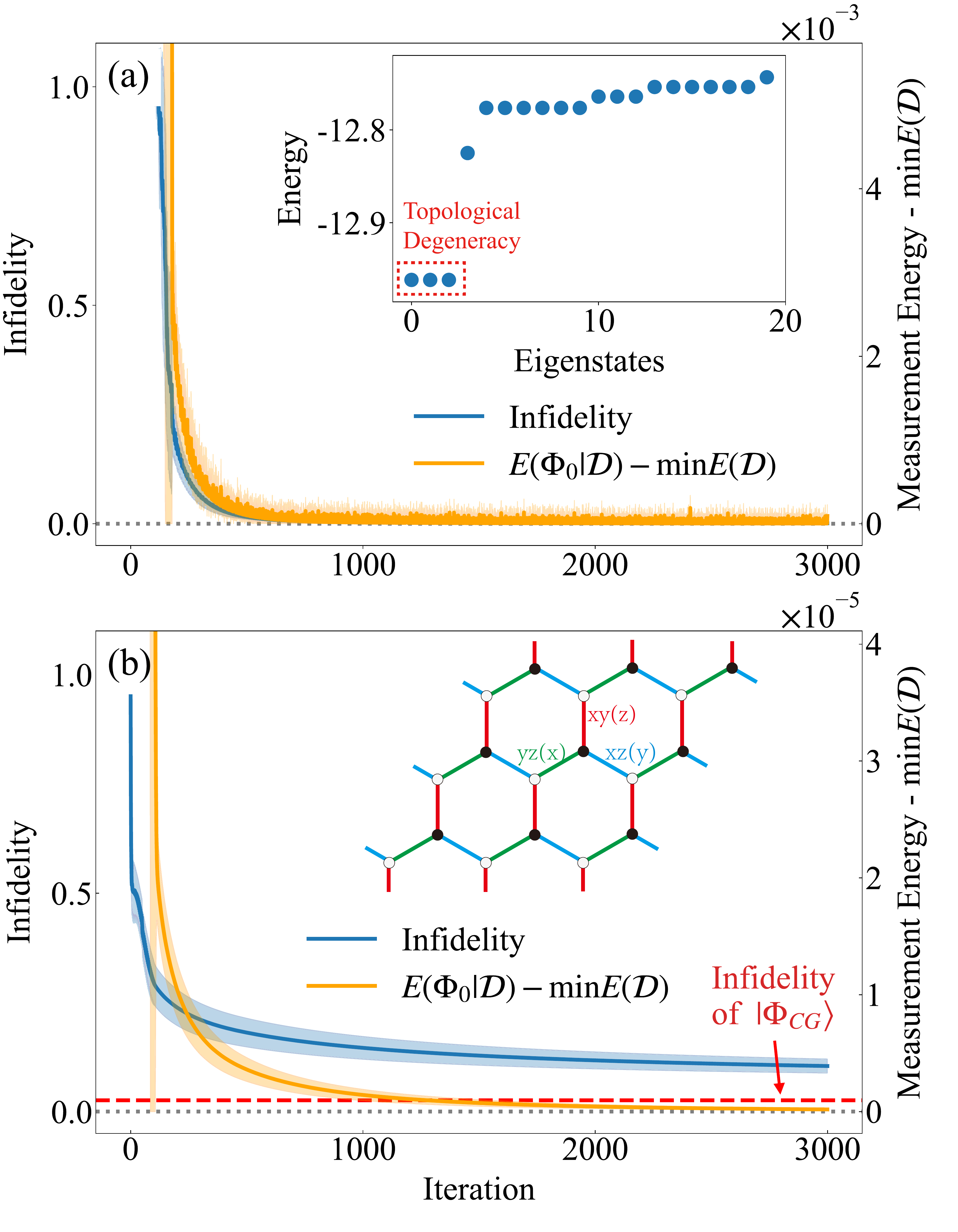}
    \caption{(a) We apply our strategy to the quantum-measurement outcomes of $\sigma_i^\lambda \sigma_j^\lambda$, $\lambda = x, y, z$, on one of the ground states of Eq. \ref{eq:KitaevHam} with $K_{ij}=-1$, $J_{ij}=0.1$, and $\Gamma_{ij}=0$ (spectrum in the inset), the measurement energy $E(\Phi_0|\mathcal{D})$ quickly saturates the lower bound, while the iteration states $|\Phi_{0}\rangle$ converge to the target state with infidelity $\sim 10^{-3}$. (b) With the same measurements on the ground states for $K_{ij}=1$, $J_{ij}=0.1$, and $\Gamma_{ij} \in [0, 0.1]$, the measurement energy $E(\Phi_0|\mathcal{D})$ still quickly saturates the lower bound, while the MLE states $|\Phi_{gs;trial}\rangle$ show slight infidelity $\sim 0.1$ with the target state and differ between trials with average overlap $\sim 0.83$. The red dashed line shows the state $\Phi_{CG}$ averaged over multiple trials \cite{SuppQMeasure}, which offers an improved approximation with infidelity $\sim 0.03$. The shades are based on multiple trials with different initializations. The inset in (b) is a sketch of the Kitaev model on a $3\times 3$ honeycomb lattice.}
    \label{fig:Kitaev12}
\end{figure}

\emph{Example: strongly-correlated Kitaev QSL state}\textemdash Let's consider the nearest-neighbor spin Hamiltonian on the honeycomb lattice:
\begin{equation}
\hat H=\sum_{\langle i j\rangle \in \alpha \beta(\gamma)}
\left[J_{ij} \vec{S}_{i} \cdot \vec{S}_{j}+K_{ij} S_{i}^{\gamma} S_{j}^{\gamma}+\Gamma_{ij}\left(S_{i}^{\alpha} S_{j}^{\beta}+S_{i}^{\beta} S_{j}^{\alpha}\right)\right],
\label{eq:KitaevHam}
\end{equation}
which potentially describes the Kitaev physics in the $A_2IrO_3$-family iridates \cite{PhysRevLett.112.077204} and Kitaev material $RuCl_3$ \cite{Banerjee2016}. $K_{ij}$, $J_{ij}$, and $\Gamma_{ij}$ are the amplitudes of the Kitaev interaction, isotropic Heisenberg interaction, and the symmetric off-diagonal interactions on bond $\langle i j\rangle$, respectively. Depending on the bond dimension, each bond is labeled by $\alpha \beta(\gamma)$, where $\gamma = x,y,z$ is the spin direction in the Kitaev term, and $\alpha, \beta$ are the two orthogonal spin directions in the $\Gamma_{ij}$ term. The pristine Kitaev model ($J_{ij}=\Gamma_{ij}=0$) is analytically solvable \cite{Kitaev2006, SuppQMeasure}. We take the ground state of $\hat H$ with a dominant Kitaev term on a $3\times 3$ system with periodic boundary condition, illustrated in the inset of Fig. \ref{fig:Kitaev12}(b), as our target quantum state. The resulting QSL states are notorious for their lack of smoking-gun signatures. Instead, we probe the target quantum states with \emph{seemingly trivial} quantum measurements. As we will see, these measurements still provide insightful information, and our strategy leads to the target states and, in turn, their abstract natures, including QSL phase \cite{smat, Grover_2013, smat2} and quantum entanglements \cite{KitaevTee, WenTee, Chen2010prb, FrankTEE, FrankCSL}.

To begin with, we set $K_{ij}=-1$, $J_{ij}=0.1$. Given the $C_3$ rotation symmetry, there are three degenerate ground states, shown in the inset of Fig. \ref{fig:Kitaev12}(a). These ground states are topologically degenerate with no quasiparticles \cite{SuppQMeasure, Wen1999gsd, Kitaev2006, Wen2005prb}. The ground states of local Hamiltonians follow the Area law, allowing us to limit to $k$-local, starting from 2-local operators. Here, we first consider quantum measurements on simple observables $\sigma_i^\lambda \sigma_j^\lambda$ of each $\langle ij\rangle$ bond on one of the ground states, $\lambda = x, y, z$. Similar quantum measurements are potentially available to QSL models in Rydberg-atom systems \cite{Rydberg2021theory, Rydberg2021, Rhine2021}, or via electron-spin-resonance scanning tunneling microscopy experiments \cite{AiP2012, ESRSTM2015}, etc. In the large $N_{\hat{O}}\rightarrow \infty$ limit, we obtain $N_{\pm}(ij,\lambda)=N_{\hat{O}} \times \langle \hat P_{\pm}(ij,\lambda) \rangle$ counts of $\pm 1$ outcomes, $\hat P_{\pm}(ij,\lambda)=(1 \pm \sigma_i^\lambda \sigma_j^\lambda)/2$, respectively. Putting these results into the iterations in Eq. \ref{eq:Heff_Menergy}, $|\Phi_0\rangle$ successfully converges to the target ground-state manifold, see Fig. \ref{fig:Kitaev12}(a). Interestingly, starting from a single ground state, we possess the entire topologically degenerate manifold with high fidelity \cite{SuppQMeasure}, with which we can achieve fundamental properties such as quasiparticle statistics \cite{smat, Grover_2013, smat2}. On the one hand, these states share identical local properties thus equal qualifications for the MLE states; on the other hand, their simultaneous presence implies that $\hat H_0$ inherits topological information already present in the target state.

Another interesting scenario is when the observables involved are insufficient to locate the target state fully, as multiple states saturate the measurement energy to the lower bound. For example, we consider the ground state of $K_{ij}=1$, $J_{ij}=0.1$, and random $\Gamma_{ij} \in [0, 0.1]$ on each bond. The system possesses a unique ground state without topological degeneracy on a $3\times 3$ system \cite{SuppQMeasure}. We keep our observables $\sigma_i^\lambda \sigma_j^\lambda, \lambda = x, y, z$ and a large number $N_{\hat{O}} \rightarrow \infty$ of quantum measurements as before, whose results on 10 independent trials are summarized in Fig. \ref{fig:Kitaev12}(b). While all trials converge fully and leave little measurement-energy residue, the obtained MLE states $|\Phi_{gs;trial} \rangle$ differ from trial to trial, with an average overlap $\sim0.83$ in between. We cannot further distinguish these states, which satisfy the quantum measurements equally, until additional observables for further information. Also, we may seek common ground $|\Phi_{CG}\rangle$ between $|\Phi_{gs;trial} \rangle$ as a contingency plan in case of limited ambiguity; see the red dashed line in Fig. \ref{fig:Kitaev12}(b) and details in Ref. \cite{SuppQMeasure}.

Finally, we consider an unprecedented scenario to showcase the adaptability of our strategy: the observables on nearest-neighbor bonds are $\sigma_i^{\hat n} \sigma_j^{\hat n}$ for random $\hat n$ directions and measured once each. Such single-shot results, a list of $\pm 1$ outcomes, are plagued with ultimate fluctuations and hard to make use of; nevertheless, our strategy can capitalize on their intrinsic information and unravel the underlying target state. To further increase the challenge, we pick disordered non-Abelian topologically ordered states by setting $K_{ij}=-1$ and random $J_{ij}\in[0,0.1]$, $\Gamma_{ij}\in[0,0.3]$ on each bond for our target quantum many-body states, whose topological properties are analyzed in detail in Ref. \cite{SuppQMeasure}. We summarize the demonstration in Fig. \ref{fig:snap}: the more single shots, the more information at disposal, and the higher the fidelity of the MLE states $|\Phi_{gs} \rangle$; based on a single state, we also obtain the degenerate manifold, even low-lying excited states \cite{SuppQMeasure}, with high fidelity \footnote{Note that a measurement-energy lower bound is no longer available for such single-shot measurements.}. We emphasize that although our setup resembles the shadow tomography \cite{huang2020predicting}, it neither satisfies nor requires the shadow's randomness prerequisite. Indeed, our strategy is generally applicable and does not rely on any scheme of measurements.

\emph{Discussions} \textemdash Considering the exponentially large Hilbert space of a quantum many-body system, we have offered a quantum strategy to interpret quantum measurements in a general and precise way. With full information and reliable convergence, our approach yields state-of-art performance, as demonstrated by several previously-intractable examples above and even for a generic quantum many-body state (in the supplemental materials \cite{SuppQMeasure}). We note that the additive form of the measurement energy in Eq. \ref{eq:m_energy} means that \emph{every single-shot quantum measurement counts}. On the other hand, for cases where the measurement outcomes $\mathcal{D}$ are not directly obtainable, we can reverse engineer values of $f_{\tau }$ from the expectation values $\langle \hat{O} \rangle$, $\langle \hat{O}^2 \rangle$, $\cdots$ \cite{Sakurai}, and $N_{\hat O}$ as a confidence measure. Our strategy also paves the way for Hamiltonian reconstruction \cite{Tianlun2022, QiHamrecon, Hamrec2019}. Generalizations on quantum measurements connecting ground state and excited states, e.g., inelastic spectroscopy experiments, remain an open question for future research.

\emph{Acknowledgement:} We acknowledge helpful discussions with Zhen-Duo Wang, Tian-Lun Zhao, Pei-Lin Zheng, Hao-Yan Chen, and Yuan Wan. We also acknowledge support from the National Key R\&D Program of China (No.2021YFA1401900) and the National Science Foundation of China (No.12174008 \& No.92270102). The calculations of this work are supported by HPC facilities at Peking University.

\bibliography{refs}

\begin{thebibliography}{89}%
\makeatletter
\providecommand \@ifxundefined [1]{%
 \@ifx{#1\undefined}
}%
\providecommand \@ifnum [1]{%
 \ifnum #1\expandafter \@firstoftwo
 \else \expandafter \@secondoftwo
 \fi
}%
\providecommand \@ifx [1]{%
 \ifx #1\expandafter \@firstoftwo
 \else \expandafter \@secondoftwo
 \fi
}%
\providecommand \natexlab [1]{#1}%
\providecommand \enquote  [1]{``#1''}%
\providecommand \bibnamefont  [1]{#1}%
\providecommand \bibfnamefont [1]{#1}%
\providecommand \citenamefont [1]{#1}%
\providecommand \href@noop [0]{\@secondoftwo}%
\providecommand \href [0]{\begingroup \@sanitize@url \@href}%
\providecommand \@href[1]{\@@startlink{#1}\@@href}%
\providecommand \@@href[1]{\endgroup#1\@@endlink}%
\providecommand \@sanitize@url [0]{\catcode `\\12\catcode `\$12\catcode
  `\&12\catcode `\#12\catcode `\^12\catcode `\_12\catcode `\%12\relax}%
\providecommand \@@startlink[1]{}%
\providecommand \@@endlink[0]{}%
\providecommand \url  [0]{\begingroup\@sanitize@url \@url }%
\providecommand \@url [1]{\endgroup\@href {#1}{\urlprefix }}%
\providecommand \urlprefix  [0]{URL }%
\providecommand \Eprint [0]{\href }%
\providecommand \doibase [0]{https://doi.org/}%
\providecommand \selectlanguage [0]{\@gobble}%
\providecommand \bibinfo  [0]{\@secondoftwo}%
\providecommand \bibfield  [0]{\@secondoftwo}%
\providecommand \translation [1]{[#1]}%
\providecommand \BibitemOpen [0]{}%
\providecommand \bibitemStop [0]{}%
\providecommand \bibitemNoStop [0]{.\EOS\space}%
\providecommand \EOS [0]{\spacefactor3000\relax}%
\providecommand \BibitemShut  [1]{\csname bibitem#1\endcsname}%
\let\auto@bib@innerbib\@empty
\bibitem [{\citenamefont {Quintanilla}\ and\ \citenamefont
  {Hooley}(2009)}]{Quintanilla2009}%
  \BibitemOpen
  \bibfield  {author} {\bibinfo {author} {\bibfnamefont {J.}~\bibnamefont
  {Quintanilla}}\ and\ \bibinfo {author} {\bibfnamefont {C.}~\bibnamefont
  {Hooley}},\ }\bibfield  {title} {\bibinfo {title} {The strong-correlations
  puzzle},\ }\href {https://doi.org/10.1088/2058-7058/22/06/38} {\bibfield
  {journal} {\bibinfo  {journal} {Physics World}\ }\textbf {\bibinfo {volume}
  {22}},\ \bibinfo {pages} {32} (\bibinfo {year} {2009})}\BibitemShut {NoStop}%
\bibitem [{\citenamefont {Kitaev}\ and\ \citenamefont
  {Preskill}(2006)}]{KitaevTee}%
  \BibitemOpen
  \bibfield  {author} {\bibinfo {author} {\bibfnamefont {A.}~\bibnamefont
  {Kitaev}}\ and\ \bibinfo {author} {\bibfnamefont {J.}~\bibnamefont
  {Preskill}},\ }\bibfield  {title} {\bibinfo {title} {{Topological
  Entanglement Entropy}},\ }\href
  {https://doi.org/10.1103/PhysRevLett.96.110404} {\bibfield  {journal}
  {\bibinfo  {journal} {Phys. Rev. Lett.}\ }\textbf {\bibinfo {volume} {96}},\
  \bibinfo {pages} {110404} (\bibinfo {year} {2006})}\BibitemShut {NoStop}%
\bibitem [{\citenamefont {Levin}\ and\ \citenamefont {Wen}(2006)}]{WenTee}%
  \BibitemOpen
  \bibfield  {author} {\bibinfo {author} {\bibfnamefont {M.}~\bibnamefont
  {Levin}}\ and\ \bibinfo {author} {\bibfnamefont {X.-G.}\ \bibnamefont
  {Wen}},\ }\bibfield  {title} {\bibinfo {title} {{Detecting Topological Order
  in a Ground State Wave Function}},\ }\href
  {https://doi.org/10.1103/PhysRevLett.96.110405} {\bibfield  {journal}
  {\bibinfo  {journal} {Phys. Rev. Lett.}\ }\textbf {\bibinfo {volume} {96}},\
  \bibinfo {pages} {110405} (\bibinfo {year} {2006})}\BibitemShut {NoStop}%
\bibitem [{\citenamefont {Chen}\ \emph {et~al.}(2010)\citenamefont {Chen},
  \citenamefont {Gu},\ and\ \citenamefont {Wen}}]{Chen2010prb}%
  \BibitemOpen
  \bibfield  {author} {\bibinfo {author} {\bibfnamefont {X.}~\bibnamefont
  {Chen}}, \bibinfo {author} {\bibfnamefont {Z.-C.}\ \bibnamefont {Gu}},\ and\
  \bibinfo {author} {\bibfnamefont {X.-G.}\ \bibnamefont {Wen}},\ }\bibfield
  {title} {\bibinfo {title} {Local unitary transformation, long-range quantum
  entanglement, wave function renormalization, and topological order},\ }\href
  {https://doi.org/10.1103/PhysRevB.82.155138} {\bibfield  {journal} {\bibinfo
  {journal} {Phys. Rev. B}\ }\textbf {\bibinfo {volume} {82}},\ \bibinfo
  {pages} {155138} (\bibinfo {year} {2010})}\BibitemShut {NoStop}%
\bibitem [{\citenamefont {Zhang}\ \emph
  {et~al.}(2011{\natexlab{a}})\citenamefont {Zhang}, \citenamefont {Grover},\
  and\ \citenamefont {Vishwanath}}]{FrankTEE}%
  \BibitemOpen
  \bibfield  {author} {\bibinfo {author} {\bibfnamefont {Y.}~\bibnamefont
  {Zhang}}, \bibinfo {author} {\bibfnamefont {T.}~\bibnamefont {Grover}},\ and\
  \bibinfo {author} {\bibfnamefont {A.}~\bibnamefont {Vishwanath}},\ }\bibfield
   {title} {\bibinfo {title} {Topological entanglement entropy of
  ${\mathbb{z}}_{2}$ spin liquids and lattice laughlin states},\ }\href
  {https://doi.org/10.1103/PhysRevB.84.075128} {\bibfield  {journal} {\bibinfo
  {journal} {Phys. Rev. B}\ }\textbf {\bibinfo {volume} {84}},\ \bibinfo
  {pages} {075128} (\bibinfo {year} {2011}{\natexlab{a}})}\BibitemShut
  {NoStop}%
\bibitem [{\citenamefont {Zhang}\ \emph
  {et~al.}(2011{\natexlab{b}})\citenamefont {Zhang}, \citenamefont {Grover},\
  and\ \citenamefont {Vishwanath}}]{FrankCSL}%
  \BibitemOpen
  \bibfield  {author} {\bibinfo {author} {\bibfnamefont {Y.}~\bibnamefont
  {Zhang}}, \bibinfo {author} {\bibfnamefont {T.}~\bibnamefont {Grover}},\ and\
  \bibinfo {author} {\bibfnamefont {A.}~\bibnamefont {Vishwanath}},\ }\bibfield
   {title} {\bibinfo {title} {Entanglement entropy of critical spin liquids},\
  }\href {https://doi.org/10.1103/PhysRevLett.107.067202} {\bibfield  {journal}
  {\bibinfo  {journal} {Phys. Rev. Lett.}\ }\textbf {\bibinfo {volume} {107}},\
  \bibinfo {pages} {067202} (\bibinfo {year} {2011}{\natexlab{b}})}\BibitemShut
  {NoStop}%
\bibitem [{\citenamefont {Zhang}\ \emph {et~al.}(2012)\citenamefont {Zhang},
  \citenamefont {Grover}, \citenamefont {Turner}, \citenamefont {Oshikawa},\
  and\ \citenamefont {Vishwanath}}]{smat}%
  \BibitemOpen
  \bibfield  {author} {\bibinfo {author} {\bibfnamefont {Y.}~\bibnamefont
  {Zhang}}, \bibinfo {author} {\bibfnamefont {T.}~\bibnamefont {Grover}},
  \bibinfo {author} {\bibfnamefont {A.}~\bibnamefont {Turner}}, \bibinfo
  {author} {\bibfnamefont {M.}~\bibnamefont {Oshikawa}},\ and\ \bibinfo
  {author} {\bibfnamefont {A.}~\bibnamefont {Vishwanath}},\ }\bibfield  {title}
  {\bibinfo {title} {Quasiparticle statistics and braiding from ground-state
  entanglement},\ }\href {https://doi.org/10.1103/PhysRevB.85.235151}
  {\bibfield  {journal} {\bibinfo  {journal} {Phys. Rev. B}\ }\textbf {\bibinfo
  {volume} {85}},\ \bibinfo {pages} {235151} (\bibinfo {year}
  {2012})}\BibitemShut {NoStop}%
\bibitem [{\citenamefont {Grover}\ \emph {et~al.}(2013)\citenamefont {Grover},
  \citenamefont {Zhang},\ and\ \citenamefont {Vishwanath}}]{Grover_2013}%
  \BibitemOpen
  \bibfield  {author} {\bibinfo {author} {\bibfnamefont {T.}~\bibnamefont
  {Grover}}, \bibinfo {author} {\bibfnamefont {Y.}~\bibnamefont {Zhang}},\ and\
  \bibinfo {author} {\bibfnamefont {A.}~\bibnamefont {Vishwanath}},\ }\bibfield
   {title} {\bibinfo {title} {Entanglement entropy as a portal to the physics
  of quantum spin liquids},\ }\href
  {https://doi.org/10.1088/1367-2630/15/2/025002} {\bibfield  {journal}
  {\bibinfo  {journal} {New Journal of Physics}\ }\textbf {\bibinfo {volume}
  {15}},\ \bibinfo {pages} {025002} (\bibinfo {year} {2013})}\BibitemShut
  {NoStop}%
\bibitem [{\citenamefont {Zhang}\ \emph {et~al.}(2015)\citenamefont {Zhang},
  \citenamefont {Grover},\ and\ \citenamefont {Vishwanath}}]{smat2}%
  \BibitemOpen
  \bibfield  {author} {\bibinfo {author} {\bibfnamefont {Y.}~\bibnamefont
  {Zhang}}, \bibinfo {author} {\bibfnamefont {T.}~\bibnamefont {Grover}},\ and\
  \bibinfo {author} {\bibfnamefont {A.}~\bibnamefont {Vishwanath}},\ }\bibfield
   {title} {\bibinfo {title} {General procedure for determining braiding and
  statistics of anyons using entanglement interferometry},\ }\href
  {https://doi.org/10.1103/PhysRevB.91.035127} {\bibfield  {journal} {\bibinfo
  {journal} {Phys. Rev. B}\ }\textbf {\bibinfo {volume} {91}},\ \bibinfo
  {pages} {035127} (\bibinfo {year} {2015})}\BibitemShut {NoStop}%
\bibitem [{\citenamefont {Haldane}(1988)}]{Haldane1988}%
  \BibitemOpen
  \bibfield  {author} {\bibinfo {author} {\bibfnamefont {F.~D.~M.}\
  \bibnamefont {Haldane}},\ }\bibfield  {title} {\bibinfo {title} {Model for a
  quantum hall effect without landau levels: Condensed-matter realization of
  the "parity anomaly"},\ }\href {https://doi.org/10.1103/PhysRevLett.61.2015}
  {\bibfield  {journal} {\bibinfo  {journal} {Phys. Rev. Lett.}\ }\textbf
  {\bibinfo {volume} {61}},\ \bibinfo {pages} {2015} (\bibinfo {year}
  {1988})}\BibitemShut {NoStop}%
\bibitem [{\citenamefont {Laughlin}(1983)}]{LaughlinFQH}%
  \BibitemOpen
  \bibfield  {author} {\bibinfo {author} {\bibfnamefont {R.~B.}\ \bibnamefont
  {Laughlin}},\ }\bibfield  {title} {\bibinfo {title} {Anomalous quantum hall
  effect: An incompressible quantum fluid with fractionally charged
  excitations},\ }\href {https://doi.org/10.1103/PhysRevLett.50.1395}
  {\bibfield  {journal} {\bibinfo  {journal} {Phys. Rev. Lett.}\ }\textbf
  {\bibinfo {volume} {50}},\ \bibinfo {pages} {1395} (\bibinfo {year}
  {1983})}\BibitemShut {NoStop}%
\bibitem [{\citenamefont {Kane}\ and\ \citenamefont {Mele}(2005)}]{QSHE2005}%
  \BibitemOpen
  \bibfield  {author} {\bibinfo {author} {\bibfnamefont {C.~L.}\ \bibnamefont
  {Kane}}\ and\ \bibinfo {author} {\bibfnamefont {E.~J.}\ \bibnamefont
  {Mele}},\ }\bibfield  {title} {\bibinfo {title} {${Z}_{2}$ topological order
  and the quantum spin hall effect},\ }\href
  {https://doi.org/10.1103/PhysRevLett.95.146802} {\bibfield  {journal}
  {\bibinfo  {journal} {Phys. Rev. Lett.}\ }\textbf {\bibinfo {volume} {95}},\
  \bibinfo {pages} {146802} (\bibinfo {year} {2005})}\BibitemShut {NoStop}%
\bibitem [{\citenamefont {Hasan}\ and\ \citenamefont {Kane}(2010)}]{TI2010}%
  \BibitemOpen
  \bibfield  {author} {\bibinfo {author} {\bibfnamefont {M.~Z.}\ \bibnamefont
  {Hasan}}\ and\ \bibinfo {author} {\bibfnamefont {C.~L.}\ \bibnamefont
  {Kane}},\ }\bibfield  {title} {\bibinfo {title} {\textit{Colloquium} :
  Topological insulators},\ }\href {https://doi.org/10.1103/RevModPhys.82.3045}
  {\bibfield  {journal} {\bibinfo  {journal} {Rev. Mod. Phys.}\ }\textbf
  {\bibinfo {volume} {82}},\ \bibinfo {pages} {3045} (\bibinfo {year}
  {2010})}\BibitemShut {NoStop}%
\bibitem [{\citenamefont {Nielsen}\ and\ \citenamefont
  {Ninomiya}(1983)}]{NIELSEN1983389}%
  \BibitemOpen
  \bibfield  {author} {\bibinfo {author} {\bibfnamefont {H.}~\bibnamefont
  {Nielsen}}\ and\ \bibinfo {author} {\bibfnamefont {M.}~\bibnamefont
  {Ninomiya}},\ }\bibfield  {title} {\bibinfo {title} {The adler-bell-jackiw
  anomaly and weyl fermions in a crystal},\ }\href
  {https://doi.org/https://doi.org/10.1016/0370-2693(83)91529-0} {\bibfield
  {journal} {\bibinfo  {journal} {Physics Letters B}\ }\textbf {\bibinfo
  {volume} {130}},\ \bibinfo {pages} {389} (\bibinfo {year}
  {1983})}\BibitemShut {NoStop}%
\bibitem [{\citenamefont {Yuan}\ \emph {et~al.}(2020)\citenamefont {Yuan},
  \citenamefont {Zhang}, \citenamefont {Zhang}, \citenamefont {Yan},
  \citenamefont {Lyu}, \citenamefont {Zhang}, \citenamefont {Li}, \citenamefont
  {Song}, \citenamefont {Zhao}, \citenamefont {Leng}, \citenamefont {Ozerov},
  \citenamefont {Chen}, \citenamefont {Wang}, \citenamefont {Shi},
  \citenamefont {Yan},\ and\ \citenamefont {Xiu}}]{Yuan2020}%
  \BibitemOpen
  \bibfield  {author} {\bibinfo {author} {\bibfnamefont {X.}~\bibnamefont
  {Yuan}}, \bibinfo {author} {\bibfnamefont {C.}~\bibnamefont {Zhang}},
  \bibinfo {author} {\bibfnamefont {Y.}~\bibnamefont {Zhang}}, \bibinfo
  {author} {\bibfnamefont {Z.}~\bibnamefont {Yan}}, \bibinfo {author}
  {\bibfnamefont {T.}~\bibnamefont {Lyu}}, \bibinfo {author} {\bibfnamefont
  {M.}~\bibnamefont {Zhang}}, \bibinfo {author} {\bibfnamefont
  {Z.}~\bibnamefont {Li}}, \bibinfo {author} {\bibfnamefont {C.}~\bibnamefont
  {Song}}, \bibinfo {author} {\bibfnamefont {M.}~\bibnamefont {Zhao}}, \bibinfo
  {author} {\bibfnamefont {P.}~\bibnamefont {Leng}}, \bibinfo {author}
  {\bibfnamefont {M.}~\bibnamefont {Ozerov}}, \bibinfo {author} {\bibfnamefont
  {X.}~\bibnamefont {Chen}}, \bibinfo {author} {\bibfnamefont {N.}~\bibnamefont
  {Wang}}, \bibinfo {author} {\bibfnamefont {Y.}~\bibnamefont {Shi}}, \bibinfo
  {author} {\bibfnamefont {H.}~\bibnamefont {Yan}},\ and\ \bibinfo {author}
  {\bibfnamefont {F.}~\bibnamefont {Xiu}},\ }\bibfield  {title} {\bibinfo
  {title} {The discovery of dynamic chiral anomaly in a weyl semimetal nbas},\
  }\href {https://doi.org/10.1038/s41467-020-14749-4} {\bibfield  {journal}
  {\bibinfo  {journal} {Nature Communications}\ }\textbf {\bibinfo {volume}
  {11}},\ \bibinfo {pages} {1259} (\bibinfo {year} {2020})}\BibitemShut
  {NoStop}%
\bibitem [{\citenamefont {Keimer}\ and\ \citenamefont
  {Moore}(2017)}]{Keimer2017}%
  \BibitemOpen
  \bibfield  {author} {\bibinfo {author} {\bibfnamefont {B.}~\bibnamefont
  {Keimer}}\ and\ \bibinfo {author} {\bibfnamefont {J.~E.}\ \bibnamefont
  {Moore}},\ }\bibfield  {title} {\bibinfo {title} {The physics of quantum
  materials},\ }\href {https://doi.org/10.1038/nphys4302} {\bibfield  {journal}
  {\bibinfo  {journal} {Nature Physics}\ }\textbf {\bibinfo {volume} {13}},\
  \bibinfo {pages} {1045} (\bibinfo {year} {2017})}\BibitemShut {NoStop}%
\bibitem [{\citenamefont {Kitaev}(2003)}]{Kitaev20032}%
  \BibitemOpen
  \bibfield  {author} {\bibinfo {author} {\bibfnamefont {A.}~\bibnamefont
  {Kitaev}},\ }\bibfield  {title} {\bibinfo {title} {Fault-tolerant quantum
  computation by anyons},\ }\href
  {https://doi.org/http://dx.doi.org/10.1016/S0003-4916(02)00018-0} {\bibfield
  {journal} {\bibinfo  {journal} {Annals of Physics}\ }\textbf {\bibinfo
  {volume} {303}},\ \bibinfo {pages} {2 } (\bibinfo {year} {2003})}\BibitemShut
  {NoStop}%
\bibitem [{\citenamefont {Kitaev}(2006)}]{Kitaev2006}%
  \BibitemOpen
  \bibfield  {author} {\bibinfo {author} {\bibfnamefont {A.}~\bibnamefont
  {Kitaev}},\ }\bibfield  {title} {\bibinfo {title} {Anyons in an exactly
  solved model and beyond},\ }\href
  {https://doi.org/http://dx.doi.org/10.1016/j.aop.2005.10.005} {\bibfield
  {journal} {\bibinfo  {journal} {Annals of Physics}\ }\textbf {\bibinfo
  {volume} {321}},\ \bibinfo {pages} {2 } (\bibinfo {year} {2006})}\BibitemShut
  {NoStop}%
\bibitem [{\citenamefont {Yan}\ \emph {et~al.}(2011)\citenamefont {Yan},
  \citenamefont {Huse},\ and\ \citenamefont {White}}]{White2010}%
  \BibitemOpen
  \bibfield  {author} {\bibinfo {author} {\bibfnamefont {S.}~\bibnamefont
  {Yan}}, \bibinfo {author} {\bibfnamefont {D.~A.}\ \bibnamefont {Huse}},\ and\
  \bibinfo {author} {\bibfnamefont {S.~R.}\ \bibnamefont {White}},\ }\bibfield
  {title} {\bibinfo {title} {Spin-liquid ground state of the s = 1/2 kagome
  heisenberg antiferromagnet},\ }\href
  {https://doi.org/10.1126/science.1201080} {\bibfield  {journal} {\bibinfo
  {journal} {Science}\ }\textbf {\bibinfo {volume} {332}},\ \bibinfo {pages}
  {1173} (\bibinfo {year} {2011})}\BibitemShut {NoStop}%
\bibitem [{\citenamefont {Banerjee}\ \emph {et~al.}(2016)\citenamefont
  {Banerjee}, \citenamefont {Bridges}, \citenamefont {Yan}, \citenamefont
  {Aczel}, \citenamefont {Li}, \citenamefont {Stone}, \citenamefont {Granroth},
  \citenamefont {Lumsden}, \citenamefont {Yiu}, \citenamefont {Knolle},
  \citenamefont {Bhattacharjee}, \citenamefont {Kovrizhin}, \citenamefont
  {Moessner}, \citenamefont {Tennant}, \citenamefont {Mandrus},\ and\
  \citenamefont {Nagler}}]{Banerjee2016}%
  \BibitemOpen
  \bibfield  {author} {\bibinfo {author} {\bibfnamefont {A.}~\bibnamefont
  {Banerjee}}, \bibinfo {author} {\bibfnamefont {C.~A.}\ \bibnamefont
  {Bridges}}, \bibinfo {author} {\bibfnamefont {J.-Q.}\ \bibnamefont {Yan}},
  \bibinfo {author} {\bibfnamefont {A.~A.}\ \bibnamefont {Aczel}}, \bibinfo
  {author} {\bibfnamefont {L.}~\bibnamefont {Li}}, \bibinfo {author}
  {\bibfnamefont {M.~B.}\ \bibnamefont {Stone}}, \bibinfo {author}
  {\bibfnamefont {G.~E.}\ \bibnamefont {Granroth}}, \bibinfo {author}
  {\bibfnamefont {M.~D.}\ \bibnamefont {Lumsden}}, \bibinfo {author}
  {\bibfnamefont {Y.}~\bibnamefont {Yiu}}, \bibinfo {author} {\bibfnamefont
  {J.}~\bibnamefont {Knolle}}, \bibinfo {author} {\bibfnamefont
  {S.}~\bibnamefont {Bhattacharjee}}, \bibinfo {author} {\bibfnamefont {D.~L.}\
  \bibnamefont {Kovrizhin}}, \bibinfo {author} {\bibfnamefont {R.}~\bibnamefont
  {Moessner}}, \bibinfo {author} {\bibfnamefont {D.~A.}\ \bibnamefont
  {Tennant}}, \bibinfo {author} {\bibfnamefont {D.~G.}\ \bibnamefont
  {Mandrus}},\ and\ \bibinfo {author} {\bibfnamefont {S.~E.}\ \bibnamefont
  {Nagler}},\ }\bibfield  {title} {\bibinfo {title} {Proximate kitaev quantum
  spin liquid behaviour in a honeycomb magnet},\ }\href
  {https://doi.org/10.1038/nmat4604} {\bibfield  {journal} {\bibinfo  {journal}
  {Nature Materials}\ }\textbf {\bibinfo {volume} {15}},\ \bibinfo {pages}
  {733} (\bibinfo {year} {2016})}\BibitemShut {NoStop}%
\bibitem [{\citenamefont {Bednorz}\ and\ \citenamefont
  {M{\"u}ller}(1986)}]{Bednorz1986}%
  \BibitemOpen
  \bibfield  {author} {\bibinfo {author} {\bibfnamefont {J.~G.}\ \bibnamefont
  {Bednorz}}\ and\ \bibinfo {author} {\bibfnamefont {K.~A.}\ \bibnamefont
  {M{\"u}ller}},\ }\bibfield  {title} {\bibinfo {title} {Possible hight c
  superconductivity in the ba- la- cu- o system},\ }\href
  {https://doi.org/10.1007/BF01303701} {\bibfield  {journal} {\bibinfo
  {journal} {Zeitschrift f{\"u}r Physik B Condensed Matter}\ }\textbf {\bibinfo
  {volume} {64}},\ \bibinfo {pages} {189} (\bibinfo {year} {1986})}\BibitemShut
  {NoStop}%
\bibitem [{\citenamefont {Kamihara}\ \emph {et~al.}(2006)\citenamefont
  {Kamihara}, \citenamefont {Hiramatsu}, \citenamefont {Hirano}, \citenamefont
  {Kawamura}, \citenamefont {Yanagi}, \citenamefont {Kamiya},\ and\
  \citenamefont {Hosono}}]{FeSC1}%
  \BibitemOpen
  \bibfield  {author} {\bibinfo {author} {\bibfnamefont {Y.}~\bibnamefont
  {Kamihara}}, \bibinfo {author} {\bibfnamefont {H.}~\bibnamefont {Hiramatsu}},
  \bibinfo {author} {\bibfnamefont {M.}~\bibnamefont {Hirano}}, \bibinfo
  {author} {\bibfnamefont {R.}~\bibnamefont {Kawamura}}, \bibinfo {author}
  {\bibfnamefont {H.}~\bibnamefont {Yanagi}}, \bibinfo {author} {\bibfnamefont
  {T.}~\bibnamefont {Kamiya}},\ and\ \bibinfo {author} {\bibfnamefont
  {H.}~\bibnamefont {Hosono}},\ }\bibfield  {title} {\bibinfo {title}
  {Iron-based layered superconductor: Laofep},\ }\href
  {https://doi.org/10.1021/ja063355c} {\bibfield  {journal} {\bibinfo
  {journal} {Journal of the American Chemical Society}\ }\textbf {\bibinfo
  {volume} {128}},\ \bibinfo {pages} {10012} (\bibinfo {year}
  {2006})}\BibitemShut {NoStop}%
\bibitem [{\citenamefont {Kamihara}\ \emph {et~al.}(2008)\citenamefont
  {Kamihara}, \citenamefont {Watanabe}, \citenamefont {Hirano},\ and\
  \citenamefont {Hosono}}]{FeSC2}%
  \BibitemOpen
  \bibfield  {author} {\bibinfo {author} {\bibfnamefont {Y.}~\bibnamefont
  {Kamihara}}, \bibinfo {author} {\bibfnamefont {T.}~\bibnamefont {Watanabe}},
  \bibinfo {author} {\bibfnamefont {M.}~\bibnamefont {Hirano}},\ and\ \bibinfo
  {author} {\bibfnamefont {H.}~\bibnamefont {Hosono}},\ }\bibfield  {title}
  {\bibinfo {title} {Iron-based layered superconductor la [o1-x f x] feas (x=
  0.05- 0.12) with t c= 26 k},\ }\href {https://doi.org/10.1021/ja800073m}
  {\bibfield  {journal} {\bibinfo  {journal} {Journal of the American Chemical
  Society}\ }\textbf {\bibinfo {volume} {130}},\ \bibinfo {pages} {3296}
  (\bibinfo {year} {2008})}\BibitemShut {NoStop}%
\bibitem [{\citenamefont {Cao}\ \emph {et~al.}(2018)\citenamefont {Cao},
  \citenamefont {Fatemi}, \citenamefont {Fang}, \citenamefont {Watanabe},
  \citenamefont {Taniguchi}, \citenamefont {Kaxiras},\ and\ \citenamefont
  {Jarillo-Herrero}}]{Cao2018}%
  \BibitemOpen
  \bibfield  {author} {\bibinfo {author} {\bibfnamefont {Y.}~\bibnamefont
  {Cao}}, \bibinfo {author} {\bibfnamefont {V.}~\bibnamefont {Fatemi}},
  \bibinfo {author} {\bibfnamefont {S.}~\bibnamefont {Fang}}, \bibinfo {author}
  {\bibfnamefont {K.}~\bibnamefont {Watanabe}}, \bibinfo {author}
  {\bibfnamefont {T.}~\bibnamefont {Taniguchi}}, \bibinfo {author}
  {\bibfnamefont {E.}~\bibnamefont {Kaxiras}},\ and\ \bibinfo {author}
  {\bibfnamefont {P.}~\bibnamefont {Jarillo-Herrero}},\ }\bibfield  {title}
  {\bibinfo {title} {Unconventional superconductivity in magic-angle graphene
  superlattices},\ }\href {https://doi.org/10.1038/nature26160} {\bibfield
  {journal} {\bibinfo  {journal} {Nature}\ }\textbf {\bibinfo {volume} {556}},\
  \bibinfo {pages} {43} (\bibinfo {year} {2018})}\BibitemShut {NoStop}%
\bibitem [{\citenamefont {Klitzing}\ \emph {et~al.}(1980)\citenamefont
  {Klitzing}, \citenamefont {Dorda},\ and\ \citenamefont {Pepper}}]{QHE1980}%
  \BibitemOpen
  \bibfield  {author} {\bibinfo {author} {\bibfnamefont {K.~v.}\ \bibnamefont
  {Klitzing}}, \bibinfo {author} {\bibfnamefont {G.}~\bibnamefont {Dorda}},\
  and\ \bibinfo {author} {\bibfnamefont {M.}~\bibnamefont {Pepper}},\
  }\bibfield  {title} {\bibinfo {title} {New method for high-accuracy
  determination of the fine-structure constant based on quantized hall
  resistance},\ }\href {https://doi.org/10.1103/PhysRevLett.45.494} {\bibfield
  {journal} {\bibinfo  {journal} {Phys. Rev. Lett.}\ }\textbf {\bibinfo
  {volume} {45}},\ \bibinfo {pages} {494} (\bibinfo {year} {1980})}\BibitemShut
  {NoStop}%
\bibitem [{\citenamefont {Tsui}\ \emph {et~al.}(1982)\citenamefont {Tsui},
  \citenamefont {Stormer},\ and\ \citenamefont {Gossard}}]{TsuiFQH}%
  \BibitemOpen
  \bibfield  {author} {\bibinfo {author} {\bibfnamefont {D.~C.}\ \bibnamefont
  {Tsui}}, \bibinfo {author} {\bibfnamefont {H.~L.}\ \bibnamefont {Stormer}},\
  and\ \bibinfo {author} {\bibfnamefont {A.~C.}\ \bibnamefont {Gossard}},\
  }\bibfield  {title} {\bibinfo {title} {Two-dimensional magnetotransport in
  the extreme quantum limit},\ }\href
  {https://doi.org/10.1103/PhysRevLett.48.1559} {\bibfield  {journal} {\bibinfo
   {journal} {Phys. Rev. Lett.}\ }\textbf {\bibinfo {volume} {48}},\ \bibinfo
  {pages} {1559} (\bibinfo {year} {1982})}\BibitemShut {NoStop}%
\bibitem [{\citenamefont {Haldane}(1983)}]{Haldane1983}%
  \BibitemOpen
  \bibfield  {author} {\bibinfo {author} {\bibfnamefont {F.~D.~M.}\
  \bibnamefont {Haldane}},\ }\bibfield  {title} {\bibinfo {title} {Nonlinear
  field theory of large-spin heisenberg antiferromagnets: Semiclassically
  quantized solitons of the one-dimensional easy-axis n\'eel state},\ }\href
  {https://doi.org/10.1103/PhysRevLett.50.1153} {\bibfield  {journal} {\bibinfo
   {journal} {Phys. Rev. Lett.}\ }\textbf {\bibinfo {volume} {50}},\ \bibinfo
  {pages} {1153} (\bibinfo {year} {1983})}\BibitemShut {NoStop}%
\bibitem [{\citenamefont {von Klitzing}(1986)}]{QHE1986}%
  \BibitemOpen
  \bibfield  {author} {\bibinfo {author} {\bibfnamefont {K.}~\bibnamefont {von
  Klitzing}},\ }\bibfield  {title} {\bibinfo {title} {The quantized hall
  effect},\ }\href {https://doi.org/10.1103/RevModPhys.58.519} {\bibfield
  {journal} {\bibinfo  {journal} {Rev. Mod. Phys.}\ }\textbf {\bibinfo {volume}
  {58}},\ \bibinfo {pages} {519} (\bibinfo {year} {1986})}\BibitemShut
  {NoStop}%
\bibitem [{\citenamefont {Wen}\ and\ \citenamefont {Niu}(1990)}]{Wen1999gsd}%
  \BibitemOpen
  \bibfield  {author} {\bibinfo {author} {\bibfnamefont {X.~G.}\ \bibnamefont
  {Wen}}\ and\ \bibinfo {author} {\bibfnamefont {Q.}~\bibnamefont {Niu}},\
  }\bibfield  {title} {\bibinfo {title} {Ground-state degeneracy of the
  fractional quantum hall states in the presence of a random potential and on
  high-genus riemann surfaces},\ }\href
  {https://doi.org/10.1103/PhysRevB.41.9377} {\bibfield  {journal} {\bibinfo
  {journal} {Phys. Rev. B}\ }\textbf {\bibinfo {volume} {41}},\ \bibinfo
  {pages} {9377} (\bibinfo {year} {1990})}\BibitemShut {NoStop}%
\bibitem [{\citenamefont {Haldane}(2004)}]{Haldane2004}%
  \BibitemOpen
  \bibfield  {author} {\bibinfo {author} {\bibfnamefont {F.~D.~M.}\
  \bibnamefont {Haldane}},\ }\bibfield  {title} {\bibinfo {title} {Berry
  curvature on the fermi surface: Anomalous hall effect as a topological
  fermi-liquid property},\ }\href
  {https://doi.org/10.1103/PhysRevLett.93.206602} {\bibfield  {journal}
  {\bibinfo  {journal} {Phys. Rev. Lett.}\ }\textbf {\bibinfo {volume} {93}},\
  \bibinfo {pages} {206602} (\bibinfo {year} {2004})}\BibitemShut {NoStop}%
\bibitem [{\citenamefont {Bernevig}\ \emph {et~al.}(2006)\citenamefont
  {Bernevig}, \citenamefont {Hughes},\ and\ \citenamefont
  {Zhang}}]{Bernevig2006}%
  \BibitemOpen
  \bibfield  {author} {\bibinfo {author} {\bibfnamefont {B.~A.}\ \bibnamefont
  {Bernevig}}, \bibinfo {author} {\bibfnamefont {T.~L.}\ \bibnamefont
  {Hughes}},\ and\ \bibinfo {author} {\bibfnamefont {S.-C.}\ \bibnamefont
  {Zhang}},\ }\bibfield  {title} {\bibinfo {title} {Quantum spin hall effect
  and topological phase transition in hgte quantum wells},\ }\href
  {https://doi.org/10.1126/science.1133734} {\bibfield  {journal} {\bibinfo
  {journal} {Science}\ }\textbf {\bibinfo {volume} {314}},\ \bibinfo {pages}
  {1757} (\bibinfo {year} {2006})}\BibitemShut {NoStop}%
\bibitem [{\citenamefont {Fu}\ \emph {et~al.}(2007)\citenamefont {Fu},
  \citenamefont {Kane},\ and\ \citenamefont {Mele}}]{3DTI}%
  \BibitemOpen
  \bibfield  {author} {\bibinfo {author} {\bibfnamefont {L.}~\bibnamefont
  {Fu}}, \bibinfo {author} {\bibfnamefont {C.~L.}\ \bibnamefont {Kane}},\ and\
  \bibinfo {author} {\bibfnamefont {E.~J.}\ \bibnamefont {Mele}},\ }\bibfield
  {title} {\bibinfo {title} {Topological insulators in three dimensions},\
  }\href {https://doi.org/10.1103/PhysRevLett.98.106803} {\bibfield  {journal}
  {\bibinfo  {journal} {Phys. Rev. Lett.}\ }\textbf {\bibinfo {volume} {98}},\
  \bibinfo {pages} {106803} (\bibinfo {year} {2007})}\BibitemShut {NoStop}%
\bibitem [{\citenamefont {Chen}\ \emph {et~al.}(2012)\citenamefont {Chen},
  \citenamefont {Gu}, \citenamefont {Liu},\ and\ \citenamefont
  {Wen}}]{ChenSPT}%
  \BibitemOpen
  \bibfield  {author} {\bibinfo {author} {\bibfnamefont {X.}~\bibnamefont
  {Chen}}, \bibinfo {author} {\bibfnamefont {Z.-C.}\ \bibnamefont {Gu}},
  \bibinfo {author} {\bibfnamefont {Z.-X.}\ \bibnamefont {Liu}},\ and\ \bibinfo
  {author} {\bibfnamefont {X.-G.}\ \bibnamefont {Wen}},\ }\bibfield  {title}
  {\bibinfo {title} {{Symmetry-Protected Topological Orders in Interacting
  Bosonic Systems}},\ }\href {https://doi.org/10.1126/science.1227224}
  {\bibfield  {journal} {\bibinfo  {journal} {Science}\ }\textbf {\bibinfo
  {volume} {338}},\ \bibinfo {pages} {1604} (\bibinfo {year}
  {2012})}\BibitemShut {NoStop}%
\bibitem [{\citenamefont {Fradkin}\ \emph {et~al.}(2015)\citenamefont
  {Fradkin}, \citenamefont {Kivelson},\ and\ \citenamefont
  {Tranquada}}]{KivelsonIntertwine}%
  \BibitemOpen
  \bibfield  {author} {\bibinfo {author} {\bibfnamefont {E.}~\bibnamefont
  {Fradkin}}, \bibinfo {author} {\bibfnamefont {S.~A.}\ \bibnamefont
  {Kivelson}},\ and\ \bibinfo {author} {\bibfnamefont {J.~M.}\ \bibnamefont
  {Tranquada}},\ }\bibfield  {title} {\bibinfo {title} {Colloquium},\ }\href
  {https://doi.org/10.1103/RevModPhys.87.457} {\bibfield  {journal} {\bibinfo
  {journal} {Rev. Mod. Phys.}\ }\textbf {\bibinfo {volume} {87}},\ \bibinfo
  {pages} {457} (\bibinfo {year} {2015})}\BibitemShut {NoStop}%
\bibitem [{\citenamefont {Proust}\ and\ \citenamefont
  {Taillefer}(2019)}]{Taillefer2019}%
  \BibitemOpen
  \bibfield  {author} {\bibinfo {author} {\bibfnamefont {C.}~\bibnamefont
  {Proust}}\ and\ \bibinfo {author} {\bibfnamefont {L.}~\bibnamefont
  {Taillefer}},\ }\bibfield  {title} {\bibinfo {title} {The remarkable
  underlying ground states of cuprate superconductors},\ }\href
  {https://doi.org/10.1146/annurev-conmatphys-031218-013210} {\bibfield
  {journal} {\bibinfo  {journal} {Annual Review of Condensed Matter Physics}\
  }\textbf {\bibinfo {volume} {10}},\ \bibinfo {pages} {409} (\bibinfo {year}
  {2019})}\BibitemShut {NoStop}%
\bibitem [{\citenamefont {Sakurai}\ and\ \citenamefont
  {Napolitano}(2011)}]{Sakurai}%
  \BibitemOpen
  \bibfield  {author} {\bibinfo {author} {\bibfnamefont {J.~J.}\ \bibnamefont
  {Sakurai}}\ and\ \bibinfo {author} {\bibfnamefont {J.}~\bibnamefont
  {Napolitano}},\ }\href {https://cds.cern.ch/record/1341875} {\emph {\bibinfo
  {title} {{Modern quantum mechanics; 2nd ed.}}}}\ (\bibinfo  {publisher}
  {Addison-Wesley},\ \bibinfo {address} {San Francisco, CA},\ \bibinfo {year}
  {2011})\BibitemShut {NoStop}%
\bibitem [{\citenamefont {Han}\ \emph {et~al.}(2012)\citenamefont {Han},
  \citenamefont {Helton}, \citenamefont {Chu}, \citenamefont {Nocera},
  \citenamefont {Rodriguez-Rivera}, \citenamefont {Broholm},\ and\
  \citenamefont {Lee}}]{Han2012}%
  \BibitemOpen
  \bibfield  {author} {\bibinfo {author} {\bibfnamefont {T.-H.}\ \bibnamefont
  {Han}}, \bibinfo {author} {\bibfnamefont {J.~S.}\ \bibnamefont {Helton}},
  \bibinfo {author} {\bibfnamefont {S.}~\bibnamefont {Chu}}, \bibinfo {author}
  {\bibfnamefont {D.~G.}\ \bibnamefont {Nocera}}, \bibinfo {author}
  {\bibfnamefont {J.~A.}\ \bibnamefont {Rodriguez-Rivera}}, \bibinfo {author}
  {\bibfnamefont {C.}~\bibnamefont {Broholm}},\ and\ \bibinfo {author}
  {\bibfnamefont {Y.~S.}\ \bibnamefont {Lee}},\ }\bibfield  {title} {\bibinfo
  {title} {Fractionalized excitations in the spin-liquid state of a
  kagome-lattice antiferromagnet},\ }\href
  {https://doi.org/10.1038/nature11659} {\bibfield  {journal} {\bibinfo
  {journal} {Nature}\ }\textbf {\bibinfo {volume} {492}},\ \bibinfo {pages}
  {406} (\bibinfo {year} {2012})}\BibitemShut {NoStop}%
\bibitem [{\citenamefont {Lvovsky}\ and\ \citenamefont
  {Raymer}(2009)}]{QSTRMP2009}%
  \BibitemOpen
  \bibfield  {author} {\bibinfo {author} {\bibfnamefont {A.~I.}\ \bibnamefont
  {Lvovsky}}\ and\ \bibinfo {author} {\bibfnamefont {M.~G.}\ \bibnamefont
  {Raymer}},\ }\bibfield  {title} {\bibinfo {title} {Continuous-variable
  optical quantum-state tomography},\ }\href
  {https://doi.org/10.1103/RevModPhys.81.299} {\bibfield  {journal} {\bibinfo
  {journal} {Rev. Mod. Phys.}\ }\textbf {\bibinfo {volume} {81}},\ \bibinfo
  {pages} {299} (\bibinfo {year} {2009})}\BibitemShut {NoStop}%
\bibitem [{\citenamefont {Torlai}\ \emph {et~al.}(2018)\citenamefont {Torlai},
  \citenamefont {Mazzola}, \citenamefont {Carrasquilla}, \citenamefont
  {Troyer}, \citenamefont {Melko},\ and\ \citenamefont {Carleo}}]{Torlai2018}%
  \BibitemOpen
  \bibfield  {author} {\bibinfo {author} {\bibfnamefont {G.}~\bibnamefont
  {Torlai}}, \bibinfo {author} {\bibfnamefont {G.}~\bibnamefont {Mazzola}},
  \bibinfo {author} {\bibfnamefont {J.}~\bibnamefont {Carrasquilla}}, \bibinfo
  {author} {\bibfnamefont {M.}~\bibnamefont {Troyer}}, \bibinfo {author}
  {\bibfnamefont {R.}~\bibnamefont {Melko}},\ and\ \bibinfo {author}
  {\bibfnamefont {G.}~\bibnamefont {Carleo}},\ }\bibfield  {title} {\bibinfo
  {title} {Neural-network quantum state tomography},\ }\href
  {https://doi.org/10.1038/s41567-018-0048-5} {\bibfield  {journal} {\bibinfo
  {journal} {Nature Physics}\ }\textbf {\bibinfo {volume} {14}},\ \bibinfo
  {pages} {447} (\bibinfo {year} {2018})}\BibitemShut {NoStop}%
\bibitem [{\citenamefont {Carrasquilla}\ \emph {et~al.}(2019)\citenamefont
  {Carrasquilla}, \citenamefont {Torlai}, \citenamefont {Melko},\ and\
  \citenamefont {Aolita}}]{Torlai2019}%
  \BibitemOpen
  \bibfield  {author} {\bibinfo {author} {\bibfnamefont {J.}~\bibnamefont
  {Carrasquilla}}, \bibinfo {author} {\bibfnamefont {G.}~\bibnamefont
  {Torlai}}, \bibinfo {author} {\bibfnamefont {R.~G.}\ \bibnamefont {Melko}},\
  and\ \bibinfo {author} {\bibfnamefont {L.}~\bibnamefont {Aolita}},\
  }\bibfield  {title} {\bibinfo {title} {Reconstructing quantum states with
  generative models},\ }\href {https://doi.org/10.1038/s42256-019-0028-1}
  {\bibfield  {journal} {\bibinfo  {journal} {Nat. Mach. Intell.}\ }\textbf
  {\bibinfo {volume} {1}},\ \bibinfo {pages} {155} (\bibinfo {year}
  {2019})}\BibitemShut {NoStop}%
\bibitem [{\citenamefont {Huang}\ \emph
  {et~al.}(2020{\natexlab{a}})\citenamefont {Huang}, \citenamefont {Kueng},\
  and\ \citenamefont {Preskill}}]{Huang2020}%
  \BibitemOpen
  \bibfield  {author} {\bibinfo {author} {\bibfnamefont {H.-Y.}\ \bibnamefont
  {Huang}}, \bibinfo {author} {\bibfnamefont {R.}~\bibnamefont {Kueng}},\ and\
  \bibinfo {author} {\bibfnamefont {J.}~\bibnamefont {Preskill}},\ }\bibfield
  {title} {\bibinfo {title} {Predicting many properties of a quantum system
  from very few measurements},\ }\href
  {https://doi.org/10.1038/s41567-020-0932-7} {\bibfield  {journal} {\bibinfo
  {journal} {Nature Physics}\ }\textbf {\bibinfo {volume} {16}},\ \bibinfo
  {pages} {1050} (\bibinfo {year} {2020}{\natexlab{a}})}\BibitemShut {NoStop}%
\bibitem [{\citenamefont {Zhao}\ \emph {et~al.}(2022)\citenamefont {Zhao},
  \citenamefont {Hu},\ and\ \citenamefont {Zhang}}]{Tianlun2022}%
  \BibitemOpen
  \bibfield  {author} {\bibinfo {author} {\bibfnamefont {T.-L.}\ \bibnamefont
  {Zhao}}, \bibinfo {author} {\bibfnamefont {S.-X.}\ \bibnamefont {Hu}},\ and\
  \bibinfo {author} {\bibfnamefont {Y.}~\bibnamefont {Zhang}},\ }\href
  {https://doi.org/10.48550/ARXIV.2212.13718} {\bibinfo {title} {Supervised
  hamiltonian learning via efficient and robust quantum descent}} (\bibinfo
  {year} {2022}),\ \Eprint {https://arxiv.org/abs/2212.13718}
  {arXiv:2212.13718} \BibitemShut {NoStop}%
\bibitem [{\citenamefont {Carleo}\ and\ \citenamefont
  {Troyer}(2017)}]{Carleo2016}%
  \BibitemOpen
  \bibfield  {author} {\bibinfo {author} {\bibfnamefont {G.}~\bibnamefont
  {Carleo}}\ and\ \bibinfo {author} {\bibfnamefont {M.}~\bibnamefont
  {Troyer}},\ }\bibfield  {title} {\bibinfo {title} {{Solving the quantum
  many-body problem with artificial neural networks}},\ }\href
  {https://doi.org/10.1126/science.aag2302} {\bibfield  {journal} {\bibinfo
  {journal} {Science}\ }\textbf {\bibinfo {volume} {355}},\ \bibinfo {pages}
  {602} (\bibinfo {year} {2017})}\BibitemShut {NoStop}%
\bibitem [{\citenamefont {Husz{\'a}r}\ and\ \citenamefont
  {Houlsby}(2012)}]{huszar2012adaptive}%
  \BibitemOpen
  \bibfield  {author} {\bibinfo {author} {\bibfnamefont {F.}~\bibnamefont
  {Husz{\'a}r}}\ and\ \bibinfo {author} {\bibfnamefont {N.~M.}\ \bibnamefont
  {Houlsby}},\ }\bibfield  {title} {\bibinfo {title} {Adaptive bayesian quantum
  tomography},\ }\href@noop {} {\bibfield  {journal} {\bibinfo  {journal}
  {Physical Review A}\ }\textbf {\bibinfo {volume} {85}},\ \bibinfo {pages}
  {052120} (\bibinfo {year} {2012})}\BibitemShut {NoStop}%
\bibitem [{Note1()}]{Note1}%
  \BibitemOpen
  \bibinfo {note} {Unlike the expectation value of a linear operator, the
  measurement energy $E(\Phi |\protect \mathcal {D})$ is explicitly nonlinear
  due to the $\protect \qopname \relax o{log}$ function. Therefore, the
  probability distribution of a quantum state with measurement outcomes offers
  realizations of exotic nonlinear-operator Hamiltonian.}\BibitemShut {Stop}%
\bibitem [{\citenamefont {Hradil}\ \emph {et~al.}(2004)\citenamefont {Hradil},
  \citenamefont {{\v{R}}eh{\'a}{\v{c}}ek}, \citenamefont {Fiur{\'a}{\v{s}}ek},\
  and\ \citenamefont {Je{\v{z}}ek}}]{hradil20043}%
  \BibitemOpen
  \bibfield  {author} {\bibinfo {author} {\bibfnamefont {Z.}~\bibnamefont
  {Hradil}}, \bibinfo {author} {\bibfnamefont {J.}~\bibnamefont
  {{\v{R}}eh{\'a}{\v{c}}ek}}, \bibinfo {author} {\bibfnamefont
  {J.}~\bibnamefont {Fiur{\'a}{\v{s}}ek}},\ and\ \bibinfo {author}
  {\bibfnamefont {M.}~\bibnamefont {Je{\v{z}}ek}},\ }\bibfield  {title}
  {\bibinfo {title} {3 maximum-likelihood methodsin quantum mechanics},\ }in\
  \href@noop {} {\emph {\bibinfo {booktitle} {Quantum state estimation}}}\
  (\bibinfo  {publisher} {Springer},\ \bibinfo {year} {2004})\ pp.\ \bibinfo
  {pages} {59--112}\BibitemShut {NoStop}%
\bibitem [{\citenamefont {Altepeter}\ \emph {et~al.}(2005)\citenamefont
  {Altepeter}, \citenamefont {Jeffrey},\ and\ \citenamefont
  {Kwiat}}]{ALTEPETER2005105}%
  \BibitemOpen
  \bibfield  {author} {\bibinfo {author} {\bibfnamefont {J.}~\bibnamefont
  {Altepeter}}, \bibinfo {author} {\bibfnamefont {E.}~\bibnamefont {Jeffrey}},\
  and\ \bibinfo {author} {\bibfnamefont {P.}~\bibnamefont {Kwiat}},\ }\bibfield
   {title} {\bibinfo {title} {Photonic state tomography}\ }(\bibinfo
  {publisher} {Academic Press},\ \bibinfo {year} {2005})\ pp.\ \bibinfo {pages}
  {105--159}\BibitemShut {NoStop}%
\bibitem [{\citenamefont {Hradil}(1997)}]{PhysRevA.55.R1561}%
  \BibitemOpen
  \bibfield  {author} {\bibinfo {author} {\bibfnamefont {Z.}~\bibnamefont
  {Hradil}},\ }\bibfield  {title} {\bibinfo {title} {Quantum-state
  estimation},\ }\href {https://doi.org/10.1103/PhysRevA.55.R1561} {\bibfield
  {journal} {\bibinfo  {journal} {Phys. Rev. A}\ }\textbf {\bibinfo {volume}
  {55}},\ \bibinfo {pages} {R1561} (\bibinfo {year} {1997})}\BibitemShut
  {NoStop}%
\bibitem [{\citenamefont {\ifmmode \check{R}\else
  \v{R}\fi{}eh\'a\ifmmode~\check{c}\else \v{c}\fi{}ek}\ \emph
  {et~al.}(2001)\citenamefont {\ifmmode \check{R}\else
  \v{R}\fi{}eh\'a\ifmmode~\check{c}\else \v{c}\fi{}ek}, \citenamefont
  {Hradil},\ and\ \citenamefont {Je\ifmmode~\check{z}\else
  \v{z}\fi{}ek}}]{PhysRevA.63.040303}%
  \BibitemOpen
  \bibfield  {author} {\bibinfo {author} {\bibfnamefont {J.}~\bibnamefont
  {\ifmmode \check{R}\else \v{R}\fi{}eh\'a\ifmmode~\check{c}\else
  \v{c}\fi{}ek}}, \bibinfo {author} {\bibfnamefont {Z.}~\bibnamefont
  {Hradil}},\ and\ \bibinfo {author} {\bibfnamefont {M.}~\bibnamefont
  {Je\ifmmode~\check{z}\else \v{z}\fi{}ek}},\ }\bibfield  {title} {\bibinfo
  {title} {Iterative algorithm for reconstruction of entangled states},\ }\href
  {https://doi.org/10.1103/PhysRevA.63.040303} {\bibfield  {journal} {\bibinfo
  {journal} {Phys. Rev. A}\ }\textbf {\bibinfo {volume} {63}},\ \bibinfo
  {pages} {040303} (\bibinfo {year} {2001})}\BibitemShut {NoStop}%
\bibitem [{\citenamefont {James}\ \emph {et~al.}(2001)\citenamefont {James},
  \citenamefont {Kwiat}, \citenamefont {Munro},\ and\ \citenamefont
  {White}}]{PhysRevA.64.052312}%
  \BibitemOpen
  \bibfield  {author} {\bibinfo {author} {\bibfnamefont {D.~F.~V.}\
  \bibnamefont {James}}, \bibinfo {author} {\bibfnamefont {P.~G.}\ \bibnamefont
  {Kwiat}}, \bibinfo {author} {\bibfnamefont {W.~J.}\ \bibnamefont {Munro}},\
  and\ \bibinfo {author} {\bibfnamefont {A.~G.}\ \bibnamefont {White}},\
  }\bibfield  {title} {\bibinfo {title} {Measurement of qubits},\ }\href
  {https://doi.org/10.1103/PhysRevA.64.052312} {\bibfield  {journal} {\bibinfo
  {journal} {Phys. Rev. A}\ }\textbf {\bibinfo {volume} {64}},\ \bibinfo
  {pages} {052312} (\bibinfo {year} {2001})}\BibitemShut {NoStop}%
\bibitem [{\citenamefont {Shang}\ \emph {et~al.}(2017)\citenamefont {Shang},
  \citenamefont {Zhang},\ and\ \citenamefont {Ng}}]{PhysRevA.95.062336}%
  \BibitemOpen
  \bibfield  {author} {\bibinfo {author} {\bibfnamefont {J.}~\bibnamefont
  {Shang}}, \bibinfo {author} {\bibfnamefont {Z.}~\bibnamefont {Zhang}},\ and\
  \bibinfo {author} {\bibfnamefont {H.~K.}\ \bibnamefont {Ng}},\ }\bibfield
  {title} {\bibinfo {title} {Superfast maximum-likelihood reconstruction for
  quantum tomography},\ }\href {https://doi.org/10.1103/PhysRevA.95.062336}
  {\bibfield  {journal} {\bibinfo  {journal} {Phys. Rev. A}\ }\textbf {\bibinfo
  {volume} {95}},\ \bibinfo {pages} {062336} (\bibinfo {year}
  {2017})}\BibitemShut {NoStop}%
\bibitem [{\citenamefont {\ifmmode \check{R}\else
  \v{R}\fi{}eh\'a\ifmmode~\check{c}\else \v{c}\fi{}ek}\ \emph
  {et~al.}(2007)\citenamefont {\ifmmode \check{R}\else
  \v{R}\fi{}eh\'a\ifmmode~\check{c}\else \v{c}\fi{}ek}, \citenamefont {Hradil},
  \citenamefont {Knill},\ and\ \citenamefont {Lvovsky}}]{PhysRevA.75.042108}%
  \BibitemOpen
  \bibfield  {author} {\bibinfo {author} {\bibfnamefont {J.}~\bibnamefont
  {\ifmmode \check{R}\else \v{R}\fi{}eh\'a\ifmmode~\check{c}\else
  \v{c}\fi{}ek}}, \bibinfo {author} {\bibfnamefont {Z.~c.~v.}\ \bibnamefont
  {Hradil}}, \bibinfo {author} {\bibfnamefont {E.}~\bibnamefont {Knill}},\ and\
  \bibinfo {author} {\bibfnamefont {A.~I.}\ \bibnamefont {Lvovsky}},\
  }\bibfield  {title} {\bibinfo {title} {Diluted maximum-likelihood algorithm
  for quantum tomography},\ }\href {https://doi.org/10.1103/PhysRevA.75.042108}
  {\bibfield  {journal} {\bibinfo  {journal} {Phys. Rev. A}\ }\textbf {\bibinfo
  {volume} {75}},\ \bibinfo {pages} {042108} (\bibinfo {year}
  {2007})}\BibitemShut {NoStop}%
\bibitem [{\citenamefont {{Michael Nielsen}}(2013)}]{MLbook}%
  \BibitemOpen
  \bibfield  {author} {\bibinfo {author} {\bibnamefont {{Michael Nielsen}}},\
  }\href@noop {} {\emph {\bibinfo {title} {{Neural Networks and Deep
  Learning}}}}\ (\bibinfo  {publisher} {{Free Online Book}},\ \bibinfo {year}
  {2013})\BibitemShut {NoStop}%
\bibitem [{\citenamefont {Carrasquilla}\ and\ \citenamefont
  {Melko}(2017)}]{Melko20161}%
  \BibitemOpen
  \bibfield  {author} {\bibinfo {author} {\bibfnamefont {J.}~\bibnamefont
  {Carrasquilla}}\ and\ \bibinfo {author} {\bibfnamefont {R.~G.}\ \bibnamefont
  {Melko}},\ }\bibfield  {title} {\bibinfo {title} {Machine learning phases of
  matter},\ }\href {http://dx.doi.org/10.1038/nphys4035} {\bibfield  {journal}
  {\bibinfo  {journal} {Nature Physics}\ }\textbf {\bibinfo {volume} {13}},\
  \bibinfo {pages} {431} (\bibinfo {year} {2017})}\BibitemShut {NoStop}%
\bibitem [{\citenamefont {Zhang}\ and\ \citenamefont {Kim}(2017)}]{qlt2016}%
  \BibitemOpen
  \bibfield  {author} {\bibinfo {author} {\bibfnamefont {Y.}~\bibnamefont
  {Zhang}}\ and\ \bibinfo {author} {\bibfnamefont {E.-A.}\ \bibnamefont
  {Kim}},\ }\bibfield  {title} {\bibinfo {title} {{Quantum Loop Topography for
  Machine Learning}},\ }\href {https://doi.org/10.1103/PhysRevLett.118.216401}
  {\bibfield  {journal} {\bibinfo  {journal} {Phys. Rev. Lett.}\ }\textbf
  {\bibinfo {volume} {118}},\ \bibinfo {pages} {216401} (\bibinfo {year}
  {2017})}\BibitemShut {NoStop}%
\bibitem [{\citenamefont {Zhang}\ \emph {et~al.}(2019)\citenamefont {Zhang},
  \citenamefont {Mesaros}, \citenamefont {Fujita}, \citenamefont {Edkins},
  \citenamefont {Hamidian}, \citenamefont {Ch'ng}, \citenamefont {Eisaki},
  \citenamefont {Uchida}, \citenamefont {Davis}, \citenamefont {Khatami} \emph
  {et~al.}}]{mlstm2019}%
  \BibitemOpen
  \bibfield  {author} {\bibinfo {author} {\bibfnamefont {Y.}~\bibnamefont
  {Zhang}}, \bibinfo {author} {\bibfnamefont {A.}~\bibnamefont {Mesaros}},
  \bibinfo {author} {\bibfnamefont {K.}~\bibnamefont {Fujita}}, \bibinfo
  {author} {\bibfnamefont {S.}~\bibnamefont {Edkins}}, \bibinfo {author}
  {\bibfnamefont {M.}~\bibnamefont {Hamidian}}, \bibinfo {author}
  {\bibfnamefont {K.}~\bibnamefont {Ch'ng}}, \bibinfo {author} {\bibfnamefont
  {H.}~\bibnamefont {Eisaki}}, \bibinfo {author} {\bibfnamefont
  {S.}~\bibnamefont {Uchida}}, \bibinfo {author} {\bibfnamefont {J.~S.}\
  \bibnamefont {Davis}}, \bibinfo {author} {\bibfnamefont {E.}~\bibnamefont
  {Khatami}}, \emph {et~al.},\ }\bibfield  {title} {\bibinfo {title} {Machine
  learning in electronic-quantum-matter imaging experiments},\ }\href
  {https://doi.org/10.1038/s41586-019-1319-8} {\bibfield  {journal} {\bibinfo
  {journal} {Nature}\ }\textbf {\bibinfo {volume} {570}},\ \bibinfo {pages}
  {484} (\bibinfo {year} {2019})}\BibitemShut {NoStop}%
\bibitem [{Sup()}]{SuppQMeasure}%
  \BibitemOpen
  \href@noop {} {}\bibinfo {note} {See examples and details on
  hyper-parameters, finite number of measurements, interacting fermion models,
  topological degeneracy of the Kitaev model, contingency plan for
  insufficiency observables, generalization to mixed states and Haar random
  states without a-priori knowledge in Supplemental Materials, which also
  include Ref. \onlinecite{PhysRevLett.105.027204, mandal2020introduction,
  PhysRevLett.73.2158, PhysRevB.84.165414, PhysRevB.92.014403,
  PhysRevB.97.241110, MLbook_ap, PhysRevA.85.042317, Biswas_2021}.}\BibitemShut
  {Stop}%
\bibitem [{\citenamefont {Fannes}\ \emph {et~al.}(1992)\citenamefont {Fannes},
  \citenamefont {Nachtergaele},\ and\ \citenamefont {Werner}}]{MPS1992}%
  \BibitemOpen
  \bibfield  {author} {\bibinfo {author} {\bibfnamefont {M.}~\bibnamefont
  {Fannes}}, \bibinfo {author} {\bibfnamefont {B.}~\bibnamefont
  {Nachtergaele}},\ and\ \bibinfo {author} {\bibfnamefont {R.~F.}\ \bibnamefont
  {Werner}},\ }\bibfield  {title} {\bibinfo {title} {Finitely correlated states
  on quantum spin chains},\ }\href {https://doi.org/10.1007/BF02099178}
  {\bibfield  {journal} {\bibinfo  {journal} {Communications in Mathematical
  Physics}\ }\textbf {\bibinfo {volume} {144}},\ \bibinfo {pages} {443}
  (\bibinfo {year} {1992})}\BibitemShut {NoStop}%
\bibitem [{\citenamefont {Schollw\"ock}(2005)}]{DMRG2005}%
  \BibitemOpen
  \bibfield  {author} {\bibinfo {author} {\bibfnamefont {U.}~\bibnamefont
  {Schollw\"ock}},\ }\bibfield  {title} {\bibinfo {title} {The density-matrix
  renormalization group},\ }\href {https://doi.org/10.1103/RevModPhys.77.259}
  {\bibfield  {journal} {\bibinfo  {journal} {Rev. Mod. Phys.}\ }\textbf
  {\bibinfo {volume} {77}},\ \bibinfo {pages} {259} (\bibinfo {year}
  {2005})}\BibitemShut {NoStop}%
\bibitem [{\citenamefont {Foulkes}\ \emph {et~al.}(2001)\citenamefont
  {Foulkes}, \citenamefont {Mitas}, \citenamefont {Needs},\ and\ \citenamefont
  {Rajagopal}}]{QMCreview2001}%
  \BibitemOpen
  \bibfield  {author} {\bibinfo {author} {\bibfnamefont {W.~M.~C.}\
  \bibnamefont {Foulkes}}, \bibinfo {author} {\bibfnamefont {L.}~\bibnamefont
  {Mitas}}, \bibinfo {author} {\bibfnamefont {R.~J.}\ \bibnamefont {Needs}},\
  and\ \bibinfo {author} {\bibfnamefont {G.}~\bibnamefont {Rajagopal}},\
  }\bibfield  {title} {\bibinfo {title} {Quantum monte carlo simulations of
  solids},\ }\href {https://doi.org/10.1103/RevModPhys.73.33} {\bibfield
  {journal} {\bibinfo  {journal} {Rev. Mod. Phys.}\ }\textbf {\bibinfo {volume}
  {73}},\ \bibinfo {pages} {33} (\bibinfo {year} {2001})}\BibitemShut {NoStop}%
\bibitem [{\citenamefont {Troyer}\ and\ \citenamefont
  {Wiese}(2005)}]{Troyer2005}%
  \BibitemOpen
  \bibfield  {author} {\bibinfo {author} {\bibfnamefont {M.}~\bibnamefont
  {Troyer}}\ and\ \bibinfo {author} {\bibfnamefont {U.-J.}\ \bibnamefont
  {Wiese}},\ }\bibfield  {title} {\bibinfo {title} {Computational complexity
  and fundamental limitations to fermionic quantum monte carlo simulations},\
  }\href {https://doi.org/10.1103/PhysRevLett.94.170201} {\bibfield  {journal}
  {\bibinfo  {journal} {Phys. Rev. Lett.}\ }\textbf {\bibinfo {volume} {94}},\
  \bibinfo {pages} {170201} (\bibinfo {year} {2005})}\BibitemShut {NoStop}%
\bibitem [{\citenamefont {Farhi}\ \emph {et~al.}(2014)\citenamefont {Farhi},
  \citenamefont {Goldstone},\ and\ \citenamefont {Gutmann}}]{QAOA2014Farhi}%
  \BibitemOpen
  \bibfield  {author} {\bibinfo {author} {\bibfnamefont {E.}~\bibnamefont
  {Farhi}}, \bibinfo {author} {\bibfnamefont {J.}~\bibnamefont {Goldstone}},\
  and\ \bibinfo {author} {\bibfnamefont {S.}~\bibnamefont {Gutmann}},\
  }\href@noop {} {\bibinfo {title} {A quantum approximate optimization
  algorithm}} (\bibinfo {year} {2014}),\ \Eprint
  {https://arxiv.org/abs/1411.4028} {arXiv:1411.4028 [quant-ph]} \BibitemShut
  {NoStop}%
\bibitem [{\citenamefont {Zhou}\ \emph {et~al.}(2020)\citenamefont {Zhou},
  \citenamefont {Wang}, \citenamefont {Choi}, \citenamefont {Pichler},\ and\
  \citenamefont {Lukin}}]{QAOA2020PRX}%
  \BibitemOpen
  \bibfield  {author} {\bibinfo {author} {\bibfnamefont {L.}~\bibnamefont
  {Zhou}}, \bibinfo {author} {\bibfnamefont {S.-T.}\ \bibnamefont {Wang}},
  \bibinfo {author} {\bibfnamefont {S.}~\bibnamefont {Choi}}, \bibinfo {author}
  {\bibfnamefont {H.}~\bibnamefont {Pichler}},\ and\ \bibinfo {author}
  {\bibfnamefont {M.~D.}\ \bibnamefont {Lukin}},\ }\bibfield  {title} {\bibinfo
  {title} {Quantum approximate optimization algorithm: Performance, mechanism,
  and implementation on near-term devices},\ }\href
  {https://doi.org/10.1103/PhysRevX.10.021067} {\bibfield  {journal} {\bibinfo
  {journal} {Phys. Rev. X}\ }\textbf {\bibinfo {volume} {10}},\ \bibinfo
  {pages} {021067} (\bibinfo {year} {2020})}\BibitemShut {NoStop}%
\bibitem [{\citenamefont {Vikst\aa{}l}\ \emph {et~al.}(2020)\citenamefont
  {Vikst\aa{}l}, \citenamefont {Gr\"onkvist}, \citenamefont {Svensson},
  \citenamefont {Andersson}, \citenamefont {Johansson},\ and\ \citenamefont
  {Ferrini}}]{QAOA2020PRApplied}%
  \BibitemOpen
  \bibfield  {author} {\bibinfo {author} {\bibfnamefont {P.}~\bibnamefont
  {Vikst\aa{}l}}, \bibinfo {author} {\bibfnamefont {M.}~\bibnamefont
  {Gr\"onkvist}}, \bibinfo {author} {\bibfnamefont {M.}~\bibnamefont
  {Svensson}}, \bibinfo {author} {\bibfnamefont {M.}~\bibnamefont {Andersson}},
  \bibinfo {author} {\bibfnamefont {G.}~\bibnamefont {Johansson}},\ and\
  \bibinfo {author} {\bibfnamefont {G.}~\bibnamefont {Ferrini}},\ }\bibfield
  {title} {\bibinfo {title} {Applying the quantum approximate optimization
  algorithm to the tail-assignment problem},\ }\href
  {https://doi.org/10.1103/PhysRevApplied.14.034009} {\bibfield  {journal}
  {\bibinfo  {journal} {Phys. Rev. Applied}\ }\textbf {\bibinfo {volume}
  {14}},\ \bibinfo {pages} {034009} (\bibinfo {year} {2020})}\BibitemShut
  {NoStop}%
\bibitem [{Note2()}]{Note2}%
  \BibitemOpen
  \bibinfo {note} {A lack of observables may lead to misleading MLE states,
  which we can identify with signatures in measurement energy: it may constrain
  the search space leading to a sub-optimal MLE state as the measurement energy
  converges above its lower bound, or end up with different MLE states
  simultaneously consistent with the measurement outcomes.}\BibitemShut {Stop}%
\bibitem [{Note3()}]{Note3}%
  \BibitemOpen
  \bibinfo {note} {An observable with more eigenvalues acts as a double-edged
  sword: they may incur cost in post-processing $\protect \hat {H}_{eff}$ due
  to more complex $\protect \hat {P}_{\tau }$, but the distribution also offer
  more information than the average in a similar spirit to shot-noise studies
  \cite {Zhou2019, Sivre2019}. We can make an observable simpler to handle by
  binning together some outcomes and giving up some information, but not vice
  versa.}\BibitemShut {Stop}%
\bibitem [{Note4()}]{Note4}%
  \BibitemOpen
  \bibinfo {note} {The spikes in the figure are mainly due to the inconsistent
  particle number the iteration state $|\Phi _0\rangle $ receives over the
  slight modifications. Better convergence largely suppresses such phenomena in
  later iterations.}\BibitemShut {Stop}%
\bibitem [{\citenamefont {Rau}\ \emph {et~al.}(2014)\citenamefont {Rau},
  \citenamefont {Lee},\ and\ \citenamefont {Kee}}]{PhysRevLett.112.077204}%
  \BibitemOpen
  \bibfield  {author} {\bibinfo {author} {\bibfnamefont {J.~G.}\ \bibnamefont
  {Rau}}, \bibinfo {author} {\bibfnamefont {E.~K.-H.}\ \bibnamefont {Lee}},\
  and\ \bibinfo {author} {\bibfnamefont {H.-Y.}\ \bibnamefont {Kee}},\
  }\bibfield  {title} {\bibinfo {title} {Generic spin model for the honeycomb
  iridates beyond the kitaev limit},\ }\href
  {https://doi.org/10.1103/PhysRevLett.112.077204} {\bibfield  {journal}
  {\bibinfo  {journal} {Phys. Rev. Lett.}\ }\textbf {\bibinfo {volume} {112}},\
  \bibinfo {pages} {077204} (\bibinfo {year} {2014})}\BibitemShut {NoStop}%
\bibitem [{\citenamefont {Hastings}\ and\ \citenamefont
  {Wen}(2005)}]{Wen2005prb}%
  \BibitemOpen
  \bibfield  {author} {\bibinfo {author} {\bibfnamefont {M.~B.}\ \bibnamefont
  {Hastings}}\ and\ \bibinfo {author} {\bibfnamefont {X.-G.}\ \bibnamefont
  {Wen}},\ }\bibfield  {title} {\bibinfo {title} {Quasiadiabatic continuation
  of quantum states: The stability of topological ground-state degeneracy and
  emergent gauge invariance},\ }\href
  {https://doi.org/10.1103/PhysRevB.72.045141} {\bibfield  {journal} {\bibinfo
  {journal} {Phys. Rev. B}\ }\textbf {\bibinfo {volume} {72}},\ \bibinfo
  {pages} {045141} (\bibinfo {year} {2005})}\BibitemShut {NoStop}%
\bibitem [{\citenamefont {Verresen}\ \emph {et~al.}(2021)\citenamefont
  {Verresen}, \citenamefont {Lukin},\ and\ \citenamefont
  {Vishwanath}}]{Rydberg2021theory}%
  \BibitemOpen
  \bibfield  {author} {\bibinfo {author} {\bibfnamefont {R.}~\bibnamefont
  {Verresen}}, \bibinfo {author} {\bibfnamefont {M.~D.}\ \bibnamefont
  {Lukin}},\ and\ \bibinfo {author} {\bibfnamefont {A.}~\bibnamefont
  {Vishwanath}},\ }\bibfield  {title} {\bibinfo {title} {Prediction of toric
  code topological order from rydberg blockade},\ }\href
  {https://doi.org/10.1103/PhysRevX.11.031005} {\bibfield  {journal} {\bibinfo
  {journal} {Phys. Rev. X}\ }\textbf {\bibinfo {volume} {11}},\ \bibinfo
  {pages} {031005} (\bibinfo {year} {2021})}\BibitemShut {NoStop}%
\bibitem [{\citenamefont {Semeghini}\ \emph {et~al.}(2021)\citenamefont
  {Semeghini}, \citenamefont {Levine}, \citenamefont {Keesling}, \citenamefont
  {Ebadi}, \citenamefont {Wang}, \citenamefont {Bluvstein}, \citenamefont
  {Verresen}, \citenamefont {Pichler}, \citenamefont {Kalinowski},
  \citenamefont {Samajdar}, \citenamefont {Omran}, \citenamefont {Sachdev},
  \citenamefont {Vishwanath}, \citenamefont {Greiner}, \citenamefont
  {Vuletić},\ and\ \citenamefont {Lukin}}]{Rydberg2021}%
  \BibitemOpen
  \bibfield  {author} {\bibinfo {author} {\bibfnamefont {G.}~\bibnamefont
  {Semeghini}}, \bibinfo {author} {\bibfnamefont {H.}~\bibnamefont {Levine}},
  \bibinfo {author} {\bibfnamefont {A.}~\bibnamefont {Keesling}}, \bibinfo
  {author} {\bibfnamefont {S.}~\bibnamefont {Ebadi}}, \bibinfo {author}
  {\bibfnamefont {T.~T.}\ \bibnamefont {Wang}}, \bibinfo {author}
  {\bibfnamefont {D.}~\bibnamefont {Bluvstein}}, \bibinfo {author}
  {\bibfnamefont {R.}~\bibnamefont {Verresen}}, \bibinfo {author}
  {\bibfnamefont {H.}~\bibnamefont {Pichler}}, \bibinfo {author} {\bibfnamefont
  {M.}~\bibnamefont {Kalinowski}}, \bibinfo {author} {\bibfnamefont
  {R.}~\bibnamefont {Samajdar}}, \bibinfo {author} {\bibfnamefont
  {A.}~\bibnamefont {Omran}}, \bibinfo {author} {\bibfnamefont
  {S.}~\bibnamefont {Sachdev}}, \bibinfo {author} {\bibfnamefont
  {A.}~\bibnamefont {Vishwanath}}, \bibinfo {author} {\bibfnamefont
  {M.}~\bibnamefont {Greiner}}, \bibinfo {author} {\bibfnamefont
  {V.}~\bibnamefont {Vuletić}},\ and\ \bibinfo {author} {\bibfnamefont
  {M.~D.}\ \bibnamefont {Lukin}},\ }\bibfield  {title} {\bibinfo {title}
  {Probing topological spin liquids on a programmable quantum simulator},\
  }\href {https://doi.org/10.1126/science.abi8794} {\bibfield  {journal}
  {\bibinfo  {journal} {Science}\ }\textbf {\bibinfo {volume} {374}},\ \bibinfo
  {pages} {1242} (\bibinfo {year} {2021})}\BibitemShut {NoStop}%
\bibitem [{\citenamefont {Samajdar}\ \emph {et~al.}(2021)\citenamefont
  {Samajdar}, \citenamefont {Ho}, \citenamefont {Pichler}, \citenamefont
  {Lukin},\ and\ \citenamefont {Sachdev}}]{Rhine2021}%
  \BibitemOpen
  \bibfield  {author} {\bibinfo {author} {\bibfnamefont {R.}~\bibnamefont
  {Samajdar}}, \bibinfo {author} {\bibfnamefont {W.~W.}\ \bibnamefont {Ho}},
  \bibinfo {author} {\bibfnamefont {H.}~\bibnamefont {Pichler}}, \bibinfo
  {author} {\bibfnamefont {M.~D.}\ \bibnamefont {Lukin}},\ and\ \bibinfo
  {author} {\bibfnamefont {S.}~\bibnamefont {Sachdev}},\ }\bibfield  {title}
  {\bibinfo {title} {Quantum phases of rydberg atoms on a kagome lattice},\
  }\href {https://doi.org/10.1073/pnas.2015785118} {\bibfield  {journal}
  {\bibinfo  {journal} {Proceedings of the National Academy of Sciences}\
  }\textbf {\bibinfo {volume} {118}},\ \bibinfo {pages} {e2015785118} (\bibinfo
  {year} {2021})},\ \Eprint
  {https://arxiv.org/abs/https://www.pnas.org/doi/pdf/10.1073/pnas.2015785118}
  {https://www.pnas.org/doi/pdf/10.1073/pnas.2015785118} \BibitemShut {NoStop}%
\bibitem [{\citenamefont {Balatsky}\ \emph {et~al.}(2012)\citenamefont
  {Balatsky}, \citenamefont {Nishijima},\ and\ \citenamefont
  {Manassen}}]{AiP2012}%
  \BibitemOpen
  \bibfield  {author} {\bibinfo {author} {\bibfnamefont {A.~V.}\ \bibnamefont
  {Balatsky}}, \bibinfo {author} {\bibfnamefont {M.}~\bibnamefont
  {Nishijima}},\ and\ \bibinfo {author} {\bibfnamefont {Y.}~\bibnamefont
  {Manassen}},\ }\bibfield  {title} {\bibinfo {title} {Electron spin
  resonance-scanning tunneling microscopy},\ }\href
  {https://doi.org/10.1080/00018732.2012.668775} {\bibfield  {journal}
  {\bibinfo  {journal} {Advances in Physics}\ }\textbf {\bibinfo {volume}
  {61}},\ \bibinfo {pages} {117} (\bibinfo {year} {2012})}\BibitemShut
  {NoStop}%
\bibitem [{\citenamefont {Ternes}(2015)}]{ESRSTM2015}%
  \BibitemOpen
  \bibfield  {author} {\bibinfo {author} {\bibfnamefont {M.}~\bibnamefont
  {Ternes}},\ }\bibfield  {title} {\bibinfo {title} {Spin excitations and
  correlations in scanning tunneling spectroscopy},\ }\href
  {https://doi.org/10.1088/1367-2630/17/6/063016} {\bibfield  {journal}
  {\bibinfo  {journal} {New Journal of Physics}\ }\textbf {\bibinfo {volume}
  {17}},\ \bibinfo {pages} {063016} (\bibinfo {year} {2015})}\BibitemShut
  {NoStop}%
\bibitem [{Note5()}]{Note5}%
  \BibitemOpen
  \bibinfo {note} {Note that a measurement-energy lower bound is no longer
  available for such single-shot measurements.}\BibitemShut {Stop}%
\bibitem [{\citenamefont {Huang}\ \emph
  {et~al.}(2020{\natexlab{b}})\citenamefont {Huang}, \citenamefont {Kueng},\
  and\ \citenamefont {Preskill}}]{huang2020predicting}%
  \BibitemOpen
  \bibfield  {author} {\bibinfo {author} {\bibfnamefont {H.-Y.}\ \bibnamefont
  {Huang}}, \bibinfo {author} {\bibfnamefont {R.}~\bibnamefont {Kueng}},\ and\
  \bibinfo {author} {\bibfnamefont {J.}~\bibnamefont {Preskill}},\ }\bibfield
  {title} {\bibinfo {title} {Predicting many properties of a quantum system
  from very few measurements},\ }\href@noop {} {\bibfield  {journal} {\bibinfo
  {journal} {Nature Physics}\ }\textbf {\bibinfo {volume} {16}},\ \bibinfo
  {pages} {1050} (\bibinfo {year} {2020}{\natexlab{b}})}\BibitemShut {NoStop}%
\bibitem [{\citenamefont {Qi}\ and\ \citenamefont {Ranard}(2019)}]{QiHamrecon}%
  \BibitemOpen
  \bibfield  {author} {\bibinfo {author} {\bibfnamefont {X.-L.}\ \bibnamefont
  {Qi}}\ and\ \bibinfo {author} {\bibfnamefont {D.}~\bibnamefont {Ranard}},\
  }\bibfield  {title} {\bibinfo {title} {Determining a local hamiltonian from a
  single eigenstate},\ }\href
  {https://doi.org/https://doi.org/10.22331/q-2019-07-08-159} {\bibfield
  {journal} {\bibinfo  {journal} {Quantum}\ }\textbf {\bibinfo {volume} {3}},\
  \bibinfo {pages} {159} (\bibinfo {year} {2019})}\BibitemShut {NoStop}%
\bibitem [{\citenamefont {Turkeshi}\ \emph {et~al.}(2019)\citenamefont
  {Turkeshi}, \citenamefont {Mendes-Santos}, \citenamefont {Giudici},\ and\
  \citenamefont {Dalmonte}}]{Hamrec2019}%
  \BibitemOpen
  \bibfield  {author} {\bibinfo {author} {\bibfnamefont {X.}~\bibnamefont
  {Turkeshi}}, \bibinfo {author} {\bibfnamefont {T.}~\bibnamefont
  {Mendes-Santos}}, \bibinfo {author} {\bibfnamefont {G.}~\bibnamefont
  {Giudici}},\ and\ \bibinfo {author} {\bibfnamefont {M.}~\bibnamefont
  {Dalmonte}},\ }\bibfield  {title} {\bibinfo {title} {Entanglement-guided
  search for parent hamiltonians},\ }\href
  {https://doi.org/10.1103/PhysRevLett.122.150606} {\bibfield  {journal}
  {\bibinfo  {journal} {Phys. Rev. Lett.}\ }\textbf {\bibinfo {volume} {122}},\
  \bibinfo {pages} {150606} (\bibinfo {year} {2019})}\BibitemShut {NoStop}%
\bibitem [{\citenamefont {Chaloupka}\ \emph {et~al.}(2010)\citenamefont
  {Chaloupka}, \citenamefont {Jackeli},\ and\ \citenamefont
  {Khaliullin}}]{PhysRevLett.105.027204}%
  \BibitemOpen
  \bibfield  {author} {\bibinfo {author} {\bibfnamefont {J.~c.~v.}\
  \bibnamefont {Chaloupka}}, \bibinfo {author} {\bibfnamefont {G.}~\bibnamefont
  {Jackeli}},\ and\ \bibinfo {author} {\bibfnamefont {G.}~\bibnamefont
  {Khaliullin}},\ }\bibfield  {title} {\bibinfo {title} {Kitaev-heisenberg
  model on a honeycomb lattice: Possible exotic phases in iridium oxides
  ${A}_{2}{\mathrm{iro}}_{3}$},\ }\href
  {https://doi.org/10.1103/PhysRevLett.105.027204} {\bibfield  {journal}
  {\bibinfo  {journal} {Phys. Rev. Lett.}\ }\textbf {\bibinfo {volume} {105}},\
  \bibinfo {pages} {027204} (\bibinfo {year} {2010})}\BibitemShut {NoStop}%
\bibitem [{\citenamefont {Mandal}\ and\ \citenamefont
  {Jayannavar}(2020)}]{mandal2020introduction}%
  \BibitemOpen
  \bibfield  {author} {\bibinfo {author} {\bibfnamefont {S.}~\bibnamefont
  {Mandal}}\ and\ \bibinfo {author} {\bibfnamefont {A.~M.}\ \bibnamefont
  {Jayannavar}},\ }\bibfield  {title} {\bibinfo {title} {An introduction to
  kitaev model-i},\ }\href@noop {} {\bibfield  {journal} {\bibinfo  {journal}
  {arXiv preprint arXiv:2006.11549}\ } (\bibinfo {year} {2020})}\BibitemShut
  {NoStop}%
\bibitem [{\citenamefont {Lieb}(1994)}]{PhysRevLett.73.2158}%
  \BibitemOpen
  \bibfield  {author} {\bibinfo {author} {\bibfnamefont {E.~H.}\ \bibnamefont
  {Lieb}},\ }\bibfield  {title} {\bibinfo {title} {Flux phase of the
  half-filled band},\ }\href {https://doi.org/10.1103/PhysRevLett.73.2158}
  {\bibfield  {journal} {\bibinfo  {journal} {Phys. Rev. Lett.}\ }\textbf
  {\bibinfo {volume} {73}},\ \bibinfo {pages} {2158} (\bibinfo {year}
  {1994})}\BibitemShut {NoStop}%
\bibitem [{\citenamefont {Pedrocchi}\ \emph {et~al.}(2011)\citenamefont
  {Pedrocchi}, \citenamefont {Chesi},\ and\ \citenamefont
  {Loss}}]{PhysRevB.84.165414}%
  \BibitemOpen
  \bibfield  {author} {\bibinfo {author} {\bibfnamefont {F.~L.}\ \bibnamefont
  {Pedrocchi}}, \bibinfo {author} {\bibfnamefont {S.}~\bibnamefont {Chesi}},\
  and\ \bibinfo {author} {\bibfnamefont {D.}~\bibnamefont {Loss}},\ }\bibfield
  {title} {\bibinfo {title} {Physical solutions of the kitaev honeycomb
  model},\ }\href {https://doi.org/10.1103/PhysRevB.84.165414} {\bibfield
  {journal} {\bibinfo  {journal} {Phys. Rev. B}\ }\textbf {\bibinfo {volume}
  {84}},\ \bibinfo {pages} {165414} (\bibinfo {year} {2011})}\BibitemShut
  {NoStop}%
\bibitem [{\citenamefont {Zschocke}\ and\ \citenamefont
  {Vojta}(2015)}]{PhysRevB.92.014403}%
  \BibitemOpen
  \bibfield  {author} {\bibinfo {author} {\bibfnamefont {F.}~\bibnamefont
  {Zschocke}}\ and\ \bibinfo {author} {\bibfnamefont {M.}~\bibnamefont
  {Vojta}},\ }\bibfield  {title} {\bibinfo {title} {Physical states and
  finite-size effects in kitaev's honeycomb model: Bond disorder, spin
  excitations, and nmr line shape},\ }\href
  {https://doi.org/10.1103/PhysRevB.92.014403} {\bibfield  {journal} {\bibinfo
  {journal} {Phys. Rev. B}\ }\textbf {\bibinfo {volume} {92}},\ \bibinfo
  {pages} {014403} (\bibinfo {year} {2015})}\BibitemShut {NoStop}%
\bibitem [{\citenamefont {Zhu}\ \emph {et~al.}(2018)\citenamefont {Zhu},
  \citenamefont {Kimchi}, \citenamefont {Sheng},\ and\ \citenamefont
  {Fu}}]{PhysRevB.97.241110}%
  \BibitemOpen
  \bibfield  {author} {\bibinfo {author} {\bibfnamefont {Z.}~\bibnamefont
  {Zhu}}, \bibinfo {author} {\bibfnamefont {I.}~\bibnamefont {Kimchi}},
  \bibinfo {author} {\bibfnamefont {D.~N.}\ \bibnamefont {Sheng}},\ and\
  \bibinfo {author} {\bibfnamefont {L.}~\bibnamefont {Fu}},\ }\bibfield
  {title} {\bibinfo {title} {Robust non-abelian spin liquid and a possible
  intermediate phase in the antiferromagnetic kitaev model with magnetic
  field},\ }\href {https://doi.org/10.1103/PhysRevB.97.241110} {\bibfield
  {journal} {\bibinfo  {journal} {Phys. Rev. B}\ }\textbf {\bibinfo {volume}
  {97}},\ \bibinfo {pages} {241110} (\bibinfo {year} {2018})}\BibitemShut
  {NoStop}%
\bibitem [{\citenamefont {Dawid}\ \emph {et~al.}(2022)\citenamefont {Dawid},
  \citenamefont {Arnold}, \citenamefont {Requena}, \citenamefont {Gresch},
  \citenamefont {Płodzień}, \citenamefont {Donatella}, \citenamefont
  {Nicoli}, \citenamefont {Stornati}, \citenamefont {Koch}, \citenamefont
  {Büttner}, \citenamefont {Okuła}, \citenamefont {Muñoz-Gil}, \citenamefont
  {Vargas-Hernández}, \citenamefont {Cervera-Lierta}, \citenamefont
  {Carrasquilla}, \citenamefont {Dunjko}, \citenamefont {Gabrié},
  \citenamefont {Huembeli}, \citenamefont {van Nieuwenburg}, \citenamefont
  {Vicentini}, \citenamefont {Wang}, \citenamefont {Wetzel}, \citenamefont
  {Carleo}, \citenamefont {Greplová}, \citenamefont {Krems}, \citenamefont
  {Marquardt}, \citenamefont {Tomza}, \citenamefont {Lewenstein},\ and\
  \citenamefont {Dauphin}}]{MLbook_ap}%
  \BibitemOpen
  \bibfield  {author} {\bibinfo {author} {\bibfnamefont {A.}~\bibnamefont
  {Dawid}}, \bibinfo {author} {\bibfnamefont {J.}~\bibnamefont {Arnold}},
  \bibinfo {author} {\bibfnamefont {B.}~\bibnamefont {Requena}}, \bibinfo
  {author} {\bibfnamefont {A.}~\bibnamefont {Gresch}}, \bibinfo {author}
  {\bibfnamefont {M.}~\bibnamefont {Płodzień}}, \bibinfo {author}
  {\bibfnamefont {K.}~\bibnamefont {Donatella}}, \bibinfo {author}
  {\bibfnamefont {K.}~\bibnamefont {Nicoli}}, \bibinfo {author} {\bibfnamefont
  {P.}~\bibnamefont {Stornati}}, \bibinfo {author} {\bibfnamefont
  {R.}~\bibnamefont {Koch}}, \bibinfo {author} {\bibfnamefont {M.}~\bibnamefont
  {Büttner}}, \bibinfo {author} {\bibfnamefont {R.}~\bibnamefont {Okuła}},
  \bibinfo {author} {\bibfnamefont {G.}~\bibnamefont {Muñoz-Gil}}, \bibinfo
  {author} {\bibfnamefont {R.~A.}\ \bibnamefont {Vargas-Hernández}}, \bibinfo
  {author} {\bibfnamefont {A.}~\bibnamefont {Cervera-Lierta}}, \bibinfo
  {author} {\bibfnamefont {J.}~\bibnamefont {Carrasquilla}}, \bibinfo {author}
  {\bibfnamefont {V.}~\bibnamefont {Dunjko}}, \bibinfo {author} {\bibfnamefont
  {M.}~\bibnamefont {Gabrié}}, \bibinfo {author} {\bibfnamefont
  {P.}~\bibnamefont {Huembeli}}, \bibinfo {author} {\bibfnamefont
  {E.}~\bibnamefont {van Nieuwenburg}}, \bibinfo {author} {\bibfnamefont
  {F.}~\bibnamefont {Vicentini}}, \bibinfo {author} {\bibfnamefont
  {L.}~\bibnamefont {Wang}}, \bibinfo {author} {\bibfnamefont {S.~J.}\
  \bibnamefont {Wetzel}}, \bibinfo {author} {\bibfnamefont {G.}~\bibnamefont
  {Carleo}}, \bibinfo {author} {\bibfnamefont {E.}~\bibnamefont {Greplová}},
  \bibinfo {author} {\bibfnamefont {R.}~\bibnamefont {Krems}}, \bibinfo
  {author} {\bibfnamefont {F.}~\bibnamefont {Marquardt}}, \bibinfo {author}
  {\bibfnamefont {M.}~\bibnamefont {Tomza}}, \bibinfo {author} {\bibfnamefont
  {M.}~\bibnamefont {Lewenstein}},\ and\ \bibinfo {author} {\bibfnamefont
  {A.}~\bibnamefont {Dauphin}},\ }\href
  {https://doi.org/10.48550/ARXIV.2204.04198} {\bibinfo {title} {Modern
  applications of machine learning in quantum sciences}} (\bibinfo {year}
  {2022})\BibitemShut {NoStop}%
\bibitem [{\citenamefont {Teo}\ \emph {et~al.}(2012)\citenamefont {Teo},
  \citenamefont {Stoklasa}, \citenamefont {Englert}, \citenamefont {\ifmmode
  \check{R}\else \v{R}\fi{}eh\'a\ifmmode~\check{c}\else \v{c}\fi{}ek},\ and\
  \citenamefont {Hradil}}]{PhysRevA.85.042317}%
  \BibitemOpen
  \bibfield  {author} {\bibinfo {author} {\bibfnamefont {Y.~S.}\ \bibnamefont
  {Teo}}, \bibinfo {author} {\bibfnamefont {B.}~\bibnamefont {Stoklasa}},
  \bibinfo {author} {\bibfnamefont {B.-G.}\ \bibnamefont {Englert}}, \bibinfo
  {author} {\bibfnamefont {J.}~\bibnamefont {\ifmmode \check{R}\else
  \v{R}\fi{}eh\'a\ifmmode~\check{c}\else \v{c}\fi{}ek}},\ and\ \bibinfo
  {author} {\bibfnamefont {Z.~c.~v.}\ \bibnamefont {Hradil}},\ }\bibfield
  {title} {\bibinfo {title} {Incomplete quantum state estimation: A
  comprehensive study},\ }\href {https://doi.org/10.1103/PhysRevA.85.042317}
  {\bibfield  {journal} {\bibinfo  {journal} {Phys. Rev. A}\ }\textbf {\bibinfo
  {volume} {85}},\ \bibinfo {pages} {042317} (\bibinfo {year}
  {2012})}\BibitemShut {NoStop}%
\bibitem [{\citenamefont {Biswas}\ \emph {et~al.}(2021)\citenamefont {Biswas},
  \citenamefont {Biswas},\ and\ \citenamefont {Sen}}]{Biswas_2021}%
  \BibitemOpen
  \bibfield  {author} {\bibinfo {author} {\bibfnamefont {G.}~\bibnamefont
  {Biswas}}, \bibinfo {author} {\bibfnamefont {A.}~\bibnamefont {Biswas}},\
  and\ \bibinfo {author} {\bibfnamefont {U.}~\bibnamefont {Sen}},\ }\bibfield
  {title} {\bibinfo {title} {Inhibition of spread of typical bipartite and
  genuine multiparty entanglement in response to disorder},\ }\href
  {https://doi.org/10.1088/1367-2630/ac37c8} {\bibfield  {journal} {\bibinfo
  {journal} {New Journal of Physics}\ }\textbf {\bibinfo {volume} {23}},\
  \bibinfo {pages} {113042} (\bibinfo {year} {2021})}\BibitemShut {NoStop}%
\bibitem [{\citenamefont {Zhou}\ \emph {et~al.}(2019)\citenamefont {Zhou},
  \citenamefont {Chen}, \citenamefont {Liu}, \citenamefont {Sochnikov},
  \citenamefont {Bollinger}, \citenamefont {Han}, \citenamefont {Zhu},
  \citenamefont {He}, \citenamefont {Bozovic},\ and\ \citenamefont
  {Natelson}}]{Zhou2019}%
  \BibitemOpen
  \bibfield  {author} {\bibinfo {author} {\bibfnamefont {P.}~\bibnamefont
  {Zhou}}, \bibinfo {author} {\bibfnamefont {L.}~\bibnamefont {Chen}}, \bibinfo
  {author} {\bibfnamefont {Y.}~\bibnamefont {Liu}}, \bibinfo {author}
  {\bibfnamefont {I.}~\bibnamefont {Sochnikov}}, \bibinfo {author}
  {\bibfnamefont {A.~T.}\ \bibnamefont {Bollinger}}, \bibinfo {author}
  {\bibfnamefont {M.-G.}\ \bibnamefont {Han}}, \bibinfo {author} {\bibfnamefont
  {Y.}~\bibnamefont {Zhu}}, \bibinfo {author} {\bibfnamefont {X.}~\bibnamefont
  {He}}, \bibinfo {author} {\bibfnamefont {I.}~\bibnamefont {Bozovic}},\ and\
  \bibinfo {author} {\bibfnamefont {D.}~\bibnamefont {Natelson}},\ }\bibfield
  {title} {\bibinfo {title} {Electron pairing in the pseudogap state revealed
  by shot noise in copper oxide junctions},\ }\href
  {https://doi.org/10.1038/s41586-019-1486-7} {\bibfield  {journal} {\bibinfo
  {journal} {Nature}\ }\textbf {\bibinfo {volume} {572}},\ \bibinfo {pages}
  {493} (\bibinfo {year} {2019})}\BibitemShut {NoStop}%
\bibitem [{\citenamefont {Sivre}\ \emph {et~al.}(2019)\citenamefont {Sivre},
  \citenamefont {Duprez}, \citenamefont {Anthore}, \citenamefont {Aassime},
  \citenamefont {Parmentier}, \citenamefont {Cavanna}, \citenamefont {Ouerghi},
  \citenamefont {Gennser},\ and\ \citenamefont {Pierre}}]{Sivre2019}%
  \BibitemOpen
  \bibfield  {author} {\bibinfo {author} {\bibfnamefont {E.}~\bibnamefont
  {Sivre}}, \bibinfo {author} {\bibfnamefont {H.}~\bibnamefont {Duprez}},
  \bibinfo {author} {\bibfnamefont {A.}~\bibnamefont {Anthore}}, \bibinfo
  {author} {\bibfnamefont {A.}~\bibnamefont {Aassime}}, \bibinfo {author}
  {\bibfnamefont {F.~D.}\ \bibnamefont {Parmentier}}, \bibinfo {author}
  {\bibfnamefont {A.}~\bibnamefont {Cavanna}}, \bibinfo {author} {\bibfnamefont
  {A.}~\bibnamefont {Ouerghi}}, \bibinfo {author} {\bibfnamefont
  {U.}~\bibnamefont {Gennser}},\ and\ \bibinfo {author} {\bibfnamefont
  {F.}~\bibnamefont {Pierre}},\ }\bibfield  {title} {\bibinfo {title}
  {Electronic heat flow and thermal shot noise in quantum circuits},\ }\href
  {https://doi.org/10.1038/s41467-019-13566-8} {\bibfield  {journal} {\bibinfo
  {journal} {Nature Communications}\ }\textbf {\bibinfo {volume} {10}},\
  \bibinfo {pages} {5638} (\bibinfo {year} {2019})}\BibitemShut {NoStop}%
\end{thebibliography}%


\begin{thebibliography}{16}%
\makeatletter
\providecommand \@ifxundefined [1]{%
 \@ifx{#1\undefined}
}%
\providecommand \@ifnum [1]{%
 \ifnum #1\expandafter \@firstoftwo
 \else \expandafter \@secondoftwo
 \fi
}%
\providecommand \@ifx [1]{%
 \ifx #1\expandafter \@firstoftwo
 \else \expandafter \@secondoftwo
 \fi
}%
\providecommand \natexlab [1]{#1}%
\providecommand \enquote  [1]{``#1''}%
\providecommand \bibnamefont  [1]{#1}%
\providecommand \bibfnamefont [1]{#1}%
\providecommand \citenamefont [1]{#1}%
\providecommand \href@noop [0]{\@secondoftwo}%
\providecommand \href [0]{\begingroup \@sanitize@url \@href}%
\providecommand \@href[1]{\@@startlink{#1}\@@href}%
\providecommand \@@href[1]{\endgroup#1\@@endlink}%
\providecommand \@sanitize@url [0]{\catcode `\\12\catcode `\$12\catcode
  `\&12\catcode `\#12\catcode `\^12\catcode `\_12\catcode `\%12\relax}%
\providecommand \@@startlink[1]{}%
\providecommand \@@endlink[0]{}%
\providecommand \url  [0]{\begingroup\@sanitize@url \@url }%
\providecommand \@url [1]{\endgroup\@href {#1}{\urlprefix }}%
\providecommand \urlprefix  [0]{URL }%
\providecommand \Eprint [0]{\href }%
\providecommand \doibase [0]{https://doi.org/}%
\providecommand \selectlanguage [0]{\@gobble}%
\providecommand \bibinfo  [0]{\@secondoftwo}%
\providecommand \bibfield  [0]{\@secondoftwo}%
\providecommand \translation [1]{[#1]}%
\providecommand \BibitemOpen [0]{}%
\providecommand \bibitemStop [0]{}%
\providecommand \bibitemNoStop [0]{.\EOS\space}%
\providecommand \EOS [0]{\spacefactor3000\relax}%
\providecommand \BibitemShut  [1]{\csname bibitem#1\endcsname}%
\let\auto@bib@innerbib\@empty
\bibitem [{\citenamefont {Zhao}\ \emph {et~al.}(2022)\citenamefont {Zhao},
  \citenamefont {Hu},\ and\ \citenamefont {Zhang}}]{Tianlun2022}%
  \BibitemOpen
  \bibfield  {author} {\bibinfo {author} {\bibfnamefont {T.-L.}\ \bibnamefont
  {Zhao}}, \bibinfo {author} {\bibfnamefont {S.-X.}\ \bibnamefont {Hu}},\ and\
  \bibinfo {author} {\bibfnamefont {Y.}~\bibnamefont {Zhang}},\ }\href
  {https://doi.org/10.48550/ARXIV.2212.13718} {\bibinfo {title} {Supervised
  hamiltonian learning via efficient and robust quantum descent}} (\bibinfo
  {year} {2022}),\ \Eprint {https://arxiv.org/abs/2212.13718}
  {arXiv:2212.13718} \BibitemShut {NoStop}%
\bibitem [{\citenamefont {Kitaev}(2006)}]{KITAEV2006}%
  \BibitemOpen
  \bibfield  {author} {\bibinfo {author} {\bibfnamefont {A.}~\bibnamefont
  {Kitaev}},\ }\bibfield  {title} {\bibinfo {title} {Anyons in an exactly
  solved model and beyond},\ }\href
  {https://doi.org/http://dx.doi.org/10.1016/j.aop.2005.10.005} {\bibfield
  {journal} {\bibinfo  {journal} {Annals of Physics}\ }\textbf {\bibinfo
  {volume} {321}},\ \bibinfo {pages} {2 } (\bibinfo {year} {2006})}\BibitemShut
  {NoStop}%
\bibitem [{\citenamefont {Chaloupka}\ \emph {et~al.}(2010)\citenamefont
  {Chaloupka}, \citenamefont {Jackeli},\ and\ \citenamefont
  {Khaliullin}}]{PhysRevLett.105.027204}%
  \BibitemOpen
  \bibfield  {author} {\bibinfo {author} {\bibfnamefont {J.~c.~v.}\
  \bibnamefont {Chaloupka}}, \bibinfo {author} {\bibfnamefont {G.}~\bibnamefont
  {Jackeli}},\ and\ \bibinfo {author} {\bibfnamefont {G.}~\bibnamefont
  {Khaliullin}},\ }\bibfield  {title} {\bibinfo {title} {Kitaev-heisenberg
  model on a honeycomb lattice: Possible exotic phases in iridium oxides
  ${A}_{2}{\mathrm{iro}}_{3}$},\ }\href
  {https://doi.org/10.1103/PhysRevLett.105.027204} {\bibfield  {journal}
  {\bibinfo  {journal} {Phys. Rev. Lett.}\ }\textbf {\bibinfo {volume} {105}},\
  \bibinfo {pages} {027204} (\bibinfo {year} {2010})}\BibitemShut {NoStop}%
\bibitem [{\citenamefont {Mandal}\ and\ \citenamefont
  {Jayannavar}(2020)}]{mandal2020introduction}%
  \BibitemOpen
  \bibfield  {author} {\bibinfo {author} {\bibfnamefont {S.}~\bibnamefont
  {Mandal}}\ and\ \bibinfo {author} {\bibfnamefont {A.~M.}\ \bibnamefont
  {Jayannavar}},\ }\bibfield  {title} {\bibinfo {title} {An introduction to
  kitaev model-i},\ }\href@noop {} {\bibfield  {journal} {\bibinfo  {journal}
  {arXiv preprint arXiv:2006.11549}\ } (\bibinfo {year} {2020})}\BibitemShut
  {NoStop}%
\bibitem [{\citenamefont {Zhu}\ \emph {et~al.}(2018)\citenamefont {Zhu},
  \citenamefont {Kimchi}, \citenamefont {Sheng},\ and\ \citenamefont
  {Fu}}]{PhysRevB.97.241110}%
  \BibitemOpen
  \bibfield  {author} {\bibinfo {author} {\bibfnamefont {Z.}~\bibnamefont
  {Zhu}}, \bibinfo {author} {\bibfnamefont {I.}~\bibnamefont {Kimchi}},
  \bibinfo {author} {\bibfnamefont {D.~N.}\ \bibnamefont {Sheng}},\ and\
  \bibinfo {author} {\bibfnamefont {L.}~\bibnamefont {Fu}},\ }\bibfield
  {title} {\bibinfo {title} {Robust non-abelian spin liquid and a possible
  intermediate phase in the antiferromagnetic kitaev model with magnetic
  field},\ }\href {https://doi.org/10.1103/PhysRevB.97.241110} {\bibfield
  {journal} {\bibinfo  {journal} {Phys. Rev. B}\ }\textbf {\bibinfo {volume}
  {97}},\ \bibinfo {pages} {241110} (\bibinfo {year} {2018})}\BibitemShut
  {NoStop}%
\bibitem [{\citenamefont {Lieb}(1994)}]{PhysRevLett.73.2158}%
  \BibitemOpen
  \bibfield  {author} {\bibinfo {author} {\bibfnamefont {E.~H.}\ \bibnamefont
  {Lieb}},\ }\bibfield  {title} {\bibinfo {title} {Flux phase of the
  half-filled band},\ }\href {https://doi.org/10.1103/PhysRevLett.73.2158}
  {\bibfield  {journal} {\bibinfo  {journal} {Phys. Rev. Lett.}\ }\textbf
  {\bibinfo {volume} {73}},\ \bibinfo {pages} {2158} (\bibinfo {year}
  {1994})}\BibitemShut {NoStop}%
\bibitem [{\citenamefont {Zschocke}\ and\ \citenamefont
  {Vojta}(2015)}]{PhysRevB.92.014403}%
  \BibitemOpen
  \bibfield  {author} {\bibinfo {author} {\bibfnamefont {F.}~\bibnamefont
  {Zschocke}}\ and\ \bibinfo {author} {\bibfnamefont {M.}~\bibnamefont
  {Vojta}},\ }\bibfield  {title} {\bibinfo {title} {Physical states and
  finite-size effects in kitaev's honeycomb model: Bond disorder, spin
  excitations, and nmr line shape},\ }\href
  {https://doi.org/10.1103/PhysRevB.92.014403} {\bibfield  {journal} {\bibinfo
  {journal} {Phys. Rev. B}\ }\textbf {\bibinfo {volume} {92}},\ \bibinfo
  {pages} {014403} (\bibinfo {year} {2015})}\BibitemShut {NoStop}%
\bibitem [{\citenamefont {Pedrocchi}\ \emph {et~al.}(2011)\citenamefont
  {Pedrocchi}, \citenamefont {Chesi},\ and\ \citenamefont
  {Loss}}]{PhysRevB.84.165414}%
  \BibitemOpen
  \bibfield  {author} {\bibinfo {author} {\bibfnamefont {F.~L.}\ \bibnamefont
  {Pedrocchi}}, \bibinfo {author} {\bibfnamefont {S.}~\bibnamefont {Chesi}},\
  and\ \bibinfo {author} {\bibfnamefont {D.}~\bibnamefont {Loss}},\ }\bibfield
  {title} {\bibinfo {title} {Physical solutions of the kitaev honeycomb
  model},\ }\href {https://doi.org/10.1103/PhysRevB.84.165414} {\bibfield
  {journal} {\bibinfo  {journal} {Phys. Rev. B}\ }\textbf {\bibinfo {volume}
  {84}},\ \bibinfo {pages} {165414} (\bibinfo {year} {2011})}\BibitemShut
  {NoStop}%
\bibitem [{\citenamefont {Dawid}\ \emph {et~al.}(2022)\citenamefont {Dawid},
  \citenamefont {Arnold}, \citenamefont {Requena}, \citenamefont {Gresch},
  \citenamefont {Płodzień}, \citenamefont {Donatella}, \citenamefont
  {Nicoli}, \citenamefont {Stornati}, \citenamefont {Koch}, \citenamefont
  {Büttner}, \citenamefont {Okuła}, \citenamefont {Muñoz-Gil}, \citenamefont
  {Vargas-Hernández}, \citenamefont {Cervera-Lierta}, \citenamefont
  {Carrasquilla}, \citenamefont {Dunjko}, \citenamefont {Gabrié},
  \citenamefont {Huembeli}, \citenamefont {van Nieuwenburg}, \citenamefont
  {Vicentini}, \citenamefont {Wang}, \citenamefont {Wetzel}, \citenamefont
  {Carleo}, \citenamefont {Greplová}, \citenamefont {Krems}, \citenamefont
  {Marquardt}, \citenamefont {Tomza}, \citenamefont {Lewenstein},\ and\
  \citenamefont {Dauphin}}]{MLbook_ap}%
  \BibitemOpen
  \bibfield  {author} {\bibinfo {author} {\bibfnamefont {A.}~\bibnamefont
  {Dawid}}, \bibinfo {author} {\bibfnamefont {J.}~\bibnamefont {Arnold}},
  \bibinfo {author} {\bibfnamefont {B.}~\bibnamefont {Requena}}, \bibinfo
  {author} {\bibfnamefont {A.}~\bibnamefont {Gresch}}, \bibinfo {author}
  {\bibfnamefont {M.}~\bibnamefont {Płodzień}}, \bibinfo {author}
  {\bibfnamefont {K.}~\bibnamefont {Donatella}}, \bibinfo {author}
  {\bibfnamefont {K.}~\bibnamefont {Nicoli}}, \bibinfo {author} {\bibfnamefont
  {P.}~\bibnamefont {Stornati}}, \bibinfo {author} {\bibfnamefont
  {R.}~\bibnamefont {Koch}}, \bibinfo {author} {\bibfnamefont {M.}~\bibnamefont
  {Büttner}}, \bibinfo {author} {\bibfnamefont {R.}~\bibnamefont {Okuła}},
  \bibinfo {author} {\bibfnamefont {G.}~\bibnamefont {Muñoz-Gil}}, \bibinfo
  {author} {\bibfnamefont {R.~A.}\ \bibnamefont {Vargas-Hernández}}, \bibinfo
  {author} {\bibfnamefont {A.}~\bibnamefont {Cervera-Lierta}}, \bibinfo
  {author} {\bibfnamefont {J.}~\bibnamefont {Carrasquilla}}, \bibinfo {author}
  {\bibfnamefont {V.}~\bibnamefont {Dunjko}}, \bibinfo {author} {\bibfnamefont
  {M.}~\bibnamefont {Gabrié}}, \bibinfo {author} {\bibfnamefont
  {P.}~\bibnamefont {Huembeli}}, \bibinfo {author} {\bibfnamefont
  {E.}~\bibnamefont {van Nieuwenburg}}, \bibinfo {author} {\bibfnamefont
  {F.}~\bibnamefont {Vicentini}}, \bibinfo {author} {\bibfnamefont
  {L.}~\bibnamefont {Wang}}, \bibinfo {author} {\bibfnamefont {S.~J.}\
  \bibnamefont {Wetzel}}, \bibinfo {author} {\bibfnamefont {G.}~\bibnamefont
  {Carleo}}, \bibinfo {author} {\bibfnamefont {E.}~\bibnamefont {Greplová}},
  \bibinfo {author} {\bibfnamefont {R.}~\bibnamefont {Krems}}, \bibinfo
  {author} {\bibfnamefont {F.}~\bibnamefont {Marquardt}}, \bibinfo {author}
  {\bibfnamefont {M.}~\bibnamefont {Tomza}}, \bibinfo {author} {\bibfnamefont
  {M.}~\bibnamefont {Lewenstein}},\ and\ \bibinfo {author} {\bibfnamefont
  {A.}~\bibnamefont {Dauphin}},\ }\href
  {https://doi.org/10.48550/ARXIV.2204.04198} {\bibinfo {title} {Modern
  applications of machine learning in quantum sciences}} (\bibinfo {year}
  {2022})\BibitemShut {NoStop}%
\bibitem [{\citenamefont {Shang}\ \emph {et~al.}(2017)\citenamefont {Shang},
  \citenamefont {Zhang},\ and\ \citenamefont {Ng}}]{PhysRevA.95.062336}%
  \BibitemOpen
  \bibfield  {author} {\bibinfo {author} {\bibfnamefont {J.}~\bibnamefont
  {Shang}}, \bibinfo {author} {\bibfnamefont {Z.}~\bibnamefont {Zhang}},\ and\
  \bibinfo {author} {\bibfnamefont {H.~K.}\ \bibnamefont {Ng}},\ }\bibfield
  {title} {\bibinfo {title} {Superfast maximum-likelihood reconstruction for
  quantum tomography},\ }\href {https://doi.org/10.1103/PhysRevA.95.062336}
  {\bibfield  {journal} {\bibinfo  {journal} {Phys. Rev. A}\ }\textbf {\bibinfo
  {volume} {95}},\ \bibinfo {pages} {062336} (\bibinfo {year}
  {2017})}\BibitemShut {NoStop}%
\bibitem [{\citenamefont {Biswas}\ \emph {et~al.}(2021)\citenamefont {Biswas},
  \citenamefont {Biswas},\ and\ \citenamefont {Sen}}]{Biswas_2021}%
  \BibitemOpen
  \bibfield  {author} {\bibinfo {author} {\bibfnamefont {G.}~\bibnamefont
  {Biswas}}, \bibinfo {author} {\bibfnamefont {A.}~\bibnamefont {Biswas}},\
  and\ \bibinfo {author} {\bibfnamefont {U.}~\bibnamefont {Sen}},\ }\bibfield
  {title} {\bibinfo {title} {Inhibition of spread of typical bipartite and
  genuine multiparty entanglement in response to disorder},\ }\href
  {https://doi.org/10.1088/1367-2630/ac37c8} {\bibfield  {journal} {\bibinfo
  {journal} {New Journal of Physics}\ }\textbf {\bibinfo {volume} {23}},\
  \bibinfo {pages} {113042} (\bibinfo {year} {2021})}\BibitemShut {NoStop}%
\bibitem [{\citenamefont {Fannes}\ \emph {et~al.}(1992)\citenamefont {Fannes},
  \citenamefont {Nachtergaele},\ and\ \citenamefont {Werner}}]{MPS1992}%
  \BibitemOpen
  \bibfield  {author} {\bibinfo {author} {\bibfnamefont {M.}~\bibnamefont
  {Fannes}}, \bibinfo {author} {\bibfnamefont {B.}~\bibnamefont
  {Nachtergaele}},\ and\ \bibinfo {author} {\bibfnamefont {R.~F.}\ \bibnamefont
  {Werner}},\ }\bibfield  {title} {\bibinfo {title} {Finitely correlated states
  on quantum spin chains},\ }\href {https://doi.org/10.1007/BF02099178}
  {\bibfield  {journal} {\bibinfo  {journal} {Communications in Mathematical
  Physics}\ }\textbf {\bibinfo {volume} {144}},\ \bibinfo {pages} {443}
  (\bibinfo {year} {1992})}\BibitemShut {NoStop}%
\bibitem [{\citenamefont {Schollw\"ock}(2005)}]{DMRG2005}%
  \BibitemOpen
  \bibfield  {author} {\bibinfo {author} {\bibfnamefont {U.}~\bibnamefont
  {Schollw\"ock}},\ }\bibfield  {title} {\bibinfo {title} {The density-matrix
  renormalization group},\ }\href {https://doi.org/10.1103/RevModPhys.77.259}
  {\bibfield  {journal} {\bibinfo  {journal} {Rev. Mod. Phys.}\ }\textbf
  {\bibinfo {volume} {77}},\ \bibinfo {pages} {259} (\bibinfo {year}
  {2005})}\BibitemShut {NoStop}%
\bibitem [{\citenamefont {Foulkes}\ \emph {et~al.}(2001)\citenamefont
  {Foulkes}, \citenamefont {Mitas}, \citenamefont {Needs},\ and\ \citenamefont
  {Rajagopal}}]{QMCreview2001}%
  \BibitemOpen
  \bibfield  {author} {\bibinfo {author} {\bibfnamefont {W.~M.~C.}\
  \bibnamefont {Foulkes}}, \bibinfo {author} {\bibfnamefont {L.}~\bibnamefont
  {Mitas}}, \bibinfo {author} {\bibfnamefont {R.~J.}\ \bibnamefont {Needs}},\
  and\ \bibinfo {author} {\bibfnamefont {G.}~\bibnamefont {Rajagopal}},\
  }\bibfield  {title} {\bibinfo {title} {Quantum monte carlo simulations of
  solids},\ }\href {https://doi.org/10.1103/RevModPhys.73.33} {\bibfield
  {journal} {\bibinfo  {journal} {Rev. Mod. Phys.}\ }\textbf {\bibinfo {volume}
  {73}},\ \bibinfo {pages} {33} (\bibinfo {year} {2001})}\BibitemShut {NoStop}%
\bibitem [{\citenamefont {Troyer}\ and\ \citenamefont
  {Wiese}(2005)}]{Troyer2005}%
  \BibitemOpen
  \bibfield  {author} {\bibinfo {author} {\bibfnamefont {M.}~\bibnamefont
  {Troyer}}\ and\ \bibinfo {author} {\bibfnamefont {U.-J.}\ \bibnamefont
  {Wiese}},\ }\bibfield  {title} {\bibinfo {title} {Computational complexity
  and fundamental limitations to fermionic quantum monte carlo simulations},\
  }\href {https://doi.org/10.1103/PhysRevLett.94.170201} {\bibfield  {journal}
  {\bibinfo  {journal} {Phys. Rev. Lett.}\ }\textbf {\bibinfo {volume} {94}},\
  \bibinfo {pages} {170201} (\bibinfo {year} {2005})}\BibitemShut {NoStop}%
\bibitem [{\citenamefont {Carleo}\ and\ \citenamefont
  {Troyer}(2017)}]{Carleo2016}%
  \BibitemOpen
  \bibfield  {author} {\bibinfo {author} {\bibfnamefont {G.}~\bibnamefont
  {Carleo}}\ and\ \bibinfo {author} {\bibfnamefont {M.}~\bibnamefont
  {Troyer}},\ }\bibfield  {title} {\bibinfo {title} {{Solving the quantum
  many-body problem with artificial neural networks}},\ }\href
  {https://doi.org/10.1126/science.aag2302} {\bibfield  {journal} {\bibinfo
  {journal} {Science}\ }\textbf {\bibinfo {volume} {355}},\ \bibinfo {pages}
  {602} (\bibinfo {year} {2017})}\BibitemShut {NoStop}%
\end{thebibliography}%
\end{document}


\title{Supplemental Materials: Single-shot quantum measurements sketch quantum many-body states}
\author{Jia-Bao Wang}
\affiliation{International Center for Quantum Materials, Peking University, Beijing, 100871, China}
\affiliation{School of Physics, Peking University, Beijing, 100871, China}
\author{Yi Zhang}
\email{frankzhangyi@gmail.com}
\affiliation{International Center for Quantum Materials, Peking University, Beijing, 100871, China}
\affiliation{School of Physics, Peking University, Beijing, 100871, China}

\maketitle

\section{Generalization to mixed state}

In this section, we generalize our strategy to mixed states and provide a brief example. Similar to the case of a pure state, after measuring observables $\{\hat{O}_1, \hat{O}_2, \dots\}$ on the target mixed state represented by the density matrix $\rho_t$, we obtain a series of results $\mathcal{D} = \{\gamma_1, \gamma_2, \dots\}$. The posterior probability for a given state $\rho$ in the Hilbert space is:
\begin{equation}
  p(\rho|\mathcal{D}) \propto \prod_{\gamma \in \mathcal{D}}p(\gamma|\rho) = \prod_{\gamma \in \mathcal{D}} \text{tr}(\rho \hat{P}_{\gamma}),
\end{equation} 
where for each event $\gamma$, $\hat{P}_{\gamma} = \hat{P}_{\tau}$ is the projection operator corresponding to the measurement outcome $a_{\tau}$. Thus, we define the measurement energy as:
\begin{equation} 
  E(\rho|\mathcal{D})=-\sum_{\gamma \in \mathcal{D}}\log[\text{tr}(\rho \hat{P}_{\gamma})].
\end{equation}
By binning the recurring events, we obtain statistics that of the $N_{\hat{O}}$ times we measured $\hat{O}$, the outcome $a_{\tau}$ repeated $N_{\tau}$ times, and re-express the measurement energy as:
\begin{equation} 
  E(\rho|\mathcal{D})= -\sum_{\hat{O}, \tau} N_{\hat{O}} f_{\tau} \log[\text{tr}(\rho \hat{P}_{\tau})],
\end{equation}
where $f_{\tau}=N_{\tau}/N_{\hat{O}}$ is the measured frequency. 

Next, we consider the canonical ensemble of a system at a constant temperature $T$, whose Helmholtz free energy achieves minimum in equilibrium:
\begin{equation}
  F(\rho|\mathcal{D}) = E(\rho|\mathcal{D}) - T S(\rho),
\end{equation}
where $S(\rho)=-\text{tr}(\rho\ln \rho)$ is the von Neumann entropy.

Given $\text{tr}(\rho)=1$,
\begin{equation}
    \begin{aligned}
        \delta F(\rho|\mathcal{D}) &= -\sum_{\hat{O}, \tau} N_{\hat{O}} f_{\tau} \frac{\text{tr}(\delta \rho \hat{P}_{\tau})}{\text{tr}(\rho \hat{P}_{\tau})} + T\text{tr}(\delta\rho\ln \rho) + T \text{tr}(\delta\rho) \\
        &= \text{tr}\{\delta \rho[T \ln \rho -\sum_{\hat{O}, \tau} N_{\hat{O}} f_{\tau} \frac{\hat{P}_{\tau}}{\text{tr}(\rho \hat{P}_{\tau})} ]\}
    \end{aligned}
\end{equation}
where $\delta F(\rho|\mathcal{D})|_{\rho_{gs}} = 0$ gives:
\begin{equation}
  \rho_{gs} = \frac{1}{Z} \exp[{-\beta\sum_{\hat{O}, \tau}\frac{-N_{\hat{O}} f_{\tau}}{\text{tr}(\rho_{gs} \hat{P}_{\tau} )}}\hat{P}_{\tau}]
  \equiv \frac{1}{Z}e^{-\beta \hat{H}_{eff}},
\label{Eq:mix_state}
\end{equation}
where:
\begin{equation}
  \hat{H}_{eff}\equiv  \sum_{\hat{O}, \tau} N_{\hat{O}} (\frac{ -f_{\tau}}{\text{tr}(\rho_{gs} \hat{P}_{\tau} )})\hat{P}_{\tau},
\label{Eq:mix_Heff}
\end{equation}
$Z = \text{tr}(e^{-\beta \hat{H}_{eff}})$ is normalization and $\beta = 1/T$. Similar to the pure-state case in the main text, we can achieve self-consistency between Eq. \ref{Eq:mix_state} and \ref{Eq:mix_Heff} and circumvent instabilities by using $\hat H_{eff}$ as a gradient and approach the target state via an iteration of updates $\hat H_0 \rightarrow \hat H_0 + \lambda \hat H_{eff}$, where $\lambda$ controls the step size / descent rate. 

In the main text, we demonstrate the strategy with an example on Gibbs states $\rho_{tar} = \frac{e^{-\beta \hat{H}}}{\text{tr}(e^{-\beta \hat{H}})}$ of fermion models on 100 sites. Without loss of generality, we set $\beta=1$ and choose $\mu_i$ and $t_{ij}$ randomly; the resulting $\beta \hat{H}$ follows a Gaussian Orthogonal Ensemble. The target state $\rho_{tar}$ is placed in a black box, and tangible only via quantum measurements, for which we choose the observables $\hat{O}_i = c^\dagger_i c_i$, $\hat{O}'_{ij} = (c^\dagger_i+c^\dagger_{j})(c_i+c_j)/2$ for every $i, j$ sites. Given the measurement outcomes with $N_{\hat{O}}\rightarrow\infty$ to fully suppress quantum fluctuations, we obtain the iterative processes' results as in the inset of Fig. 2 in the main text. We observe a quick and unambiguous convergence of the iteration state $\rho_0$ towards the target mixed state, indicated by the vanishing infidelity $1-\text{tr}(\rho_{0}\rho_{tar})$. The average measurement energy $E(\rho_0|\mathcal{D})$ also tends toward its lower bound. Therefore, we have achieved mixed states entirely consistent with the targets. 

Further examples and discussions on mixed states, including a rigorous proof of convergence, are available in Ref. \onlinecite{Tianlun2022}.

\section{Random long-range fermion model}

\subsection{Quantum fluctuations and variable numbers of measurements}

In the main text, we demonstrate the excellent convergence given a large number of quantum measurements $N_{\hat O} \rightarrow \infty$. Here, we show that our strategy also works for data laden with quantum fluctuations from a finite number of quantum measurements. We retain the same target states and observables as in the main text. However, when $N_{\hat O}$ is finite, the measurement outcomes $ N_{\tau} =f_{\tau} N_{\hat{O}}$ may deviate from $\langle\hat{P}_{\tau}\rangle \times N_{\hat{O}}$, following a binomial distribution. Also, the iterations may not reach an apparent convergence if $N_{\hat O}$ is too small, and therefore we pick the iteration state $|\Phi_0 \rangle$ with the lowest measurement energy within $1000$ iterations. 

\begin{figure}
  \centering  
  \includegraphics[width=0.4\textwidth]{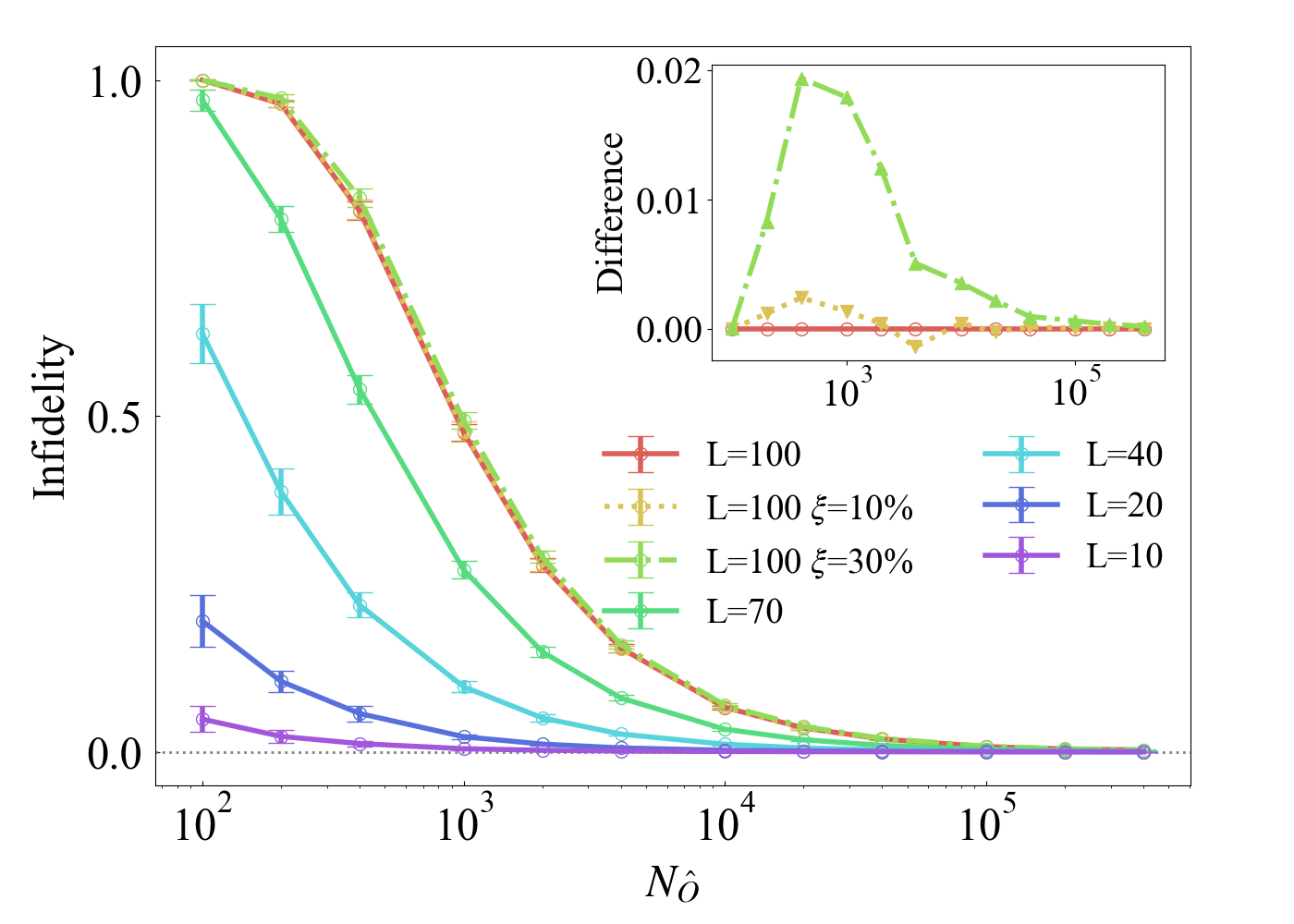} 
  \includegraphics[width=0.4\textwidth]{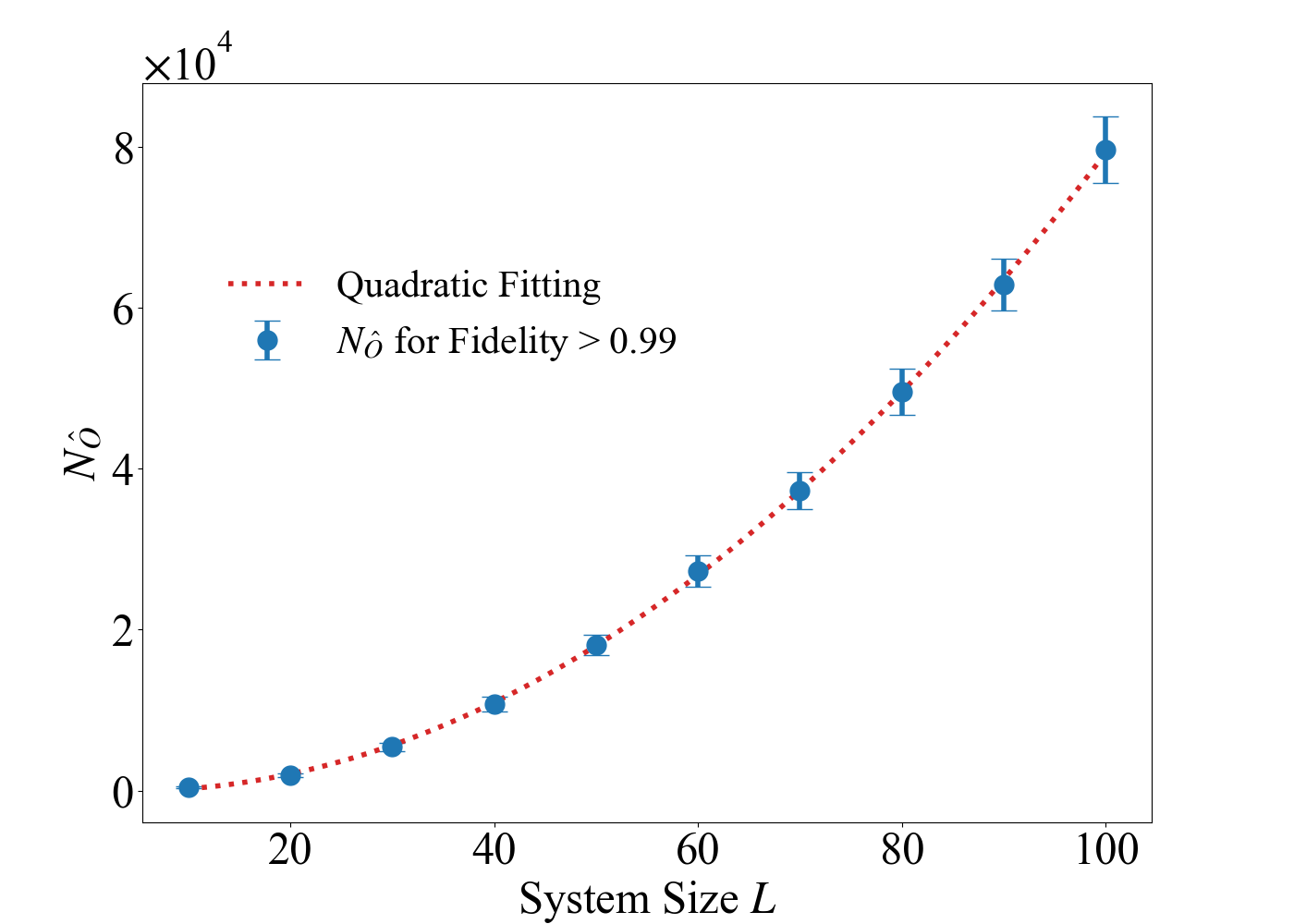}
  \caption{Upper: The MLE state $|\Phi_{gs}\rangle$'s infidelity to the target state within 1000 iterations diminishes as the (average) number of measurements $N_{\hat O}$ for each observable increases. The dashed and dotted curves show the impact if the number of measurements on each observable suffers some variance, and we zoom in on their differences in the inset. Lower: The number of measurement $N_{\hat O}$ necessary for fidelity $>0.99$ in different system size fits well to a quadratic scaling (dotted curve). The results and error bars for each system size $L$ are based upon 100 trials on different target states.}\label{fig:freefermion} 
\end{figure}

We summarize the results in Fig. \ref{fig:freefermion}. There is a clear tendency for better convergence to the target states as the number of measurements increases, and more measurements are necessary for larger systems, consistent with our expectations. We also count $N_{\hat O}$ as the infidelity crosses below $\sim 1\%$ (fidelity above $\sim 99\%$) for different system sizes $L$, which contributed to Fig. \ref{fig:freefermion}. Considering the vast $2^L$-dimensional Hilbert space, we believe that our strategy offers drastic improvements in precision and practicality. 

Another interesting topic is quantum-measurement placement, where observables may receive different priorities. For example, we consider the case where the numbers of measurements differ randomly and follow a Gaussian distribution of relative variance $\xi$ around its average $N_{\hat O}$. We summarize the results on $L=100$ systems in Fig. \ref{fig:freefermion} and its inset: while such randomness is more of a toll than a bonus, its effect is small, especially when there are sufficient measurements overall. Indeed, this conclusion is consistent with our intuition on the current setup where the observables $\hat{O}_i$ and $\hat{O}'_{ij}$ are equivalent upon permutations. It is conceivable that designed or adaptive quantum measurement schemes achieve better efficiency than the flat placement we are testing. While our strategy covers the analysis under such scenarios as well, we leave the pursuit of optimal measurement placement from the current perspective to future research.

\subsection{Comparison with the projector strategy}

Our strategy has much success in the target state as the ground state of a random long-range fermion model. Alternatively, another well-known strategy is to construct a projector to the target quantum state given all the two-point correlators. For a general fermion-bilinear tight-binding Hamiltonian:
\begin{equation}
  \hat{H} = -\sum_{i,j} t_{ij} c^{\dagger}_i c_j  = (c^{\dagger} Q) \Lambda (Q^{-1} c) = \alpha^{\dagger} \Lambda \alpha,
\end{equation} 
where $Q$ is unitary, $\Lambda$ is diagonal, and $\alpha_n^{\dagger}=\sum_{m}  c_m^{\dagger}Q_{mn}$ is the eigenstate with respect to the energy $\epsilon_n$, the $n^{th}$ element of $\Lambda$. The ground state occupies all the single-particle states with negative energy: 
\begin{equation}
  |\Phi\rangle = \prod_{\epsilon_n<0} \alpha_n^{\dagger} |0\rangle,
\end{equation}
where $|0\rangle$ is the vacuum state. 

It is straightforward to see that the two-point correlator $C_{ij} = \langle c_i^{\dagger} c_j \rangle$ behaves as $C^T = Q \tilde{\Lambda} Q^{-1}$, where $\tilde{\Lambda}=\langle \Phi|\hat{n}_m|\Phi\rangle \delta_{m,m'}$, $\hat{n}_m=\alpha ^{\dagger}_m \alpha_m$. Therefore, we can treat $C^T$ as a projector onto (the occupied single-particle states in) the state $|\Phi\rangle$. Given $C$, we can build an effective Hamiltonian following the projector to retrieve $|\Phi\rangle$:
\begin{equation}
  \hat{H}_{proj} =\sum_{i,j} \tilde{t}_{ij} c_i^{\dagger}  c_j,
\end{equation} 
where $\tilde{t}_{ij} = \delta_{i,j} - 2 \langle c_j^{\dagger} c_i \rangle$. In this way, the ground state of $\hat{H}_{proj}$ gives the target state $|\Phi\rangle$.

\begin{figure}
  \centering  
  \includegraphics[width=0.45\textwidth]{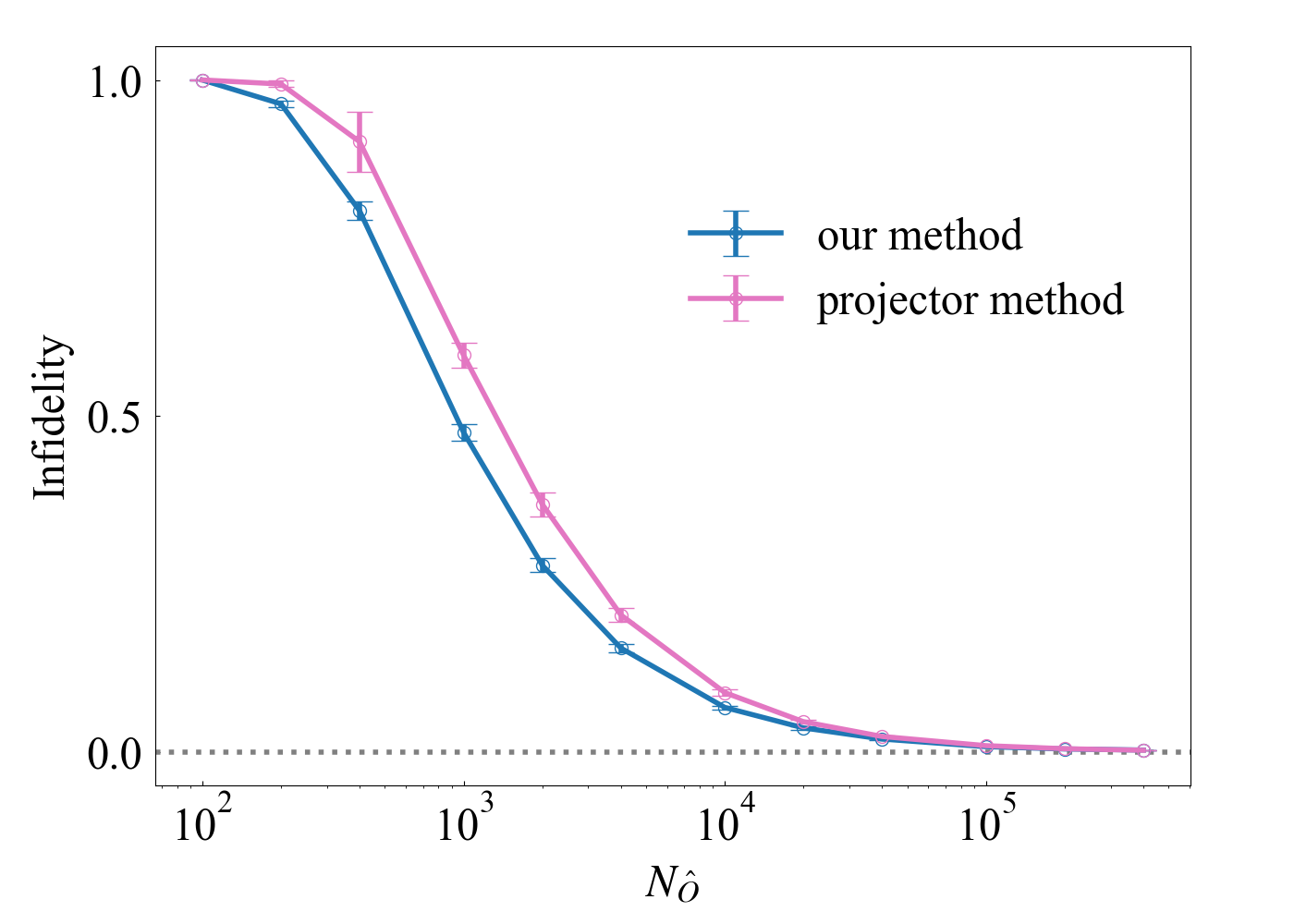} \caption{Despite using less information, our strategy performs better than the projector strategy on finite quantum measurements on random long-range fermion models. With an increasing number $N_{\hat O}$ of quantum measurements, both the MLE state and the solution of the projector approach the target quantum state asymptotically. The error bars are based on 100 trials on different target states, and the system size is $L=100$.}
  \label{fig:compare}
\end{figure}

Such a projector strategy for identifying target states is straightforward and well-known. However, even for a real target state, it requires all $Re(\langle c_i^{\dagger} c_j \rangle)$ - every term is indispensable. In comparison, while the observables $\hat{O}_i$ and $\hat{O}'_{ij}$ we consider are equivalent to $Re(\langle c^{\dagger}_i c_j \rangle)$, the absence of observables only causes partial loss of information and minor annoyance, which certainly does not constitute a roadblock. Also, our strategy generally outperforms the projector strategy (Fig. \ref{fig:compare}) and is, therefore, more accurate, more robust, and less dependent on observables than the projector strategy.

\subsection{Slater-Jastrow quantum many-body states}
\begin{figure}
  \centering  
  \includegraphics[width=0.5\textwidth]{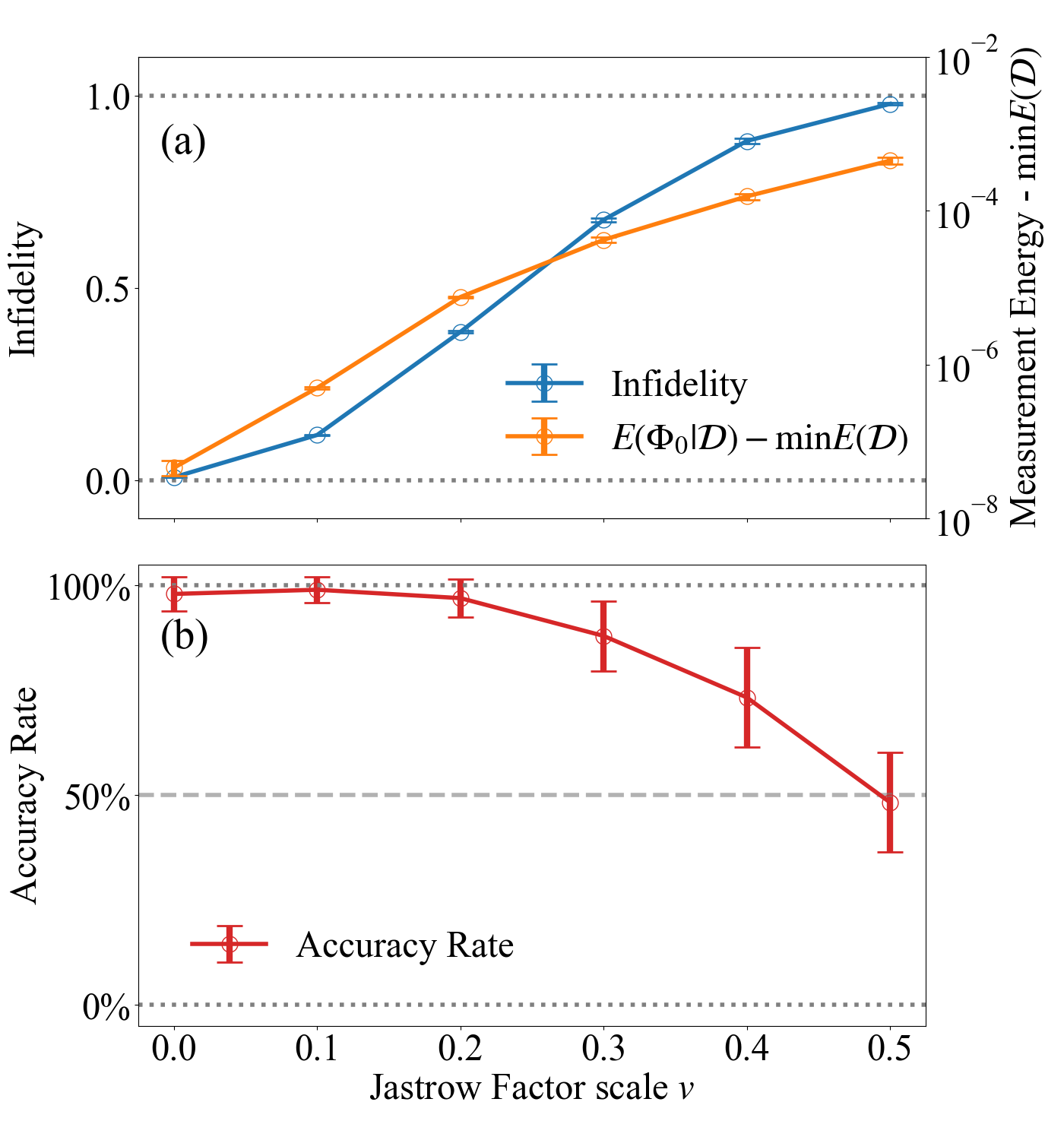} 
  \caption{(a) The observables $\hat{O}_i$ and $\hat{O}'_{ij}$ limit the search to Hartree-Fock states, and our strategy's performance becomes worse as the target state, a Slater-Jastrow quantum many-body state, ventures beyond the Hartree-Fock paradigm with an increasing Jastrow factor. (b) The accuracy of the most dominant two-point correlators also descends with an increasing Jastrow factor, yet its robustness is notably higher than that of the state (fidelity). The error bars are based on 30 different trials, and each trial consists of $10^5$ Monte Carlo samples. The system size is $10 \times 10$.}
  \label{fig:jastrow}
\end{figure}

If we broaden our scope to a generic quantum many-body state and keep the same observables $\hat{O}_i$ and $\hat{O}'_{ij}$, the search space is limited to Hartree-Fock states and away from the target state, characterized by the extent of measurement energy above its lower bound and even non-convergence (as no Hartree-Fock state satisfies the measurement result of the generic quantum many-body state). For instance, we consider the Slater-Jastrow quantum many-body state as the target state, which takes the following form:
\begin{equation}
  |\Phi \rangle = e^{-v \sum_{\langle ij \rangle} n_i n_j} |\phi \rangle, \label{eq:jastrow}
\end{equation}
where $|\phi \rangle$ is a Slater-determinant state on a two-dimensional lattice. $e^{-v \sum_{\langle ij \rangle} n_i n_j}$ is the Jastrow factor, suppressing the configuration by the factor of $e^{-v}$, $v \geqslant 0$, when pairs of electrons occupy the nearest-neighbor sites $\langle ij \rangle$. When $v=0$, $|\Phi \rangle$ reduces to the Slater-determinant state $|\phi \rangle$. Increasing $v$ effectively introduces more repulsive interactions and correlations and drives the target quantum many-body state further beyond the Hartree-Fock paradigm. 

The expectation values of observables $\hat{O}_i$ and $\hat{O}'_{ij}$ and the corresponding quantum probability $\langle \hat{P}_{\tau} \rangle$ of the quantum many-body state $|\Phi\rangle$ in Eq. \ref{eq:jastrow} can be obtained straightforwardly via variational Monte Carlo calculations. In the limit of large number $N_{\hat O}$ of quantum measurements, the number of corresponding outcomes should approach $N_{\tau} \rightarrow N_{\hat O} \times \langle \hat{P}_{\tau} \rangle$. We summarize the performance of the obtained MLE state in Fig. \ref{fig:jastrow}(a), which becomes worse as the Jastrow factor $v$ increases and the target state moves further away from the Hartree-Fock regime. Better analysis of such generic a quantum many-body quantum state requires additional observables that break the fermion-bilinear form of $\hat{H}_{eff}$ and expand the search space. The following $\hat{H}_0$ may become strongly-correlated and difficult to solve and need numerical methods such as density-matrix renormalization group (DMRG), quantum Monte Carlo (QMC) method, and exact diagonalization. 

Nevertheless, a preliminary examination such as in the efficient Hartree-Fock state space with the $\hat{O}_i$ and $\hat{O}'_{ij}$ observables may give insights such as the locality information useful for DMRG and some clues about the trial states useful for QMC. For example, we evaluate the two-point correlator $\langle c_i^{\dagger}c_j +c_j^{\dagger}c_i \rangle$ of the MLE Hartree-Fock state and the target state and summarize the interaction of the ten dominant pairs with the largest (amplitude) correlators in Fig. \ref{fig:jastrow}(b) as a measure of the accuracy on locality. Even for relatively larger $v$, e.g., fidelity $\sim 0.02$ at $v=0.5$, the accuracy on the dominant two-point correlators is still $\sim 50\%$, indicating that the MLE Hartree-Fock state, though qualitatively dissimilar, retain some information on the locality of the target state. 

\section{Strongly-correlated non-Abelian topological order - the Kitaev quantum-spin-liquid state}

\subsection{Majorana-fermion representation of the Kitaev model}

The pristine Kitaev model is an interacting spin model defined on the honeycomb lattice: 
\begin{equation}
  \hat H=\sum_{\langle i j\rangle}
  K_{ij}^{\alpha} \hat{\sigma}_{i}^{\alpha_{ij}} \hat{\sigma}_{j}^{\alpha_{ij}}
  \label{eq:pristine_Kitaev}
\end{equation}
where $K_{ij}^{\alpha}$ is the amplitude of the Kitaev interaction between sites $i$ and $j$ with the type of interaction $\alpha=x,y,z$ depending on the bond direction; see Fig. 3(b) inset in the main text. $\hat{\sigma}_i^{\alpha}$ are the Pauli spin operators.

Proposed in Kitaev's seminal work \cite{KITAEV2006}, the Kitaev model is a realization of nontrivial quantum spin liquid (QSL) states. On the material side, the exchange interactions between the $Ir^{4+}$ ions in a family of layered iridates $A_2LrO_3$ have been proposed to realize a Kitaev-type exchange Hamiltonian \cite{PhysRevLett.105.027204}. Following Kitaev's analytical solution \cite{KITAEV2006,mandal2020introduction}, we introduce four (real) Majorana fermions: $\hat{b}^x$, $\hat{b}^y$, $\hat{b}^z$ and $\hat{c}$, defined as $\hat{\sigma}_i^{\alpha} = i \hat{b}_i^{\alpha} \hat{c}_i$, then the original Hamiltonian is mapped to $\tilde{H}$ in an extended Hilbert space:
\begin{equation}
  \tilde{H}  =i\sum_{\langle i j\rangle}
  K_{ij}^{\alpha} \hat{u}_{ij} \hat{c}_i\hat{c}_j,
\end{equation}
where $\hat{u}^{\alpha}_{ij} \equiv i\hat{b}_i^{\alpha_{ij}}\hat{b}_j^{\alpha_{ij}}$. Because $\hat{u}_{ij}=-\hat{u}_{ji}$, we can specify $\hat{u}_{ij}^{\alpha}$ with a convention that $i$ is on the $A$ sublattice, and neglect the bond-dependent label $\alpha$ for simplicity. Clearly, 
\begin{equation}
    \left[\tilde{H}, \hat{u}_{ij}\right]=0, \quad
    \left[\hat{u}_{ij},\hat{u}_{kl}\right]=0,
\end{equation}
which indicates that $\hat{u}_{ij}$ are conserved quantities and split the extended Hilbert space into $\tilde{\mathcal{L}} = \oplus _u \tilde{\mathcal{L}}_u$, where $u$ represents a configuration of $\hat{u}_{ij}=\pm 1$, eigenvalues fixed by the requirement $(\hat{u}_{ij})^2=1$. The Hamiltonian $\tilde{H}_u$ is a tight-binding model of Majorana fermion $\hat{c}$, with hopping matrix elements coupled to a $\mathbf{Z}_2$ gauge field $\{\hat{u}_{ij}\}$. Therefore, we can factorize the extended Hilbert space as:
\begin{equation}
  |\tilde{\Phi} \rangle \equiv |\mathcal{M}_{\mathcal{G}} \rangle |\mathcal{G}\rangle
\end{equation}
where $|\mathcal{M}_{\mathcal{G}}\rangle$ is a many-body state of $\hat{c}$ in $\tilde{\mathcal{L}}_u$, determined by a certain $\mathbf{Z}_2$ gauge field configuration in $|\mathcal{G}\rangle$.

It is straightforward to obtain the eigenstates $|\mathcal{M}_{\mathcal{G}}\rangle$ with a canonical transformation $Q^{u}$ to new Majorana operators:
\begin{equation}
  (\hat{b}_1^{'},\hat{b}_1^{''},...\hat{b}_N^{'},\hat{b}_N^{''}) = (\hat{c}_1,...,\hat{c}_{2N})Q^{u},
\end{equation}
which brings $\tilde{H}_u$ to a decoupled form:
\begin{equation}
  \tilde{H}_u = \frac{i}{2}\sum_m \epsilon_m \hat{b}_m^{'}\hat{b}_m^{''} =  \sum_m \epsilon_m(2 \hat{n}_m -1),
\label{Eq:Hu_energy}
\end{equation}
where $\epsilon_m$ are the eigenvalues, $\hat{a}_m = 1/2(\hat{b}_m^{'}+i\hat{b}_m^{''})$ is a complex-fermion operator, and $\hat{n}_m = \hat{a}_m^{\dagger} \hat{a}_m$. since $\left[\tilde{H}_u, \hat{n}_m \right]=0$, we can neglect the operator hat of $\hat{n}_m$ and consider it as its eigenvalue $n_m=0,1$.

\subsection{Loop operators and topological degeneracy}

\begin{figure}
  \centering  
  \includegraphics[width=0.45\textwidth]{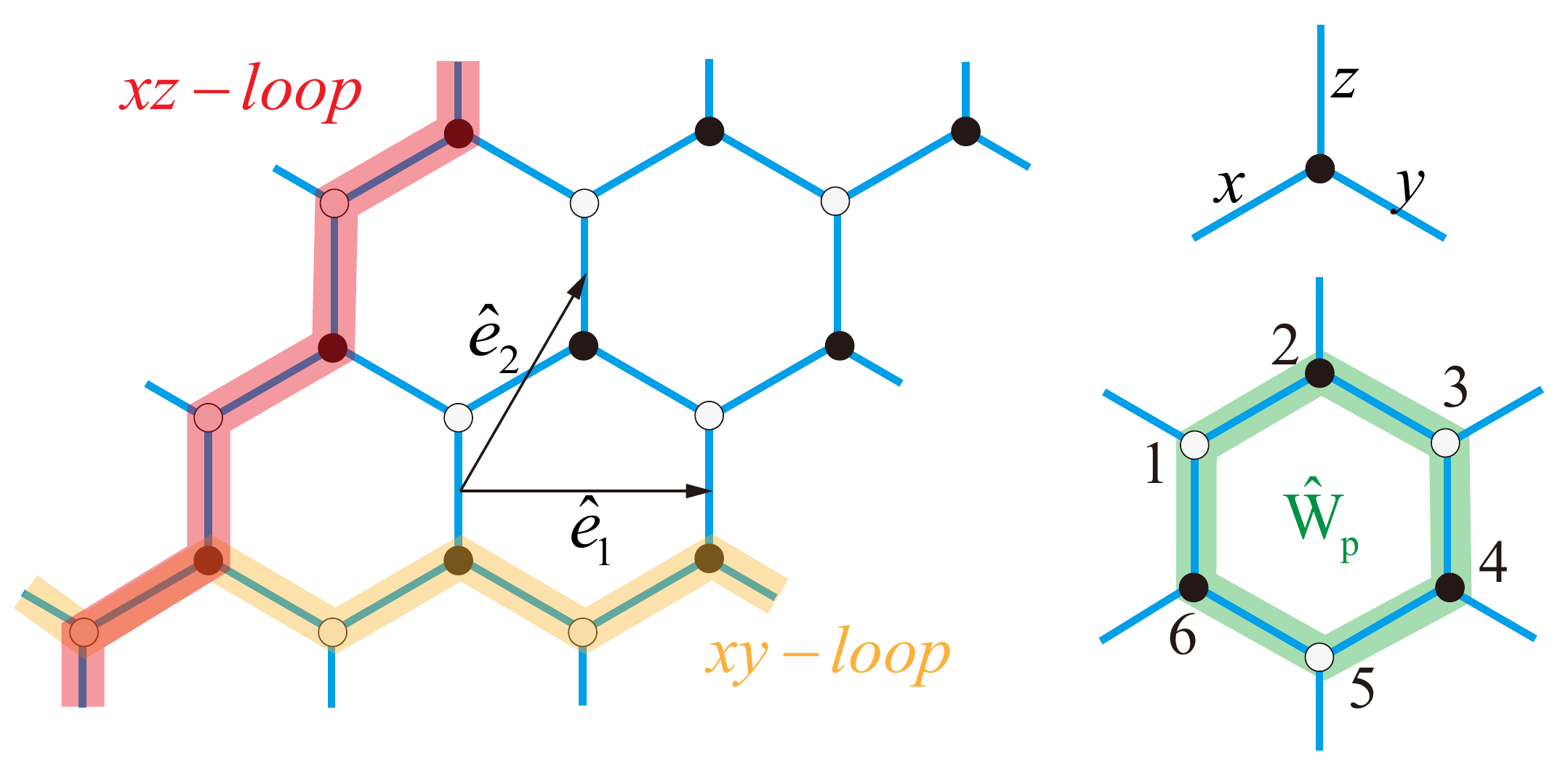} 
  \caption{Illustration for the loop operators $\hat{W}_{1,2}$ around a $3 \times 3$ system and ${\hat{W}_p}$ on a elementary plaquette on honeycomb lattice. $\hat{e}_1$ and $\hat{e}_2$ are the lattice vectors.}
  \label{fig:operator_illustration}
\end{figure}

For every closed loop $C$ on the lattice, the Kitaev model features a conserved quantity $\hat{W}_C$. It is convenient to introduce loop operators $\hat{W}_p$ on each elementary plaquette of the lattice, as demonstrated in Fig. \ref{fig:operator_illustration}:
\begin{equation}
  \hat{W}_p = \hat{\sigma}_1^x \hat{\sigma}_2^y \hat{\sigma}_3^z \hat{\sigma}_4^x \hat{\sigma}_5^y \hat{\sigma}_6^z,
\end{equation} 
whose eigenvalue is $W_p = \pm 1$. Also, the product over all plaquettes $\Pi_p \hat W_p = 1$; therefore, for a model with $2N$ spin, there are $N$ plaquettes yet only $N-1$ independent $\hat W_p$. For periodic boundary conditions, there are two addition global loop operators $\hat{W}_{1,2}$:
\begin{equation}
  \hat{W}_{1(2)} = -\prod_{xy(xz)-loop} \sigma_i^{z(y)},
\end{equation}
where the $xy-loop$ and $xz-loop$ are shown in Fig. \ref{fig:operator_illustration}. $\hat{W}_{1,2}$ cannot be expressed as a product of ${\hat{W}_p}$. The eigenvalues of $\hat{W}_{1(2)}$ are also $\pm 1$, introducing a plus/minus sign for the $\mathbf{Z}_2$ gauge field around the two respective loops of the torus \cite{PhysRevB.97.241110} and the periodic/anti-periodic boundary conditions for the Majorana fermion $\hat{c}$.

The loop operators $\hat{W}$ are gauge-invariant quantities and expressible through the bond variable ${\hat{u}_{ij}}$ in the Majorana-fermion representation. In the thermodynamic limit, the states with different $\hat W_{1,2}$ are locally indistinguishable, and the energy can only depend on the local $\{\hat{W}_p\}$. Therefore, there exist four states of identical $\{\hat{W}_p\}$ but different $\hat{W}_{1,2}$ as:
\begin{equation}
  \begin{aligned}
    &|\Phi_1 \rangle = |\mathcal{M_G}\rangle|\mathcal{G}(W_p);W_1 = 1,W_2 = 1 \rangle, \\
    &|\Phi_2 \rangle = |\mathcal{M_G}\rangle|\mathcal{G}(W_p);W_1 = -1,W_2 = 1 \rangle, \\
    &|\Phi_3 \rangle = |\mathcal{M_G}\rangle|\mathcal{G}(W_p);W_1 = 1,W_2 = -1 \rangle, \\
    &|\Phi_4 \rangle = |\mathcal{M_G}\rangle|\mathcal{G}(W_p);W_1 = -1,W_2 = -1 \rangle,
  \end{aligned}
  \label{Eq:topo state}
\end{equation}
which share the same energy in the thermodynamic limit and form a manifestation of topological degeneracy. However, we note that there will be energy splittings in a finite-size system, and not all four states may survive the physical projections - topics we will discuss in later sections. Despite these deviations, we follow the conventions and call these states connected via different $\hat{W}_{1,2}$ as ``topologically degenerate". 

For a translation-invariant system of sufficiently large size, the ground states are in the flux-free sector (i.e., all $\hat{W}_p = 1$) due to Lieb's theorem \cite{PhysRevLett.73.2158}. However, there are caveats for small system sizes: the ground states of certain small systems may not belong to the flux-free sector \cite{PhysRevB.92.014403}.

\subsection{Ground state of the flux-free sector}

For a translation-invariance system ($K^{\alpha}_{ij} =K^{\alpha}$), we consider one special flux-free scenario of $\hat{u}_{ij}=1$ for all links. The corresponding Hamiltonian $\tilde{H}_{u}$ can be solved via a Fourier transformation: 
\begin{equation}
  \hat{c}_{k,A(B)} = \sum_i \frac{1}{\sqrt{MN}}e^{-i\vec{k}\cdot \vec{r}} \hat{c}_{i,A(B)},
\end{equation}
where $A$, $B$ label the honeycomb sublattices. Subsequently, we obtain the Hamiltonian in the $k$-space:
\begin{equation}
  H_k = \sum_{k \in HBZ} \left(\hat{c}_{k,A}^{\dagger} \hat{c}_{k,B}^{\dagger} \right)
  \begin{pmatrix}
    0& i f^*(k) \\
    -i f(k) & 0 \\ 
  \end{pmatrix}
  \begin{pmatrix}
    \hat{c}_{k,A}\\
    \hat{c}_{k,B} \\ 
  \end{pmatrix},
\end{equation}
where `HBZ' stands for half Brillouin zone due to $\hat{c}_{k} = \hat{c}^{\dagger}_{-k}$, and $f(k)=2(K^x e^{i\vec{k}\cdot \hat{e}_1} +K^y e^{i\vec{k}\cdot \hat{e}_2} + K^z )$. $\hat{e}_{1,2}$ are the lattice vectors, see Fig. \ref{fig:operator_illustration}. Finally, we fully diagonalize the Hamiltonian:
\begin{equation}
  H_k = \sum_k \epsilon(\vec{k})(\eta_k^{\dagger}\eta_k - \xi_k^{\dagger} \xi_k),
\end{equation}
with \begin{equation}
  \begin{pmatrix}
    \hat{c}_{k,A}\\
    \hat{c}_{k,B} \\ 
  \end{pmatrix}=\frac{1}{\sqrt{2}}
  \begin{pmatrix}
    v_k &-v_k\\
    1 &1 \\
  \end{pmatrix}
  \begin{pmatrix}
    \eta_k\\
    \xi_k 
  \end{pmatrix},
\end{equation}
and $v_k = if^{*}_k/|f_k|$. The quasi-particle energy is $\epsilon(\vec{k}) = |f(\vec{k})|$ and the ground state in the flux-free is obtained by filling up all the negative energy states of quasi particles $\xi_k$:
\begin{equation}
  |GS\rangle = \Pi_{k,HBZ} \xi_k^{\dagger}|0\rangle,
\end{equation}
where $|0\rangle$ represents the vacuum state. Importantly, we have established the state $|GS\rangle$ in the extended Hilbert space, and need to verify that it survives the projection back to the physical Hilbert space of the original Kitaev model.

\subsection{Physical Projection}

For the Kitaev model with $2N$ spins, the Hilbert space is $2^{2N}$-dimensional. The extended Hilbert space $\tilde{\mathcal{L}}$ with four types of Majorana fermions is $4^{2N}$-dimensional and over-complete. Therefore, as detailed in Ref. \cite{PhysRevB.92.014403}, a projection $\hat{\mathcal{P}}$ to the physical subspace is necessary to avoid unphysical states:
\begin{equation}
  \hat{\mathcal{P}} = \hat{\mathcal{S}}\hat{\mathcal{P}}_0 
  = \hat{\mathcal{S}}\left( \frac{1+\hat{\mathcal{D}}}{2}\right),
\end{equation}
where $\hat{\mathcal{S}}$ symmetrizes all gauge-equivalent subspaces and $\hat{\mathcal{P}}_0$ is a projection operator onto the physical states. The vital $\hat{\mathcal{D}}$ operator takes the form \cite{PhysRevB.92.014403,PhysRevB.84.165414}:
\begin{equation}
  \hat{\mathcal{D}} = \hat{\pi} \cdot  (-1)^{\theta} \cdot \Pi_{\langle ij \rangle} \hat{u}_{ij} \cdot \det(Q^u),
\end{equation}
where $\hat{\pi}=(-1)^{n_a} $ is the parity of $\hat{a}$ fermions (in Eq. \ref{Eq:Hu_energy}). $\theta$ is the geometry factor associated with the system size and boundary condition. $\det(Q^u)=(-1)^{\gamma + N^2}$ for a translation-invariant system \cite{PhysRevB.84.165414}, where $\gamma$ is the number of reciprocal lattice vector $\mathbf{q}$ with $f(\mathbf{q})<0$ and $\pm \mathbf{q}$ equals (up to a reciprocal lattice vector).

Therefore, for a given $\{\hat{u}_{ij}\}$ sector, not all states $|\mathcal{M_G} \rangle \in \tilde{\mathcal{L}}_u$ are physical. The projection operator constrains the parity $\hat{\pi}$ of the Majorana fermion excitations in physical $|\mathcal{M_G} \rangle$.

\subsection{Kitaev model on a $3\times 3$ system}

Here, we present a detailed analysis of the Kitaev model in Eq. \ref{eq:pristine_Kitaev} on a $3\times 3$ system with periodic boundary conditions, see Fig. 3(b) inset in the main text. It is worth noting that when $K^{x}=K^y=K^z$, the Kitaev model has a $C_3$ symmetry of combined real-space and spin rotations. Thus, the topologically degenerate states $|\Phi_{1}\rangle$ and $|\Phi_{2,3,4}\rangle$ form a singlet and a triplet, following the irreducible representations of the $C_3$ group, respectively. Using exact diagonalization, we obtain the ground states of $K=\pm 1$ models, whose energies' and loop operators' expectation values are summarized in Tab. \ref{table:ED_result}. Both $K=\pm 1$ models have three-fold degenerate ground states with identical energy; however, their natures are very different: the $K=-1$ model's ground states are in the flux-free sector with different $\hat{W}_{1(2)}$, while the $K=+1$ model's ground states possess nontrivial $\hat{W}_p$ flux patterns, as we explain next. 

\begin{table}
  \renewcommand\arraystretch{1.2}
  \begin{tabular}{c|cccc p{0.4cm} c|cccc}
    \cline{1-5} \cline{7-11}
   \multicolumn{5}{c}{$K=1$ ground states} && \multicolumn{5}{c}{$K=-1$ ground states}\\ 
   \cline{1-5} \cline{7-11}
   & $W_1$ & $W_2$ & $\bar{W}_{p}$ & Energy && & $W_1$ & $W_2$ & $\bar{W}_{p}$ & Energy \\ 
   \cline{1-5} \cline{7-11}
  $|\chi_1\rangle$ &1  & 1 &-0.33  &-14.29 & &$|\phi_2\rangle$ &-1  & 1 &1  &-14.29\\
  $|\chi_2\rangle$ &1  &1  &-0.33  &-14.29 & &$|\phi_3\rangle$ &1  & -1 &1  &-14.29 \\
  $|\chi_3\rangle$ & 1 &1  &-0.33 &-14.29 &&$|\phi_4\rangle$ &-1  & -1 &1 &-14.29 \\
  \cline{1-5} \cline{7-11}
  \end{tabular}
  \caption{We obtain the ground states of the Kitaev model in Eq. \ref{eq:pristine_Kitaev} with $K=\pm 1$ on a $3\times 3$ system via exact diagonalization. Here are the energies' and loop operators' expectation values of the three-fold degenerate ground states. }
  \label{table:ED_result}
  \end{table}

For insights from the Majorana-fermion representation, we first examine the states $|\phi_{1,2,3,4} \rangle$ without Majorana fermion excitations in the flux-free sectors (all $\hat{W}_p=+1$). If the physical projector allows, these states are the ground states in the thermodynamic limit. However, there is no guarantee in small systems, and an energy comparison with other flux sectors is necessary: the energy is $\sim -13.39$ for $|\phi_{1} \rangle$ and $\sim -14.29$ for $|\phi_{2,3,4} \rangle$. 

Next, we evaluate the constraint from the physical projector, which is summarized in Tab. \ref{table:projection}: $\theta$ is even for a $3\times 3$ Kitaev model with the periodic boundary conditions; $\hat{\pi}=1$ with no Majorana fermion excitation; in the end, for $K=+1$, $\mathcal{P}_0 = 1$ for physical $|\phi_{1} \rangle$ and $\mathcal{P}_0 = 0$ for unphysical $|\phi_{2,3,4} \rangle$; on the contrary, for $K=-1$, which can be mapped to $K=+1$ with all $\hat u_{ij}$ receiving the opposite sign, $\mathcal{P}_0 = 0$ for unphysical $|\phi_{1} \rangle$ and $\mathcal{P}_0 = 1$ for physical $|\phi_{2,3,4} \rangle$. These $|\phi_{2,3,4} \rangle$ states are consistent with the three-fold flux-free topological degeneracy for $K=-1$ in Tab. \ref{table:ED_result}, while the flux-free singlet $|\phi_{1} \rangle$ state for $K=1$ is physical yet at a higher energy. 

\begin{table}
  \renewcommand\arraystretch{1.2}
  \begin{tabular}{p{0.6cm}<{\centering}|p{0.4cm}<{\centering}|p{0.9cm}<{\centering}|p{1.1cm}<{\centering}|p{0.9cm}<{\centering}|p{0.8cm}<{\centering}|p{0.8cm}<{\centering}|p{0.8cm}<{\centering}|p{0.8cm}<{\centering}}
  \hline
    &\multicolumn{4}{c|}{}& \multicolumn{2}{c|}{$K=1$}& \multicolumn{2}{c}{$K=-1$}\\
    \hline
    &$\hat{\pi}$&$(-1)^{\theta}$ &$(-1)^{N^2}$&$(-1)^{\gamma}$ & $\Pi u_{ij}$ & $\mathcal{P}_0$&$\Pi u_{ij}$ & $\mathcal{P}_0$\\ 
   \hline
  $|\phi_1\rangle$ &\multirow{4}{*}{1}  &\multirow{4}{*}{1}&\multirow{4}{*}{-1} & 1 &1  &1 &-1&0 \\
  $|\phi_2\rangle$ &  & && 1 &-1  &0 &1&1 \\
  $|\phi_3\rangle$ &  && & 1 &-1  &0 &1&1 \\
  $|\phi_4\rangle$ &  && & -1 &1  &0 &-1&1 \\
  \hline
  \end{tabular}
  \caption{Here list the factors of the projection operator of the four topologically degenerate states in Eq. \ref{Eq:topo state} in the flux-free sectors for the Kitaev model with $K=\pm 1$ on a $3 \times 3$ system. }
  \label{table:projection}
  \end{table}

\begin{figure}
    \centering  
    \includegraphics[width=0.35\textwidth]{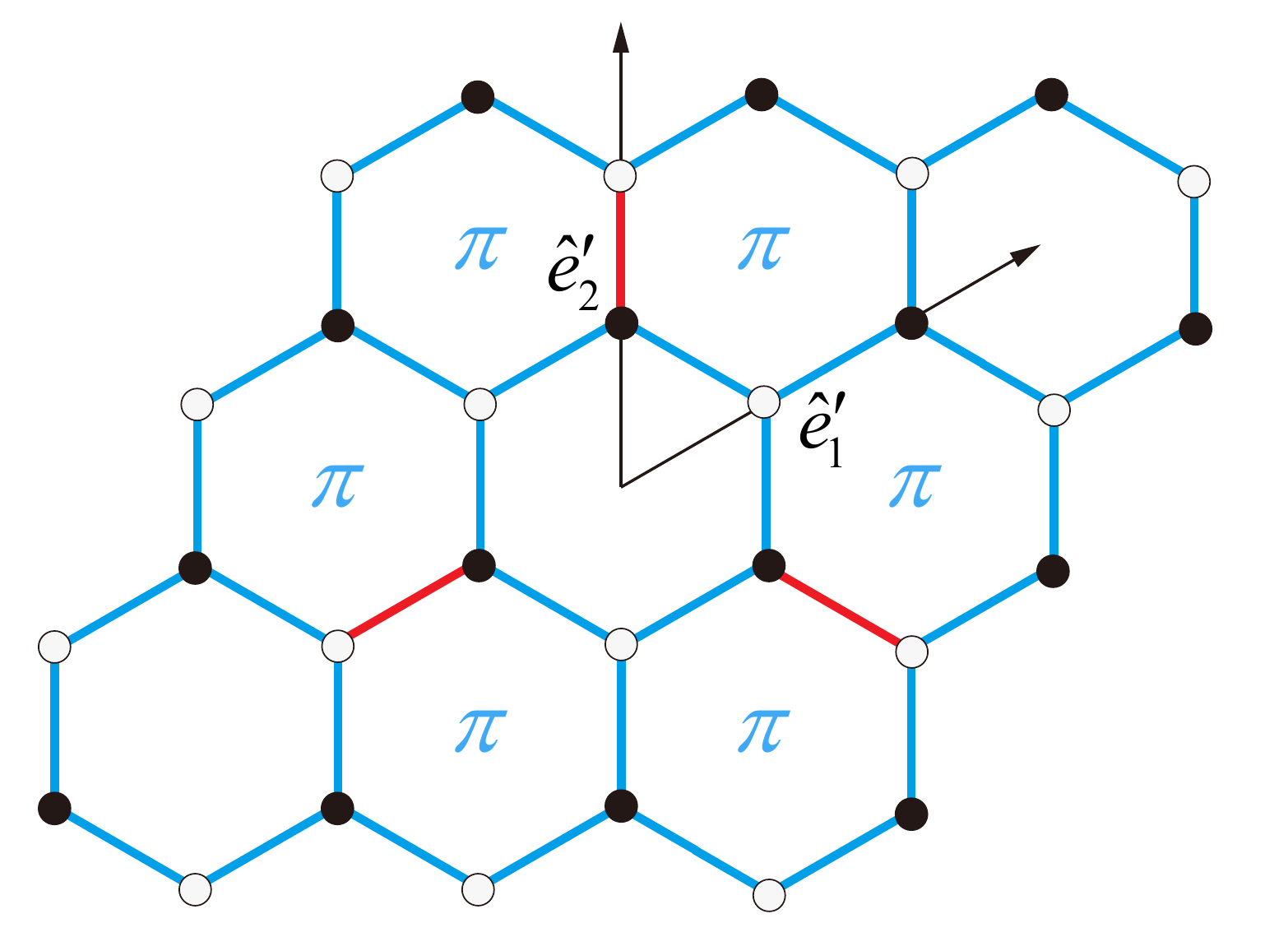} 
    \caption{The ground states of the $K=+1$ Kitaev model on a $3 \times 3$ system are not in the flux-free sector but possess six $\hat{W}_p = -1$ out of the nine plaquettes, indicated by the $\pi$ symbols. $\hat{e}'_1$ and $\hat{e}'_2$ are the new lattice vectors of the enlarged unit cell, and the red bonds correspond to $\hat u_{ij}=-1$ that realize the flux pattern. }
    \label{fig:K1_illustration}
\end{figure}

What remains to be explained is the nature of the ground states $|\chi_{1,2,3}\rangle$ of the $K=+1$ model. It turns out that they are not in the flux-free sector but possess six out of nine plaquettes with $\hat{W}_p = -1$; see Fig. \ref{fig:K1_illustration} for illustration. The flux pattern enlarges the unit cell three times, leading to a three-fold ground-state degeneracy from the translation symmetry breaking instead of topology. The resulting energy and loop operator expectation values are fully consistent with the exact diagonalization results in Tab. \ref{table:ED_result}.

\subsection{Energy flow for $3\times 3$ nearest-neighbor spin model}

In the main text, we consider the following nearest-neighbor spin model with a dominant Kitaev term, also known as the generalized Kitaev model:
\begin{equation}
  \hat H=\sum_{\langle i j\rangle \in \alpha \beta(\gamma)}
  \left[J_{ij} \vec{S}_{i} \cdot \vec{S}_{j}+K_{ij} S_{i}^{\gamma} S_{j}^{\gamma}+\Gamma_{ij}\left(S_{i}^{\alpha} S_{j}^{\beta}+S_{i}^{\beta} S_{j}^{\alpha}\right)\right],
  \label{eq:KitaevHam}
  \end{equation}
which coincides with Eq. 12 in the main text. 

With the addition of the Heisenberg and the off-diagonal interactions, we can no longer solve the model analytically in the Majorana-fermion representation. This reduces the number of conserved quantities and increases the complexity and generalizability of the model. On the other hand, we can trace the adiabatic flows of the lowest eigenstates from the pristine Kitaev model ($\zeta  = 0$) to more general models ($\zeta  = 1$) using exact diagonalization:
\begin{equation}
  \hat H=\sum
  \left[J_{ij} \vec{S}_{i} \cdot \vec{S}_{j}+\zeta \left\{K_{ij} S_{i}^{\gamma} S_{j}^{\gamma}+\Gamma_{ij}\left(S_{i}^{\alpha} S_{j}^{\beta}+S_{i}^{\beta} S_{j}^{\alpha}\right) \right\}\right], 
  \end{equation}
where we interpolate $\zeta$ between $[0, 1]$ with sufficiently fine resolution. The connectivity allows us to interpret the topological sectors of the $\zeta = 1$ general states from the corresponding exactly solvable $\zeta=0$ limit.

\begin{figure}
  \centering  
  \includegraphics[width=0.45\textwidth]{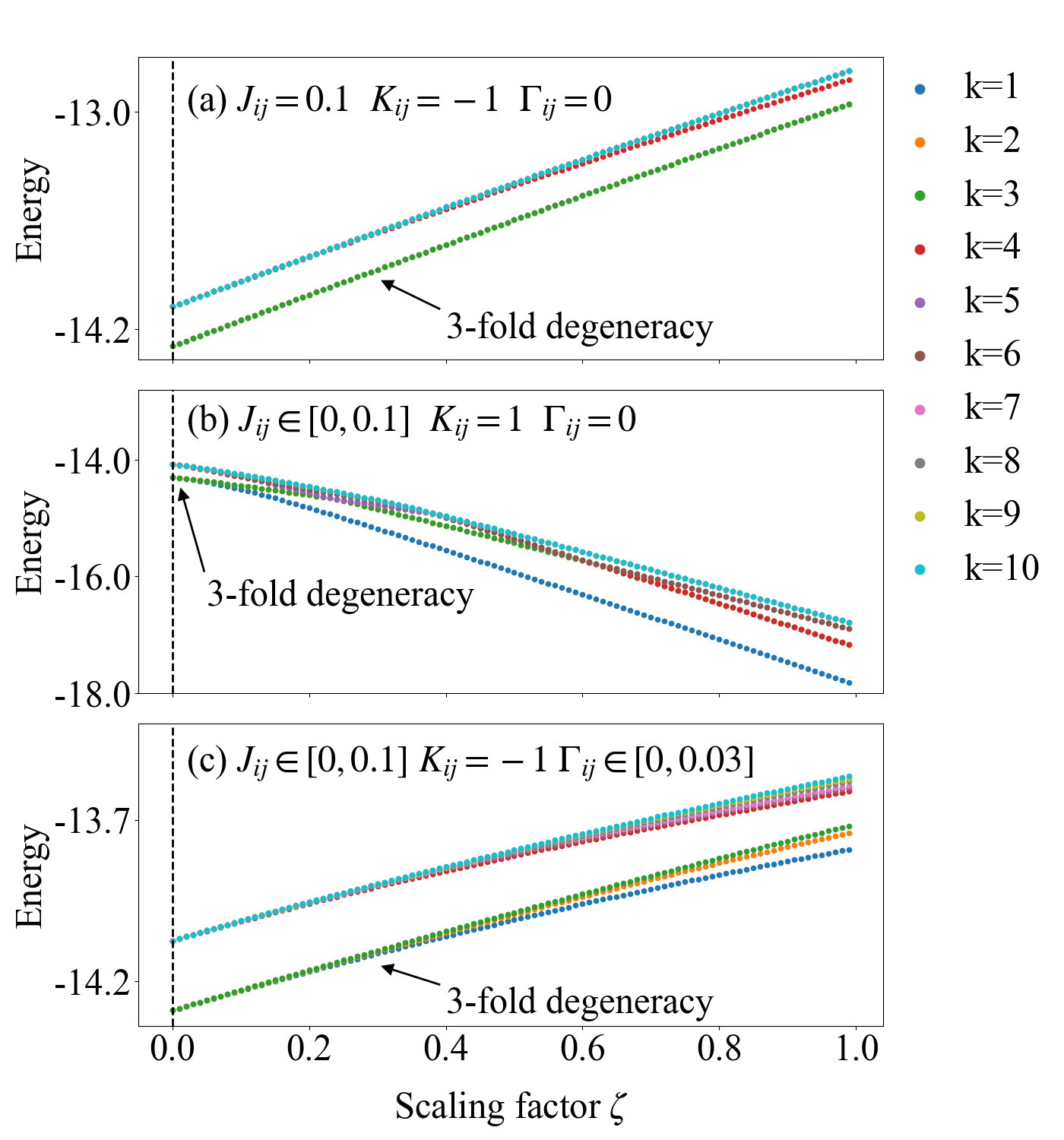} \caption{Adiabatic flow of the lowest eigenstates from the pristine Kitaev model to the generalized Kitaev model offers insights into the latter's quantum many-body states. The target models on the right hand side with full tuning $\zeta=1$ are (a) $K_{ij}=-1$, $J_{ij}=0.1$, $\Gamma_{ij}=0$ in connection to Fig. 3(a) in the main text, (b) $K_{ij}=+1$, $J_{ij}\in[0,0.1]$, $\Gamma_{ij}=0$ in connection to Fig. 3(b) in the main text, and (c) $K_{ij}=-1$, $J_{ij} \in [0,0.1]$, $ \Gamma_{ij} \in [0,0.03]$ in connection to Fig. 4 in the main text. $k$ is the index for eigenstates.}
  \label{fig:energy_flow}
\end{figure}

In Fig. \ref{fig:energy_flow}, we show the adiabatic flows to the three $(J, K, \Gamma)$ parameter settings used in Fig. 3 and Fig. 4 in the main text.
In Figs. \ref{fig:energy_flow}(a) and (c), the three-fold topological degeneracy robustly holds and separates from the rest of the spectrum with a persisting gap, and the expectation values of the loop operators $\hat{W}_{1,2}$ is also close to $\pm 1$, consistent with the dominant Kitaev interaction $K_{ij}=-1$. Indeed, we choose such $K_{ij}=-1$ models in the main text to probe the impact of topological degeneracy on our strategy. On the other hand, consistent with the non-topological nature of the $K_{ij}=+1$ ground states, the three-fold degeneracy splits and leaves a unique ground state in Fig. \ref{fig:energy_flow}(b), simplifying our discussions on the consequences of insufficient observables. 

\subsection{Principal component analysis as a contingency plan for insufficient observables}

In the main text, we discuss that when the observables do not provide sufficient distinctions, different MLE states with minimum measurement energy may exist simultaneously, especially when we employ different initialization $\hat{H}_0$. The primary solution is, of course, to incorporate quantum measurements on additional observables that can bring in new information. When the ambiguity is not overwhelming, however, there is a contingency plan for a common ground state $|\Phi_{CG}\rangle$ via principal component analysis (PCA) between the MLE states $|\Phi_{gs;trial}\rangle$ from multiple trials.

PCA is a standard statistical method for dimension reduction. It identifies orthogonal directions in the data space, called principal components, along which the data exhibits the largest variance and thus contains the most distinctive information \cite{MLbook_ap}. In our case, each $|\Phi_{gs;trial}\rangle$ contains parts that the quantum measurements can pinpoint, as well as parts that are ambiguous and differ from trial to trial due to a lack of information. Our contingency plan focuses on the former parts, which make good use of the quantum measurements and hopefully close in on the target state, especially in a higher-dimensional Hilbert space. We can carry out such extractions via PCA. In practice, we start with the correlation matrix between the $|\Phi_{gs;trial}\rangle$ states:
\begin{equation}
  X=
  \begin{bmatrix}
    \langle \Phi_1 | \Phi_1 \rangle &
     \langle \Phi_1 | \Phi_2 \rangle & 
     \cdots & 
     \langle \Phi_1 | \Phi_n \rangle\\
     \langle \Phi_2 | \Phi_1 \rangle &
     \langle \Phi_2 | \Phi_2 \rangle & 
     \cdots & 
     \langle \Phi_2 | \Phi_n \rangle\\
     \vdots & \vdots & \ddots & \vdots\\
     \langle \Phi_n | \Phi_1 \rangle &
     \langle \Phi_n | \Phi_2 \rangle & 
     \cdots & 
     \langle \Phi_n | \Phi_n \rangle\\
  \end{bmatrix}.
\end{equation}

$X$ is Hermitian and diagonalizable via the eigenvalue problem $Xv_i = \lambda_i v_i$, where $\lambda_i \geq 0$ is the eigenvalue and $v_i$ is the corresponding eigenstate. The first principal component is the direction of $v_{PC}=\arg\max_{\lambda_i} v_i$, and therefore, the corresponding state is:
\begin{equation}
  |\Phi_{CG} \rangle = \frac{1}{\sqrt{\mathcal{R}}}\left(v_{PC}\cdot  
  \begin{bmatrix}
    |\Phi_1 \rangle\\
    |\Phi_2 \rangle\\
    \vdots\\
    |\Phi_n \rangle
  \end{bmatrix}
  \right),
\end{equation}
where $\mathcal{R}$ is a normalization constant. An example of the performance of such $|\Phi_{CG} \rangle$ is the red dashed line in Fig. 3(b) in the main text.

\subsection{Beyond the topological ground states}

In Fig. 4 inset in the main text, we show the fidelity between the three lowest-energy states of $\hat{H}_0$ at convergence and the topological degeneracy around the target state. The results indicate that we can obtain the other two topologically degenerate states as by-products starting from a single topological ground state. Therefore, our strategy successfully extracts the topological information in the target ground state. Here, we further show that not only the ground-state manifold but also a series of additional low-lying excited states emerge from our strategy based upon the quantum measurements of a single topological ground state. Such generalizability does not carry over to a non-topological quantum many-body state, e.g., the random long-range fermion model, where our strategy identifies the target state and only the target state. 

\begin{figure}
  \centering  
  \includegraphics[width=0.45\textwidth]{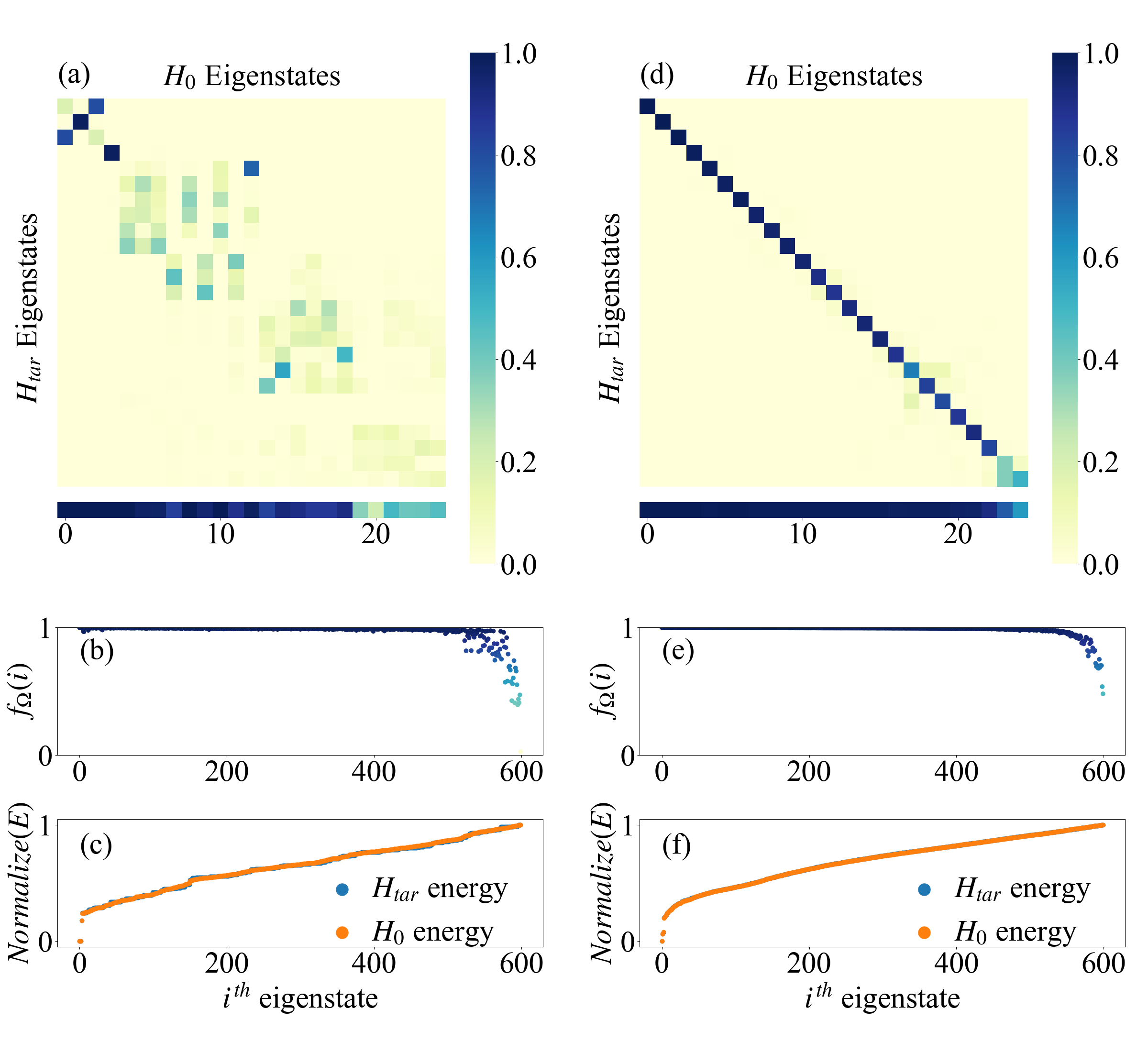} \caption{(a-c) The results are based upon observables $\sigma^\lambda_i \sigma^\lambda_j$, $\lambda=x, y, z$ from target ground state of Eq. \ref{eq:KitaevHam} with $K_{ij}=-1$, $J_{ij}=0.1$, and $\Gamma_{ij}=0$ in connection with Fig. 3(a) in the main text and Fig. \ref{fig:energy_flow}(a). (d-f) The results are based upon observables $\sigma^{\hat{n}}_i \sigma^{\hat{n}}_j$ for random $\hat{n}$ directions from target ground state of Eq. \ref{eq:KitaevHam} with $K_{ij}=-1$, $J_{ij}\in [0,0.1]$, and $\Gamma_{ij} \in [0,0.03]$ in connection with Fig. 4 in the main text and Fig. \ref{fig:energy_flow}(c). (a) and (d) are the fidelity between the 25 lowest eigenstates of $\hat H_0$ at convergence and the first 25 states of the target topological model. (b) and (e) ((c) and (f)) are the presenting weights $f_{\Omega}$ (normalized spectrums) of the first 600 eigenstates.}
  \label{fig:more states}
\end{figure}

We summarize the relevant results in Fig. \ref{fig:more states}, which suggests that our strategy extracts many low-lying states of the original topological model based upon the quantum measurements on a single topological ground state. The $(i,j)$ element in Fig. \ref{fig:more states}(a,d) shows the fidelity between the $\rm i^{th}$ eigenstate of target model and $\rm j^{th}$ eigenstate of $\hat{H}_0$. The clear block structures of the fidelity data in Fig. \ref{fig:more states}(a) is due to the corresponding spectral degeneracy. We also define a presenting weight: 
\begin{equation}
  f_{\Omega}(i) = \sum_{x=0}^{cutoff} |\langle\Phi_t^x|\Phi_0^i\rangle |^2,
\end{equation}
which describes the presence of the $i^{th}$ eigenstate of $\hat H_0$ in the eigenstate subspace of the target topological model $\hat H_{tar}$ and is demonstrated in Fig. \ref{fig:more states}(b,e). $\Phi_t^i$ ($\Phi_0^i$) is the $i^{th}$ eigenstate of $\hat{H}_{tar}$ ($\hat{H}_0$). In addition, we plot the normalized spectra $(\text{E}-\min(\text{E}))/(\max(\text{E})-\min(\text{E}))$ for both $\hat{H}_{tar}$ and $\hat{H}_0$ in Fig. \ref{fig:more states}(c,f). 

These results reveal close and deep connections between the original topological model and the iteration model $\hat{H}_0$ in our strategy, especially from a low-energy theory standpoint. Interestingly, the latter roots simply from a single topological ground state, which implies much topological information, e.g., topological quasiparticle excitations and beyond, is available state-wise. Such conjecture requires more rigorous and targeted research in the future. Again, we emphasize that there is no such emergence for the random long-range fermion models.

\section{An example of Haar random quantum states without a-priori knowledge}

Compared to the conventional maximum likelihood estimation approaches (MLE), one of the advantages of our strategy is the utilization of available a-priori knowledge of the target state. For example, the target states in random long-range fermion models are fermion direct-product states; therefore, it suffices to consider only two-point correlators. Also, the target states in generalized Kitaev models are the ground states of local quantum spin models; therefore, the locality (and Area law) entitles us to focus on $k$-local operators. These choices also direct the search to the relevant parts of the Hilbert space, where we know the target states settle - this is more efficient, and we can reach much larger systems. Such physical a-priori knowledge is widely available in condensed matter and quantum information experiments, yet most approaches previously had difficulty taking advantage of them. 

On the other hand, our strategy also applies to general quantum states without any a-priori knowledge. We advise initially focusing on lower-order and local operators for higher relevance and lower cost. Whatever the case, our strategy can locate the MLE state and tell whether the information is incomplete, and decide whether to resort to additional, potentially higher orders and more nonlocal operators. Progressively, we can obtain more information and nail down the target quantum state more accurately. Notably, for a fully global and general quantum state, the resulting measurement Hamiltonian may consist of nonlocal high-order operators, making solutions difficult other than brute-force exact diagonalization, which will restrict the feasible system to approximately the state-of-art size in conventional MLE approaches \cite{PhysRevA.95.062336}.

Here, we take the uniform Haar random quantum states \cite{Biswas_2021} on an $n$-qubit system as an example:
\begin{equation}
  |\psi\rangle = \sum_{j=1}^{2^n}(c_{1j}+ic_{2j})|j\rangle,
\end{equation}
where $|j\rangle$ represents the $j^{th}$ orthonormal basis vector in the $n$-qubit system's $2^n$-dimensional Hilbert space. $c_{1j}$ and $c_{2j}$ are random real numbers following Gaussian distribution with vanishing mean and finite variance. We also normalize the generated many-body state as $\langle \psi |\psi\rangle=1$.

Given such a target state $|\psi\rangle$ without any useful a-priori knowledge, we consider measurements of random $k$-local Pauli operators $\hat{O}=\otimes_i o_i$, $o_i \in \{1, \sigma_x, \sigma_y, \sigma_z \}$. We randomly choose $5\%$ of the observables and add them to the set of measurement operators in an increasing order of $k$ step by step. With such an increasing set of measurement operators, we repeatedly implement our strategy and monitor the performance. Here, we take the number of measurements $N_{\hat{O}} \to \infty$ to suppress the effects of quantum fluctuations for simplicity.

\begin{figure}
  \centering
  \includegraphics[width=0.4\textwidth]{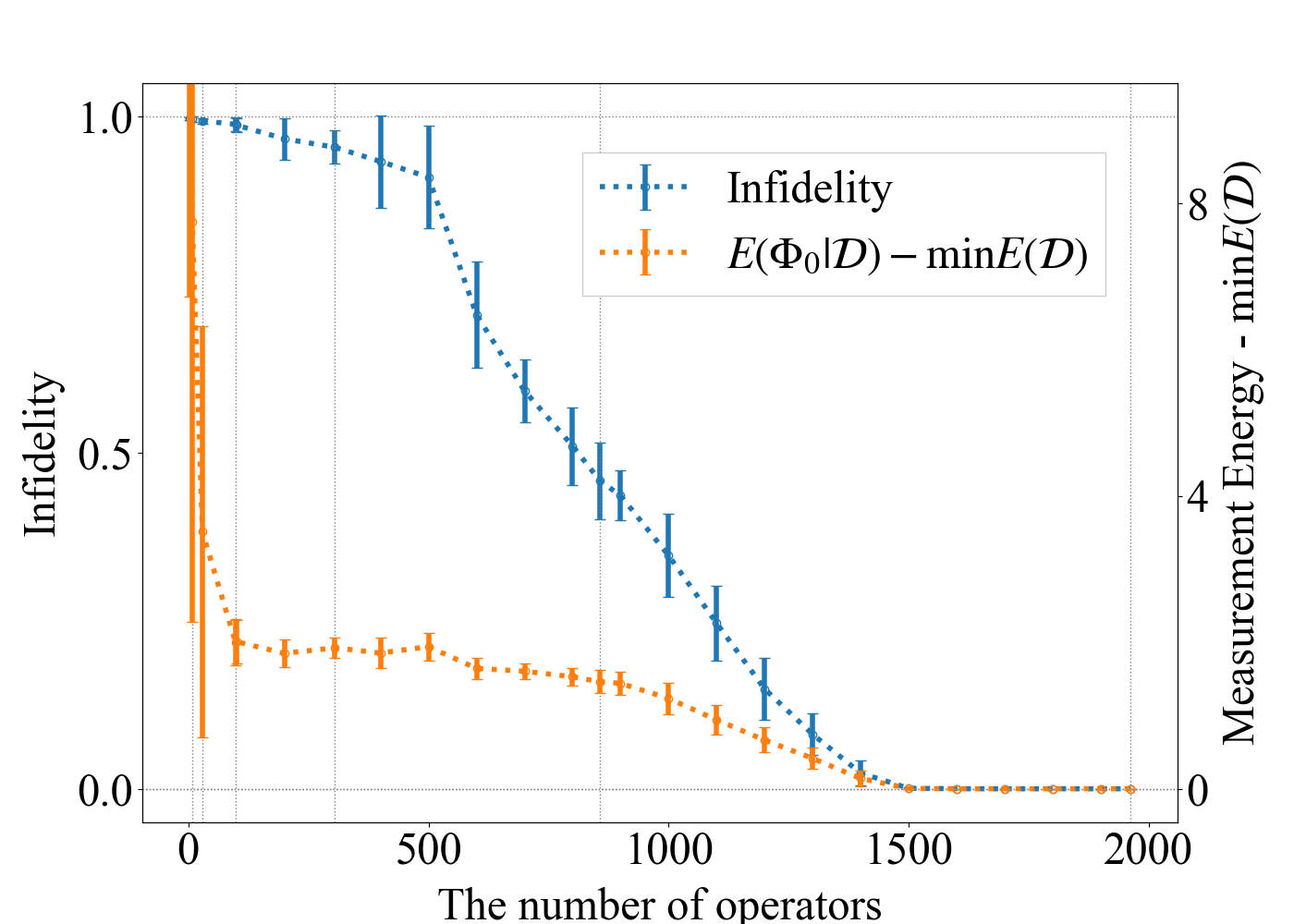}
  \caption{As the number and nonlocality of the measurement operators increase, the obtained MLE states asymptotically approach the target quantum states - Haar random quantum states, and the measurement energy converges to its lower bound. $n=8$. The vertical dotted lines denote the inclusion of $k=3, 4, 5, 6, 7, 8$, respectively, and the error bars are based on ten trials on different target states.  }\label{fig:Haar state}
\end{figure}

We summarize the results in Fig. \ref{fig:Haar state}. Our strategy always nails the MLE states, which only get the hang of the target states with local observables, and approach the target states asymptotically with more non-local observables included. Meanwhile, we resort to exact diagonalization for evaluations of $\hat H_0$, which becomes the major cost and limitation to our system sizes. Nevertheless, we can consistently achieve a state-of-art fidelity of $\sim 0.9999$ for 12-qubit Haar random states ($n=12$ and measurements on $1 \%$ of the Pauli operators). Further, we can still achieve a good fidelity $\sim 0.9975$ with a finite number of measurements ($N_{\hat{O}}=10^5$ measurements on $1 \%$ of the Pauli operators for a 12-qubit Haar random state) and a rapid convergence within 5 iterations.

\section{Ground state of random transverse field Ising model on large systems}

\begin{figure}
  \centering
  \includegraphics[width=0.4\textwidth]{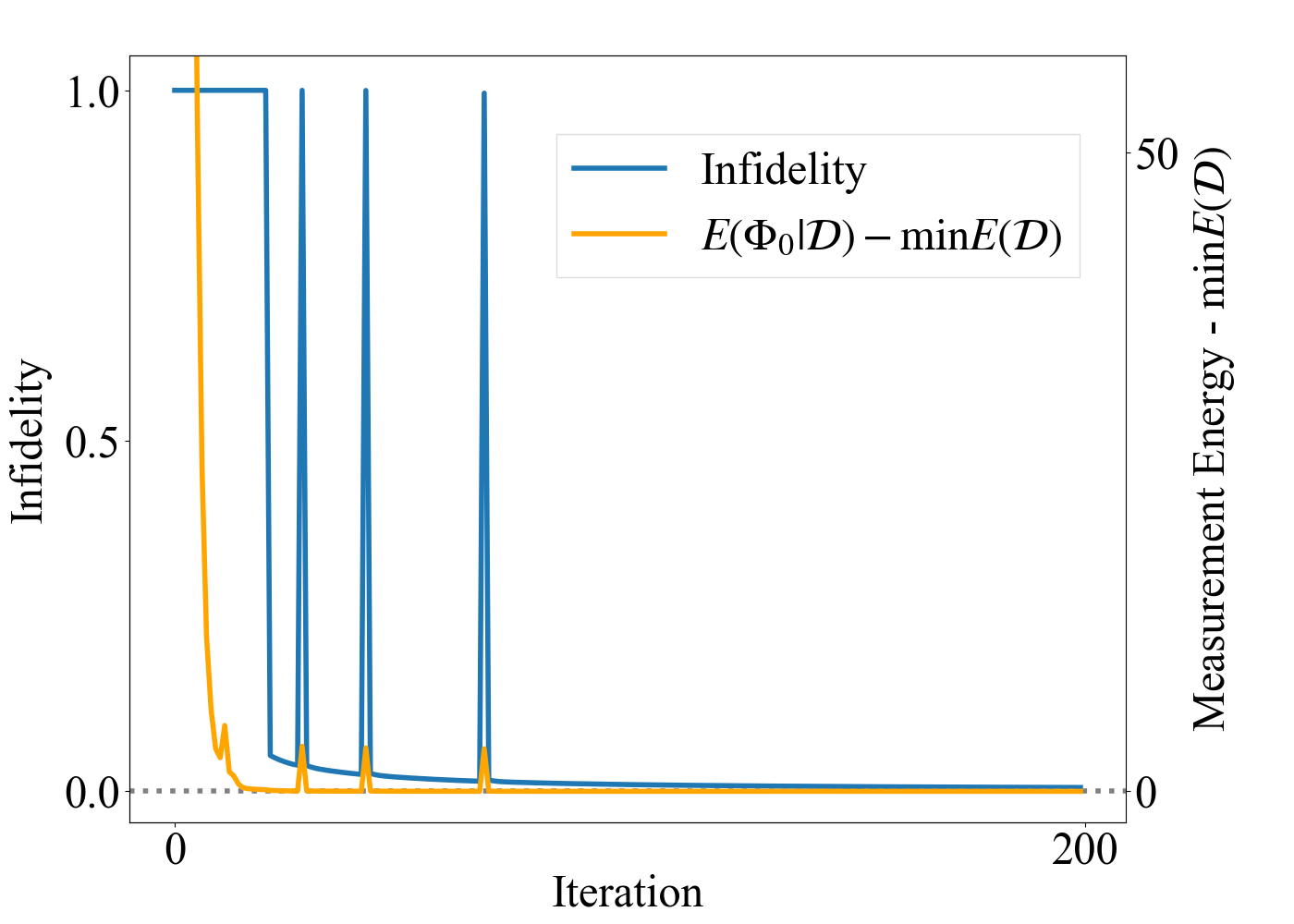}
  \caption{The iteration state quickly converges to the target quantum state, the ground state of a random TFIM on a large system of $N=200$, and the average measurement energy converges to its lower bound. The final infidelity is $\sim 5 \times 10^{-3}$. }\label{fig:TFIM state}
\end{figure}

We have proposed an efficient strategy to determine the MLE states, which may work together straightforwardly with a variety of numerical quantum many-body algorithms, such as exact diagonalization, density-matrix renormalization group (DMRG) \cite{MPS1992, DMRG2005}, quantum Monte Carlo methods \cite{QMCreview2001, Troyer2005}, and neural network states \cite{Carleo2016}. These efficient algorithms allow us to handle quantum states on unprecedented system sizes. 

For example, we consider the ground state of a random 1D transverse field Ising model (TFIM) on a larger system with $N=200$ sites:
\begin{equation}
 \hat H = \sum_{
    \langle i,j\rangle} J_{ij} \sigma_i^z \sigma_j^z -\sum_i h_i \sigma_i^x,
\end{equation}
where we randomly set the nearest-neighbor Ising interaction $J_{ij}>0$ and the transverse field $h_i$. We consider an infinite number of measurements for the observables $\hat{\sigma}_i^z \hat{\sigma}_j^z$ on every bond and $\hat{\sigma}_i^x$ on every site. 

We combine our strategy with DMRG, an efficient quantum many-body ansatz based upon the tensor-network-state representation readily applicable to such a large system size, and summarize the results in Fig. \ref{fig:TFIM state}. We achieve a satisfactory convergence with an infidelity of $\sim 5 \times 10^{-3}$. 

\section{Optimization hyper-parameters: $\hat{H}_0$ initialization and gradient-descent rate $\lambda$}

We point out in the main text that we can achieve the maximum likelihood estimate (MLE) state via solving the self-consistent equations:
\begin{equation}
  \hat{H}_{eff} = \sum_{\hat{O}}\sum_{\tau} N_{\hat{O}}\alpha_{\tau}  \hat{P}_{\tau}, \; \alpha_{\tau} = -\frac{f_{\tau}}{\langle \Phi_{gs}| \hat{P}_{\tau} | \Phi_{gs} \rangle}.
  \label{eq:alpha}
\end{equation}
Note that the MLE state has $\langle \hat{P}_{\tau}\rangle$ close to $f_{\tau}$ and renders $\hat H_{eff}$ close to a trivial constant after summing over $\tau$. Therefore, Eq. \ref{eq:alpha} is intractable via conventional iterations, which becomes unstable and bounces around, preventing a good convergence. 

We receive intuitions from the optimizations of artificial neural networks (ANNs) in supervised machine learning, where the ANN's parameters are randomly initialized and updated step-wise, $w\rightarrow w - \lambda \partial_w C$ following its gradients with respect to a cost function $C$ until convergence. $\lambda$ is a rate-controlling parameter, also called the step size / descent rate. Similarly, we treat $\hat{H}_{eff}$ as a gradient and update $\hat{H}_0$ in iterations:
\begin{equation}   
  \hat{H}_0 \to \hat{H}_0 + \lambda \hat{H}_{eff}.  \label{eq:update}
\end{equation}
Such a setting has the following advantages: (1) $\hat{H}_{eff}$'s positive contributions to the optimization accumulate in $\hat{H}_{0}$, while its small errors and noises average out over iterations. (2) The smallness of $\hat{H}_{eff}$ shall not impact $\hat{H}_{0}$'s stability as we get closer to the target state; in fact, such small gradients naturally allow finer resolutions. The initialization $\hat{H}_0$ and the gradient-descent rate $\lambda$ are also called the hyper-parameters of the optimization.

\begin{figure}
  \centering  
  \includegraphics[width=0.45\textwidth]{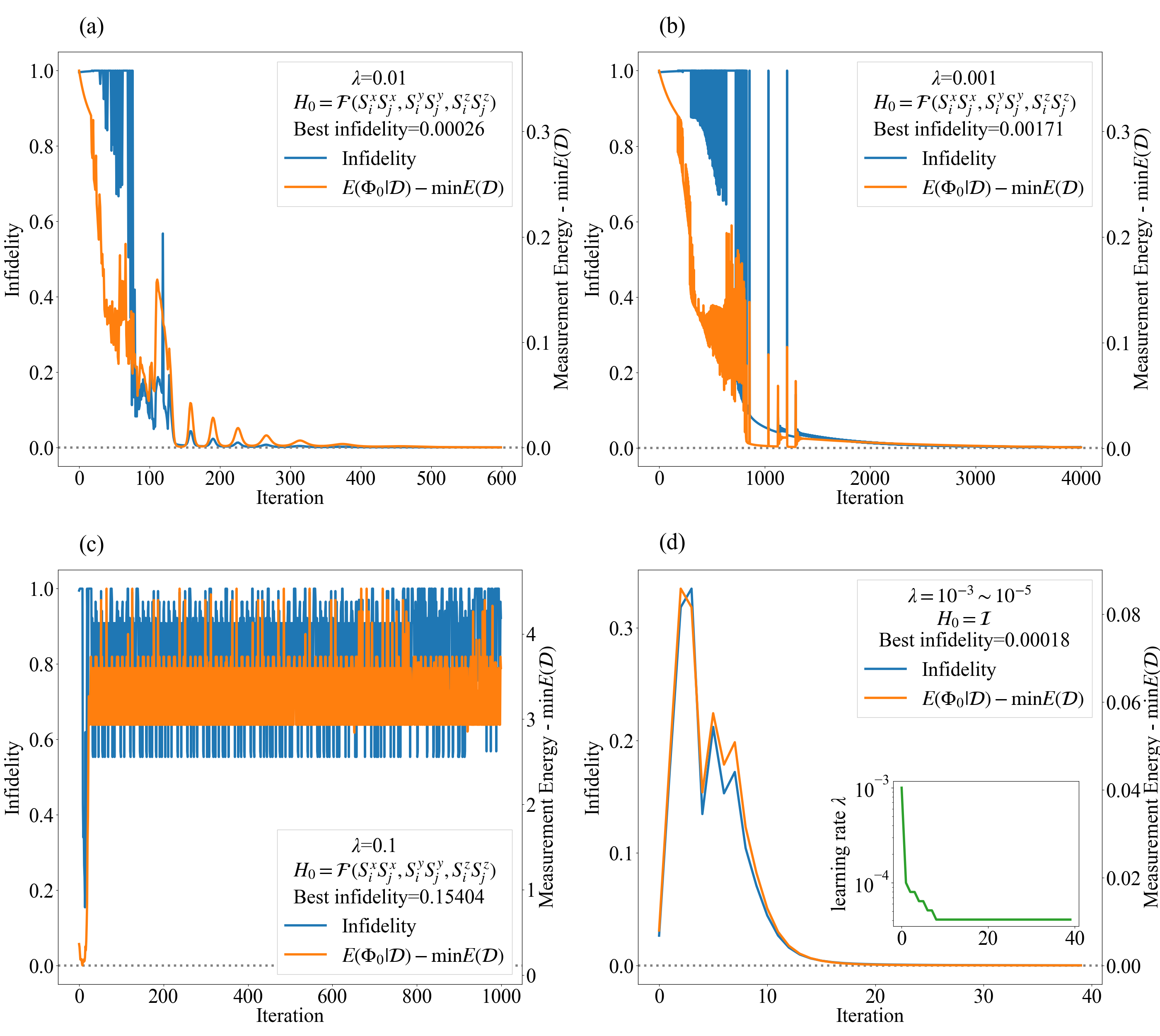} \caption{We compare the iterations and convergences of different hyper-parameter settings following Eq. \ref{eq:update}. The target state is the ground state of the Kitaev model in Eq. 12 in the main text with $K_{ij}=-1$, $J_{ij}=0.1$, and $2\times 2$ system size. (a-c) With random initializations $H_0$ with $S^x_iS^x_j$, $S^y_iS^y_j$, and $S^z_i S^z_j$ over the nearest neighbors and random weights $\in [-1,1]$, the iterations (a) converge quickly within 600 steps with a suitable $\lambda=0.01$, (b) need 4000 steps with a $\lambda=0.001$ too small, and (c) bounce around without convergence  with a  $\lambda=0.1$ too large. (d) The iterations converge rapidly within 40 steps with a continuously adjusted step size $\lambda=10^{-3} \sim 10^{-5}$ and an identity matrix as the initialization $H_0$.}
  \label{fig:hyperpara}
\end{figure}

Similar to the training of ANN in supervised machine learning, 
the gradient-descent rate $\lambda$ has a significant impact on the iterations, and a suitable choice renders a rapid and accurate convergence to the target state, as in Fig. \ref{fig:hyperpara}. Generally speaking, a larger $\lambda$ leads to faster convergence and vice versa. It also helps to avoid various local minimums in the search space. On the contrary, a smaller $\lambda$ helps avoid unstable iterations that bounce around the target. It is helpful to schedule a progressive decrease of $\lambda$ so that the procedure is more efficient in the initial stage and gradually slows down for the fine-tuning at convergence. Besides, we may further improve the stability of $\hat{H}_0$ by limiting the divergence of $\hat{H}_{eff}$: we scale $\hat{H}_{eff}$ so that the range of its eigenvalues does not exceed 1, which we implement in our random long-range fermion model examples.

The initialization $\hat{H}_0$ matters as well. Usually, a trivial initial $\hat{H}_0$ that equals an identity matrix helps a rapid convergence, demonstrated in Fig. \ref{fig:hyperpara}. We note that even a trivial initial $\hat{H}_0$ is not the same as the iterative approach based solely on $\hat{H}_{eff}$. Here, the accumulations of $\lambda \hat{H}_{eff}$ lead to some drastic progress at first and smoother performance in the later stages, which could use some carefully scheduled gradient-descent rate $\lambda$. We mainly apply such initialization in our random long-range fermion model examples. In some cases where we need other initializations, e.g., to examine whether we may receive contradicting MLE states, it helps to keep the initialization $\hat{H}_0$ physical. We note that $\hat{H}_0$ with random matrix elements is usually unphysical as its operators are commonly non-local and of high orders. A more proper initialization $\hat{H}_0$ should be random weights on operators that are local and connected to the observables so that $\hat{H}_{eff}$ can subsequently amend it. For example, the observables in Fig. \ref{fig:hyperpara} are $S^x_iS^x_j,S^y_iS^y_j,S^z_i S^z_j$; therefore, it is natural to set  the initialization $\hat{H}_0$ as the sum of these terms with random weights, denoted as the function $\mathcal{F}$.

It is also important to devise a way to monitor the suitability of hyper-parameters, especially for realistic scenarios where fidelity is generally unavailable. Instead, we can rely on the measurement energy $E(\Phi|\mathcal{D}) = - \sum_{\hat{O}} \sum_{\tau}N_{\hat O}  f_{\tau} \log{\langle \Phi|\hat{P}_{\tau} |\Phi \rangle}$ at each iteration, which is readily available. Its relative stance to the lower bound $\min E(\mathcal{D})=- \sum_{\hat{O}} \sum_{\tau} N_{\hat O} f_{\tau} \log f_{\tau}$ shows how well the iteration state $|\Phi_0 \rangle$ fits the given quantum-measurement data and whether the fitting is improving or saturating. Commonly, the ideal case of $E(\Phi_0|\mathcal{D}) \rightarrow \min E(\mathcal{D})$ will not occur unless we have a sufficiently large number of quantum measurements to suppress fluctuations fully; nevertheless, we can still use $E(\Phi_0|\mathcal{D})-\min E(\mathcal{D})$ to evaluate the iterative progress.

Our discussions on hyper-parameters offer only preliminary and qualitative guidance. Like the hyper-parameters in ANN and supervised machine learning, it is often a complicated and situation-dependent matter still under active research. On the other hand, some basic attempts over the broad directions, such as the different scales of $\lambda$ and the different types of initialization $\hat{H}_0$, can already turn out highly rewarding.

\bibliography{refs}